\documentclass[longauth, traditabstract]{aa} 
\usepackage{graphicx}
\usepackage{txfonts}
\usepackage{fixltx2e}
\usepackage{amssymb}
\usepackage{natbib}
\usepackage{rotating}
\usepackage{sidecap}
\bibpunct{(}{)}{;}{a}{}{,}

\begin{document}
\title{\textit{Herschel}-HIFI observations of high-\textit{J} CO and isotopologues in star-forming 
  regions: from low- to high-mass\thanks{Herschel is an ESA space observatory with science instruments 
provided by European-led Principal Investigator consortia and with important participation from NASA.}
}
\titlerunning{HIFI CO observations of YSOs: from low- to high-mass}

\author{
  I.~San~Jos\'{e}-Garc\'{i}a\inst{\ref{inst1}}
  \and J.~C.~Mottram\inst{\ref{inst1}}
  \and L.~E.~Kristensen\inst{\ref{inst1}}
  \and E.~F.~van Dishoeck\inst{\ref{inst1},\ref{inst3}}
  \and U.~A.~Y{\i}ld{\i}z\inst{\ref{inst1}}
  \and F.~F.~S.~van~der~Tak\inst{\ref{inst10},\ref{inst11}}
  \and F.~Herpin\inst{\ref{inst6},\ref{inst46}}
  \and R.~Visser\inst{\ref{inst14}}
  \and C.~M$^{\textrm c}$Coey\inst{\ref{inst21}}
  \and F.~Wyrowski\inst{\ref{inst30}}
  \and J.~Braine\inst{\ref{inst6},\ref{inst46}}
  \and D.~Johnstone\inst{\ref{inst7},\ref{inst8}}
}

\institute{
  Leiden Observatory, Leiden University, PO Box 9513, 2300 RA Leiden, 
  The Netherlands. \label{inst1} \\
  \email{sanjose@strw.leidenuniv.nl}
  \and
   Max Planck Institut f{\"u}r Extraterrestrische Physik, Giessenbachstrasse 2, 
  85478 Garching, Germany. \label{inst3}
  \and
  SRON Netherlands Institute for Space Research, PO Box 800, 9700 AV 
  Groningen, The Netherlands\label{inst10}
  \and
  Kapteyn Astronomical Institute, University of Groningen, PO Box 800, 
  9700 AV Groningen, The Netherlands\label{inst11}
  \and
  Universit\'{e} de Bordeaux, Observatoire Aquitain des Sciences de l'Univers, 
  2 rue de l'Observatoire, BP 89, F-33270 Floirac Cedex, France \label{inst6}
  \and
  CNRS, LAB, UMR 5804, Laboratoire d'Astrophysique de Bordeaux, 2 rue de l'Observatoire, 
  BP 89, F-33270 Floirac Cedex, France\label{inst46} 
  \and
  Department of Astronomy, The University of Michigan, 500 Church Street, 
  Ann Arbor, MI 48109-1042, USA\label{inst14}
  \and
  University of Waterloo, Department of Physics and Astronomy, Waterloo, 
  Ontario, Canada\label{inst21}
  \and
  Max-Planck-Institut f\"{u}r Radioastronomie, Auf dem H\"{u}gel 69, 53121 Bonn, 
  Germany\label{inst30}
  \and
  National Research Council Canada, Herzberg Institute of Astrophysics, 
  5071 West Saanich Road, Victoria, BC V9E 2E7, Canada\label{inst7}
  \and
  Department of Physics and Astronomy, University of Victoria, Victoria, 
  BC V8P 1A1, Canada\label{inst8}
}

\date{January 17, 2013}


\def\placeFigDecomposition{
  \begin{figure*}[!t]
    \sidecaption
    \includegraphics[scale=0.4, angle=0]{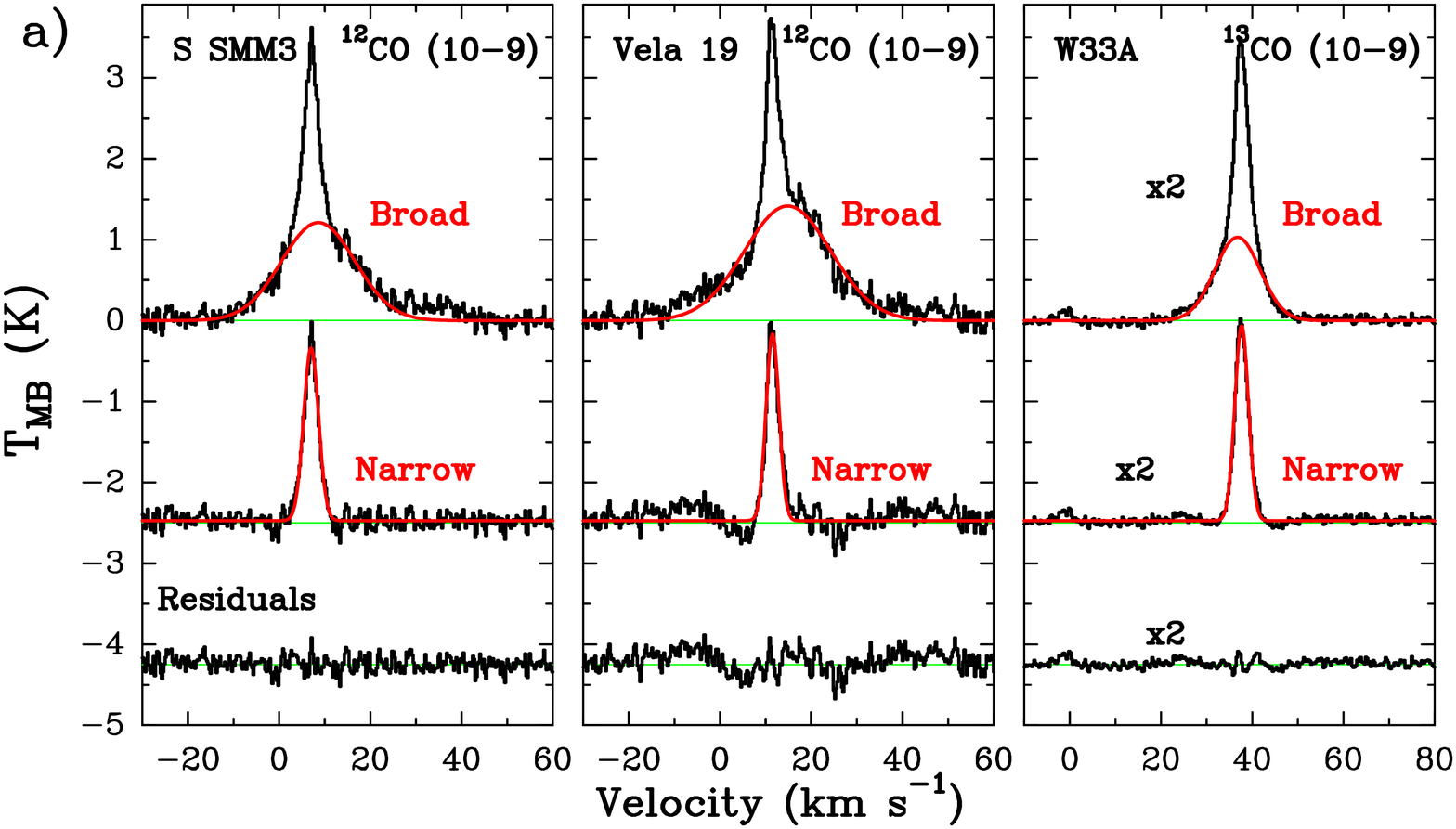}
    \includegraphics[scale=0.4, angle=0]{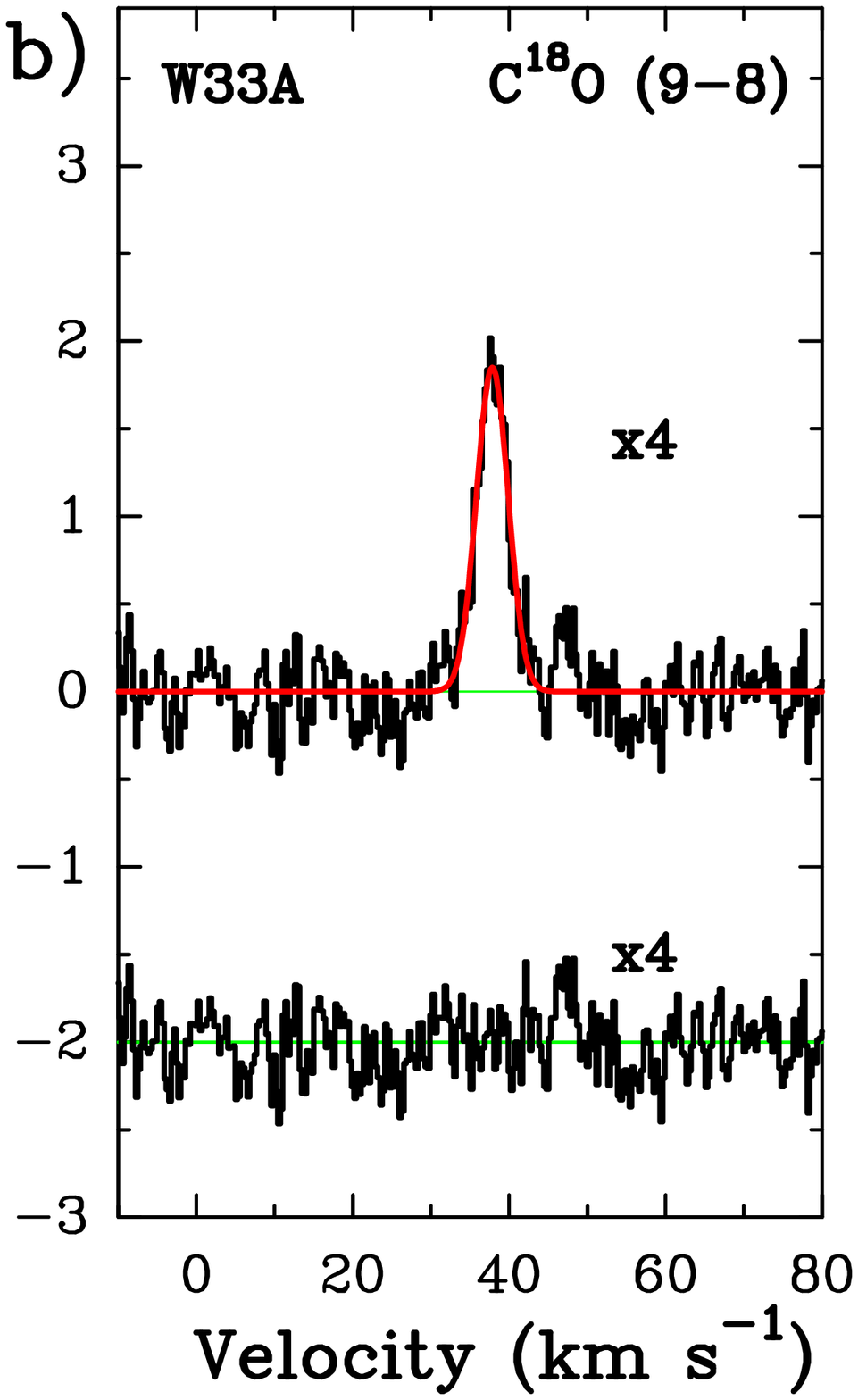}    
    \caption{ Gaussian decomposition for the CO and isotopologues line profiles:
      a) two Gaussian fit for the line profiles with two different velocity 
      components identified, such as the $^{12}$CO~$J$=10--9 spectra
      for the low-mass YSOs Ser\,SMM3 (left), the intermediate-mass Vela\,19 (centre) and
      the $^{13}$CO~$J$=10--9 spectrum of the high-mass W33A (right). b) Single Gaussian fit of sources 
      characterised by one component profile, such as the C$^{18}$O~$J$=9--8 spectra of W33A.
      The red lines show the Gaussian fits and the green lines the baseline.}
    \label{fig:decomposition}
  \end{figure*}
}

\def\placeFigRepresentadoceco{
  \begin{figure}[!t]
    \centering
    \includegraphics[scale=0.45, angle=0]{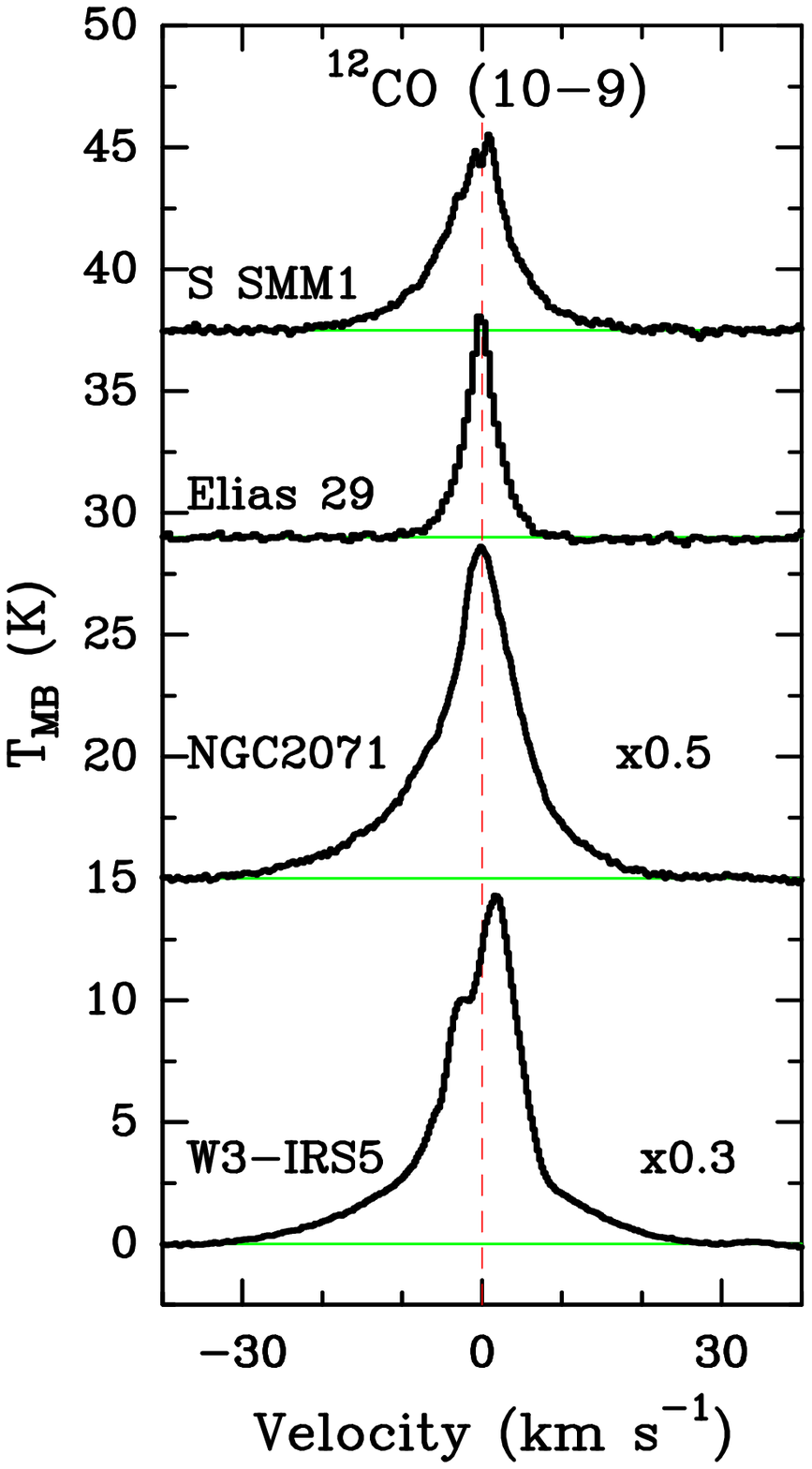}
    \hspace{0.1cm}
    \includegraphics[scale=0.45, angle=0]{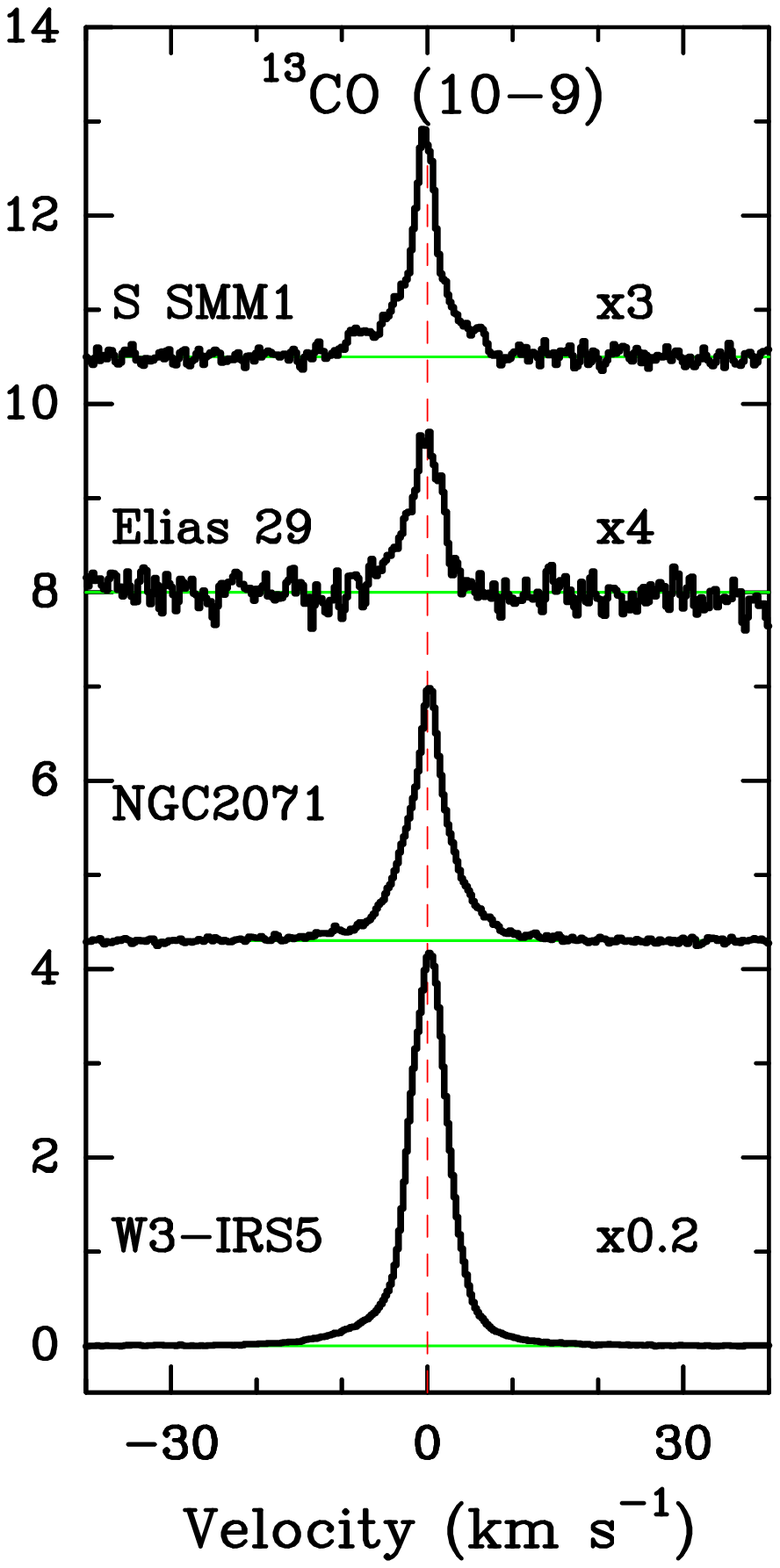}
    \caption{ $^{12}$CO~$J$=10--9 (left) and $^{13}$CO~$J$=10--9 (right) spectra for a low-mass 
      Class~0 protostar (top, Ser\,SMM1),
      low-mass Class~I source (Elias\,29), intermediate-mass object (NGC\,2071) and 
      high-mass YSO (bottom, W3-IRS5). The green line indicates the baseline level and the
      red dashed line the 0~km\,s$^{-1}$ value. All spectra have been re-binned to 0.27~km\,s$^{-1}$ 
      and shifted with respect to their relative local standard of rest velocity.}
    \label{fig:1213co}
  \end{figure}
}

\def\placeFigRepresentacDieciochoo{
  \begin{figure}[!t]
    \centering
    \includegraphics[scale=0.32, angle=0]{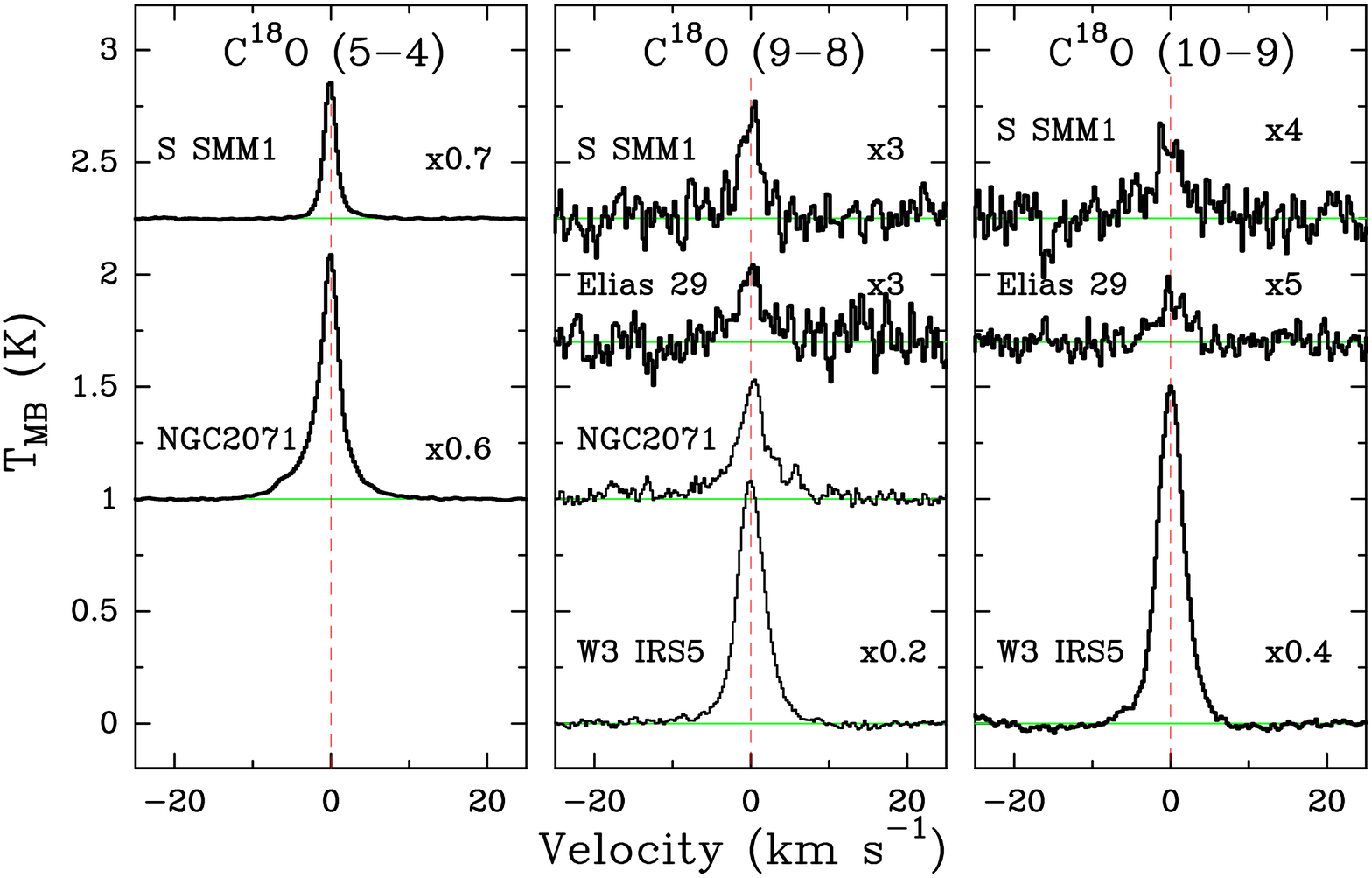}
    \caption{ Same as Fig. \ref{fig:1213co} but for the C$^{18}$O spectra from the 
      observed transitions: $J$=5--4 (left), $J$=9--8 (centre) and $J$=10--9 (right). For
      details about these objects see Appendix~\ref{Data}.}
    \label{fig:c18o}
  \end{figure}
}


\def\placeFigHistogramaBroad{
  \begin{figure}[!t]
    \centering
    \includegraphics[scale=0.5, angle=0]{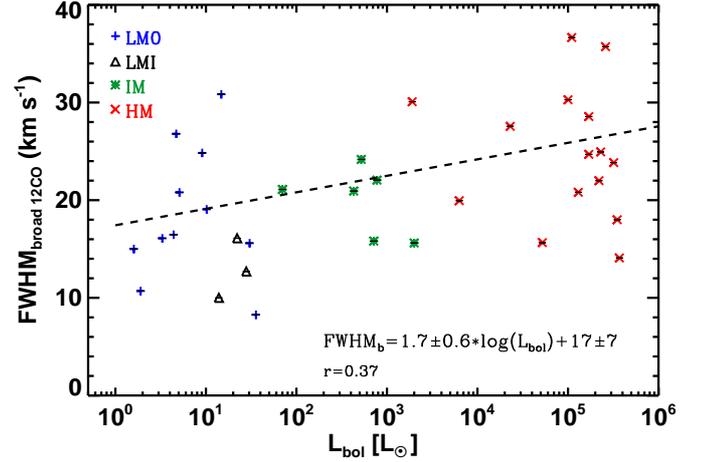}
    \caption{ FWHM of the broad velocity component identified in the $^{12}$CO~$J$=10--9 
      line profiles 	for each type of YSOs versus their bolometric luminosities. 
      Low-mass Class~0 (LM0) sources are indicated with blue pluses, low-mass Class~I (LMI) with black triangles,  
      intermediate-mass (IM) YSOs with green asterisks and high-mass (HM) objects with red crosses. 
      For the high-mass sources, 
      the $^{12}$CO~$J$=3--2 width is used instead. The black dashed line indicates the
      linear function that fits the relation between the FWHM and the logarithm of $L_{\rm{bol}}$.}
    \label{fig:HistogrBroad}
  \end{figure}
}

\def\placeFigHistogramaCdiecioHIFI{
  \begin{figure}[!t]
    \centering
    \includegraphics[scale=0.5, angle=0]{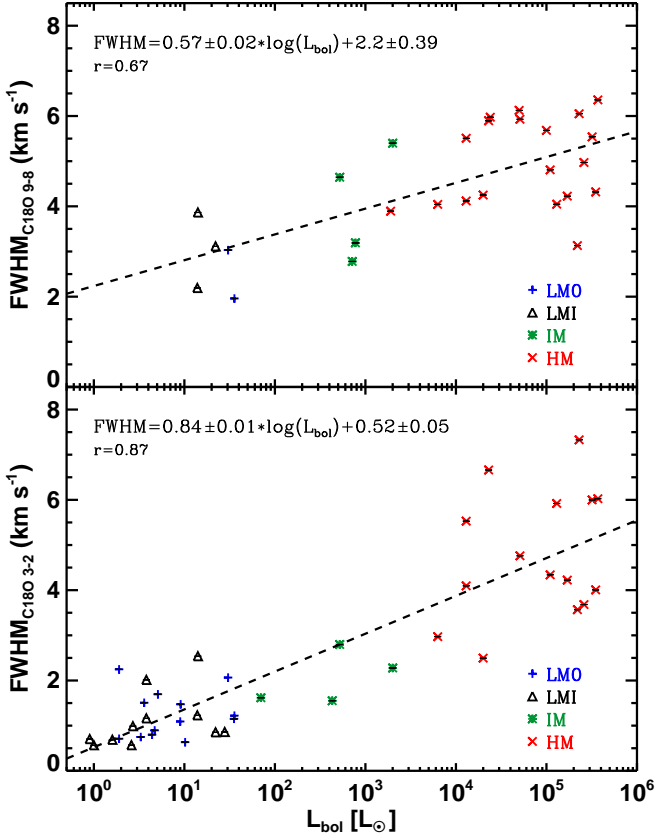}
    \caption{ Same as Fig.~\ref{fig:HistogrBroad} but for the narrow C$^{18}$O~$J$=9--8 line profiles (top) and
      C$^{18}$O~$J$=3--2 (bottom). Note that only a few low-mass YSOs have been detected in C$^{18}$O~$J$=9--8.}
    \label{fig:HistogrC18O-HIFI}
  \end{figure}
}

\def\placeFigCorrelationCO{
  \begin{figure}[!t]
    \centering
    \includegraphics[scale=0.5, angle=0]{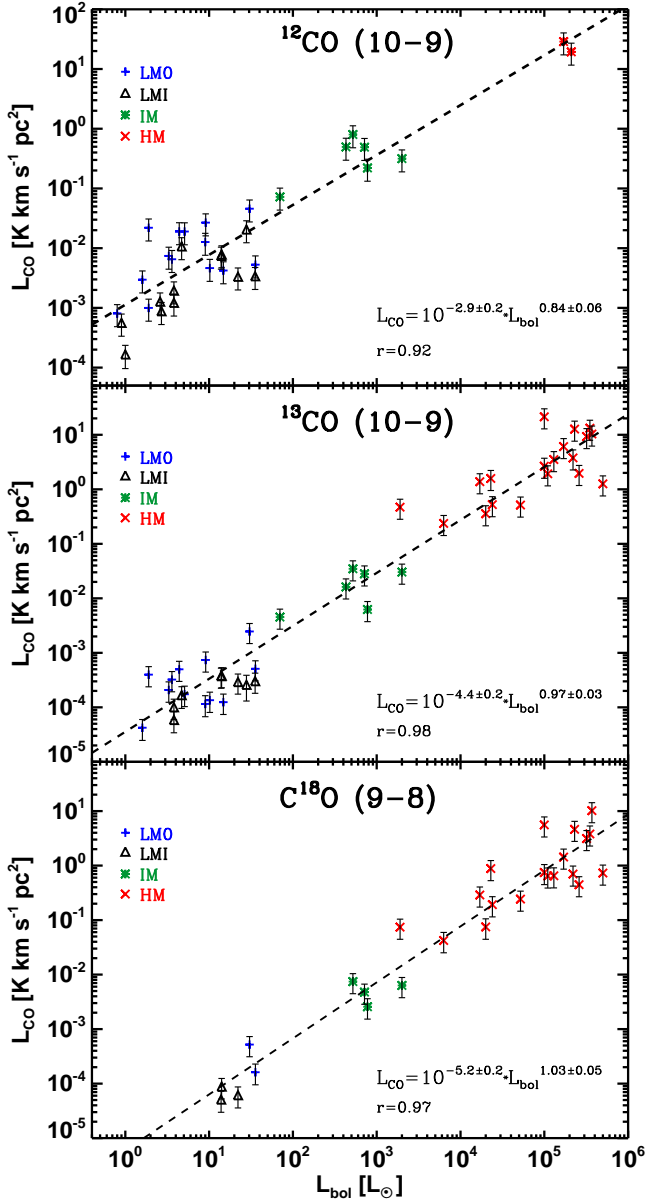}
    \caption{ Line luminosity of the $^{12}$CO~$J$=10--9 (top),  
      $^{13}$CO~$J$=10--9 (middle) and C$^{18}$O~$J$=9--8 (bottom) spectra 
      for low-mass Class~0 (LM0; blue pluses), low-mass Class~I (LMI; black triangles), 
      intermediate-mass (IM; green asterisks) and high-mass (HM; red crosses)
      YSOs versus their bolometric luminosity. The black dashed line represents the linear function that
      fits the logarithm of the plotted quantities.}
    \label{fig:Correlation12CO}
  \end{figure}
}

\def\placeFigCorrelationdoceTreceCO{
  \begin{figure}[!t]
    \centering
    \includegraphics[scale=0.5, angle=0]{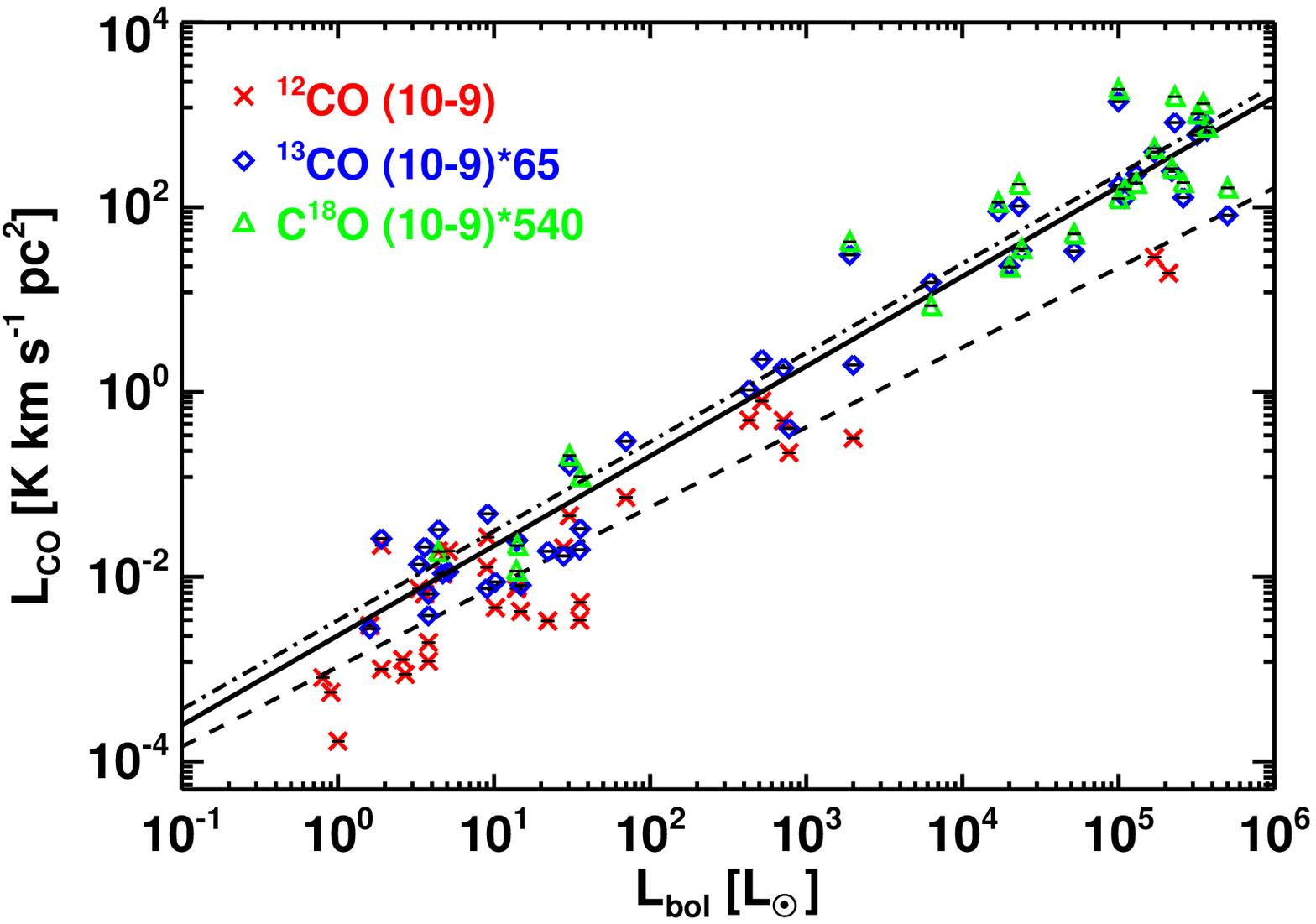}
    \caption{ Line luminosity of the $^{12}$CO~$J$=10--9 emission lines, red
      crosses, versus their bolometric luminosity, together with the line luminosity of the $^{13}$CO~$J$=10--9
      spectra, blue diamonds, multiply by the assumed abundance ratio of $^{12}$C/$^{13}$C
      for the entire WISH sample of YSOs. The line luminosity
      of the C$^{18}$O~$J$=10--9 lines, green triangles, multiply by the assumed 
      abundance ratio of $^{16}$O/$^{18}$O is plotted together with the previous values. 
      The dashed line represents the linear fit of the $^{12}$CO~$J$=10--9 spectra, the full line
      that for the $^{13}$CO~$J$=10--9 transition and the 
      dash-dot line indicates the fit for the C$^{18}$O~$J$=10--9 data.}
    \label{fig:Correlation1213CO}
  \end{figure}
}

\def\placeFigCorrelationCOextragalactic{
  \begin{figure}[!t]
    \centering
    \includegraphics[scale=0.5, angle=0]{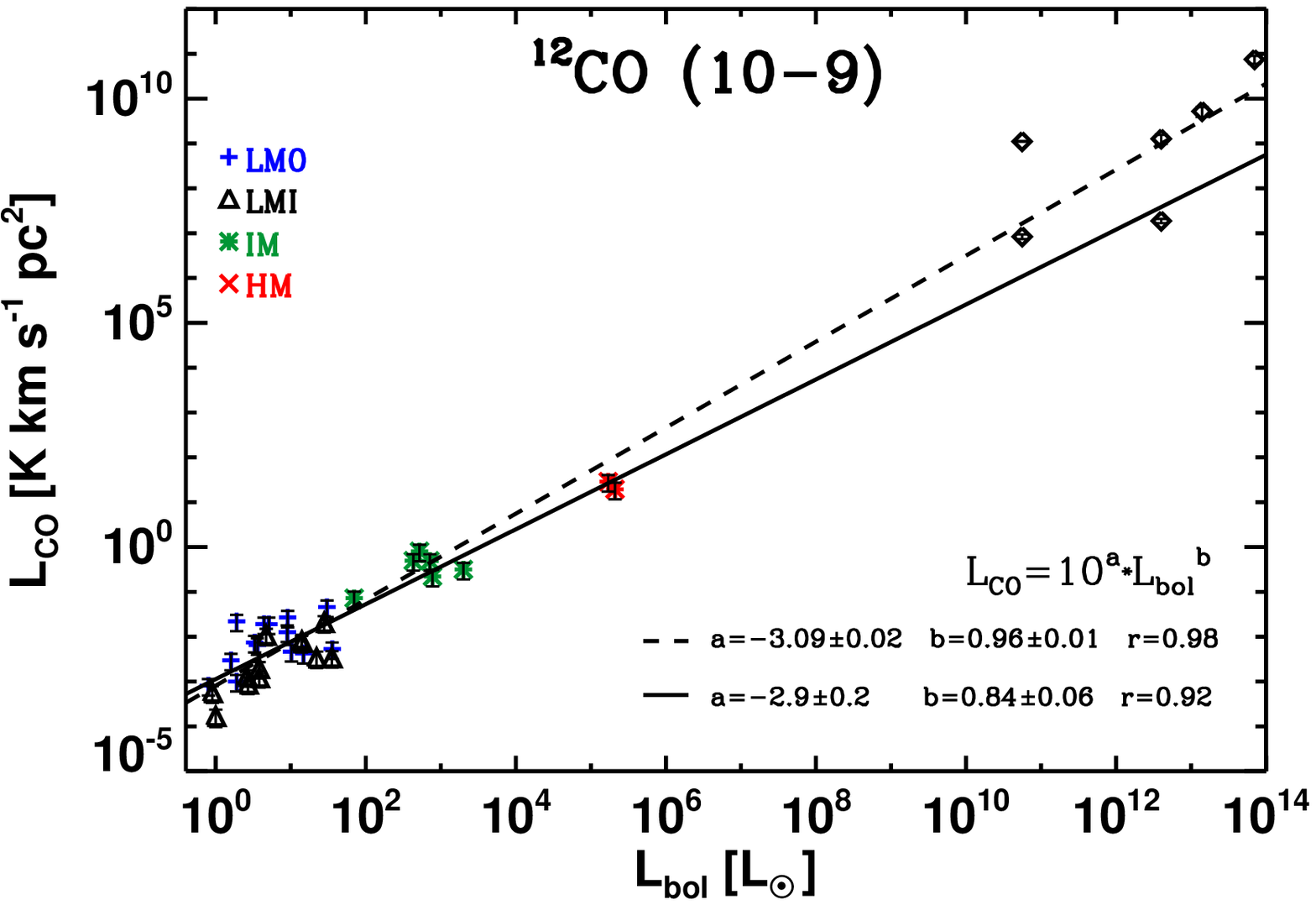}
    \caption{ Line luminosity of the $^{12}$CO~$J$=10--9 emission versus their bolometric luminosity 
      for the entire WISH sample of YSOs (same notation as Fig.~\ref{fig:Correlation12CO}) 
      and for four galaxies in our local universe and two at high redshifts (black diamonds). 
      The solid line represents the linear fit of the galactic $^{12}$CO~$J$=10--9 spectra and the 
      dashed line the linear fit including the values of the six galaxies.}
    \label{fig:Correlationextra}
  \end{figure}
}


\def\placeFigCorrelationTreceCOMenv{
  \begin{figure}[!t]
    \centering
    \includegraphics[scale=0.5, angle=0]{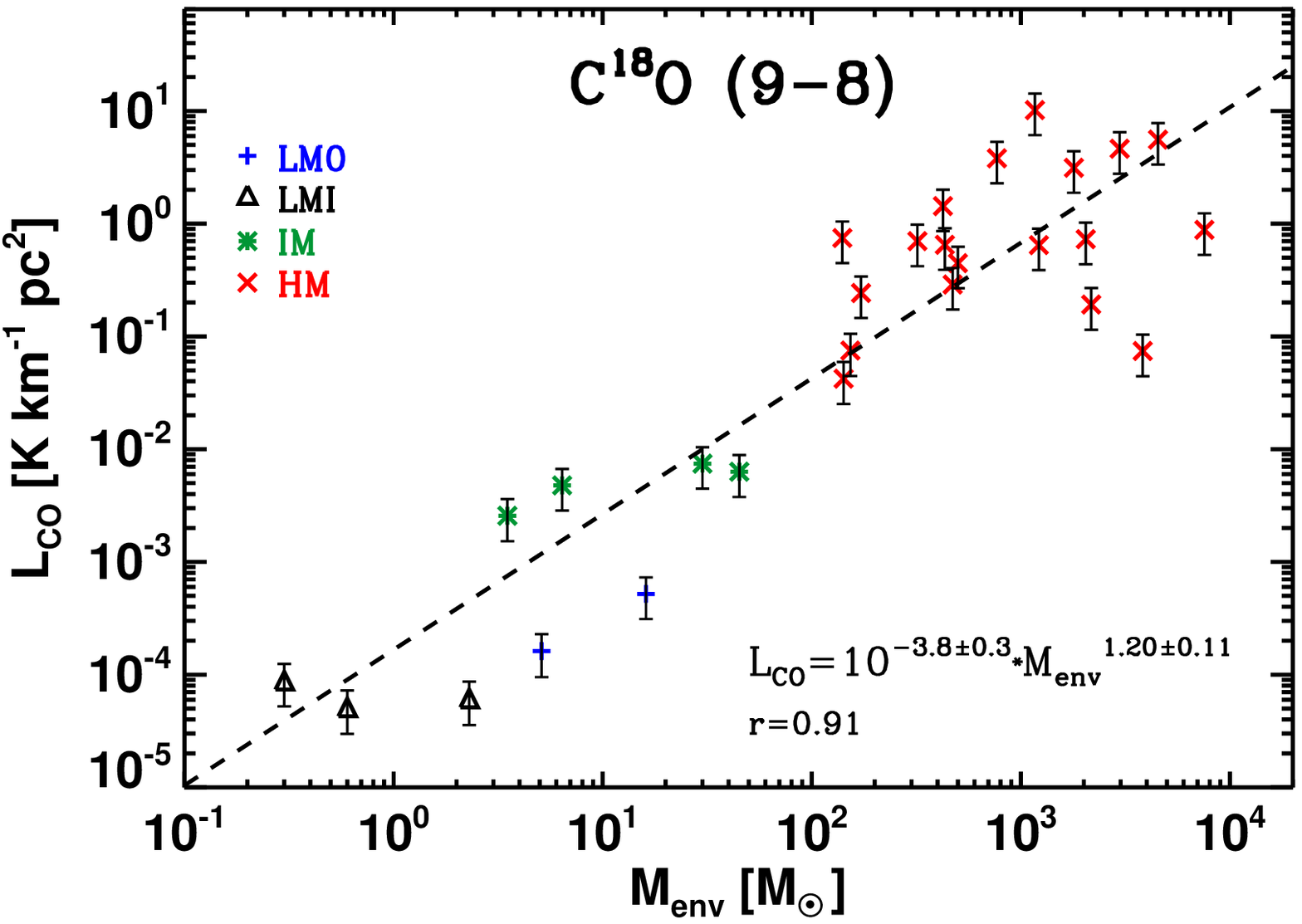}
    \caption{ C$^{18}$O~$J$=9--8 line luminosity
      for low-mass Class~0 (LM0; blue pluses), low-mass Class~I (LMI; black triangles), 
      intermediate-mass (IM; green asterisks) and high-mass (HM; red crosses)
      YSOs versus their envelope masses, $M_{\rm{env}}$. The dash black line represents the linear function that
      fits the logarithm of the plotted quantities.}
    \label{fig:CorrelationC18OMenv}
  \end{figure}
}

\def\placeFigCorrelationDeltaVBandDeltaVCdieciocho{
  \begin{figure}[!t]
    \centering
    \includegraphics[scale=0.5, angle=0]{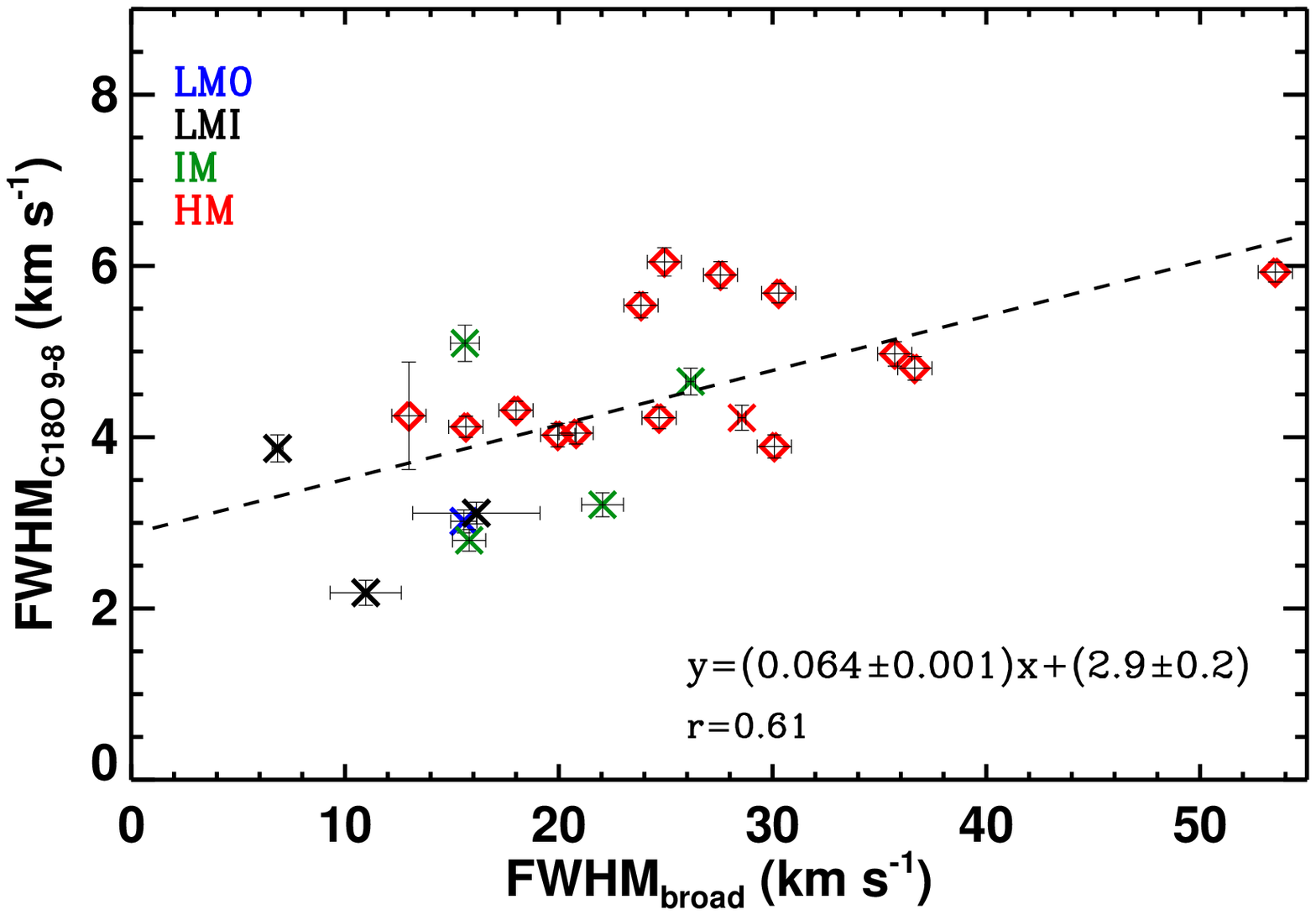}
    \caption{ Width of the C$^{18}$O~$J$=9--8 line profile versus the 
      width of the broader component of $^{12}$CO~$J$=10--9 emission lines (crosses) and for the 
      $^{12}$CO~$J$=3--2 line profiles (diamonds). The black dashed line represents the linear function that
      fits the logarithm of the plotted quantities.}
    \label{fig:CorrelationDbDvDvC18O}
  \end{figure}
}

\def\placeFigTkin{
  \begin{figure}[!t]
    \centering
    \includegraphics[scale=0.5, angle=0]{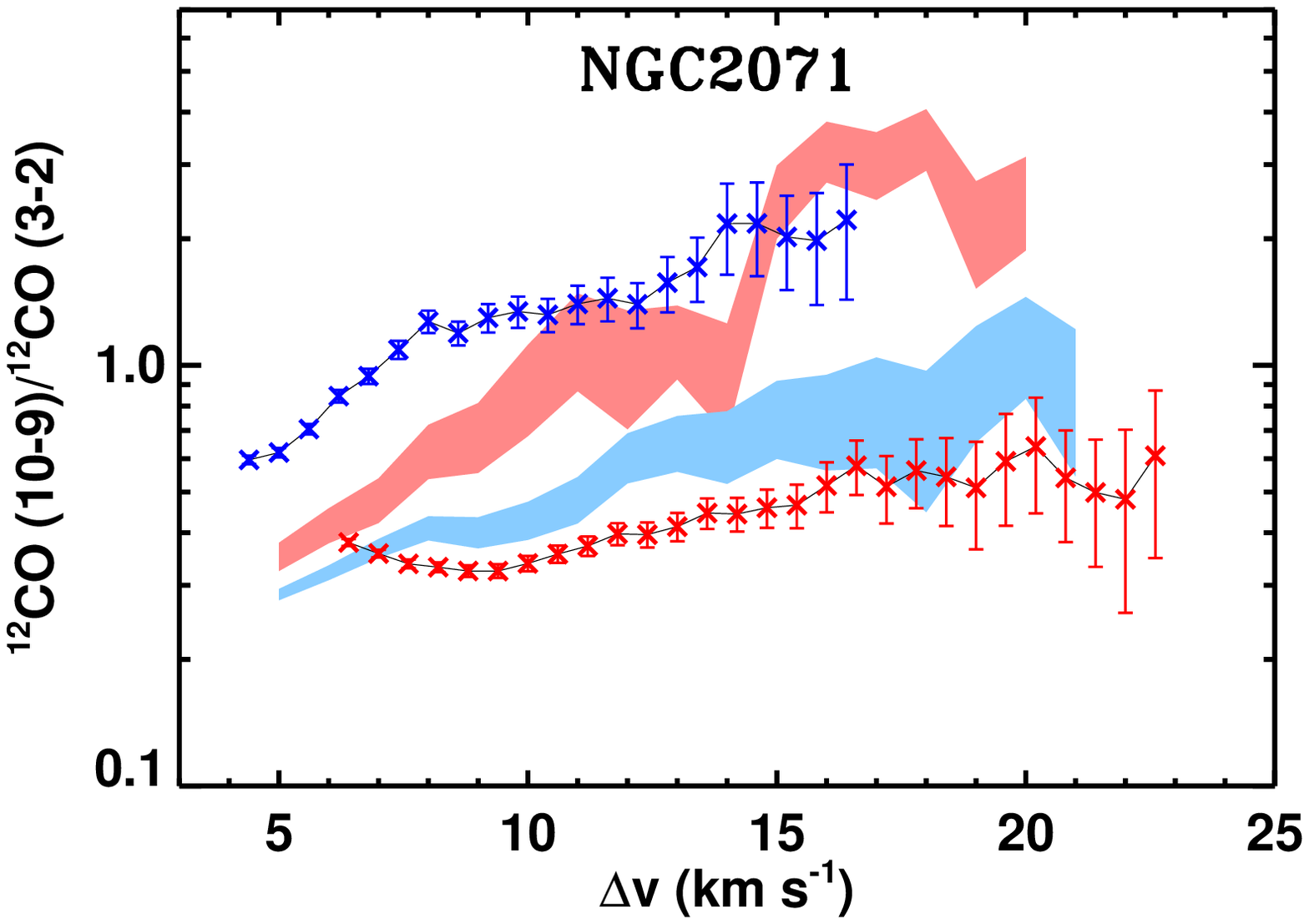}
    \includegraphics[scale=0.5, angle=0]{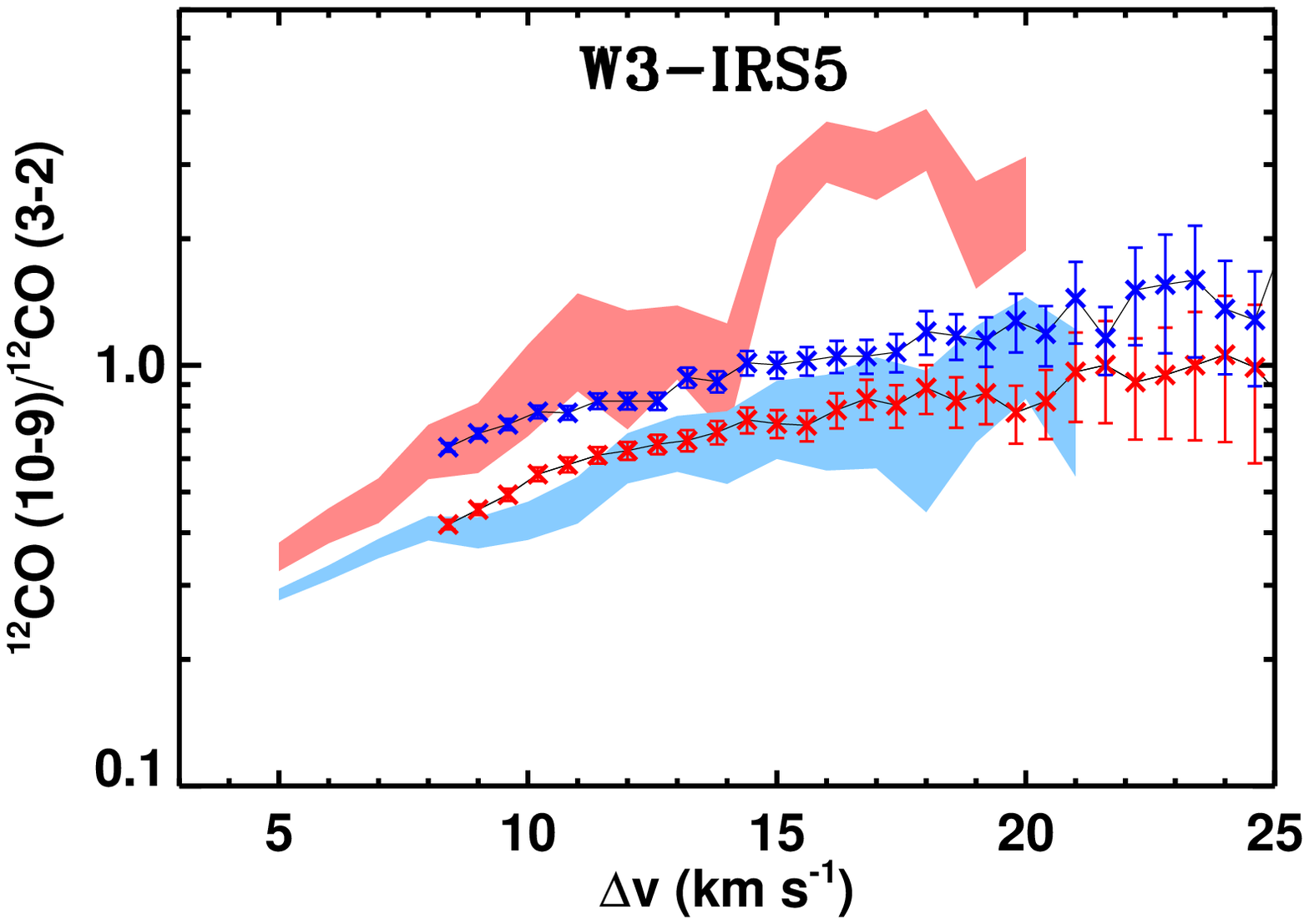}
    \caption{ Line-wing ratios of $^{12}$CO~$J$=10--9/$J$=3--2 as a function of absolute offset 
    from the source velocity for the sources NGC2071 (top) and W3-IRS5 (bottom) for the red and blue wings. 
    The shaded regions are obtained from the average spectra of 
    the low-mass sample for these transitions (see \citealt{Yildiz13sub}). 
    The spectra have been resampled to 0.6~km\,s$^{-1}$ bin and shifted to 0~km\,s$^{-1}$.}
    \label{fig:line_ratio}
  \end{figure}
}

\def\placeFigAveragedoceco{
  \begin{figure}[!t]
    \centering
    \includegraphics[scale=0.44, angle=0]{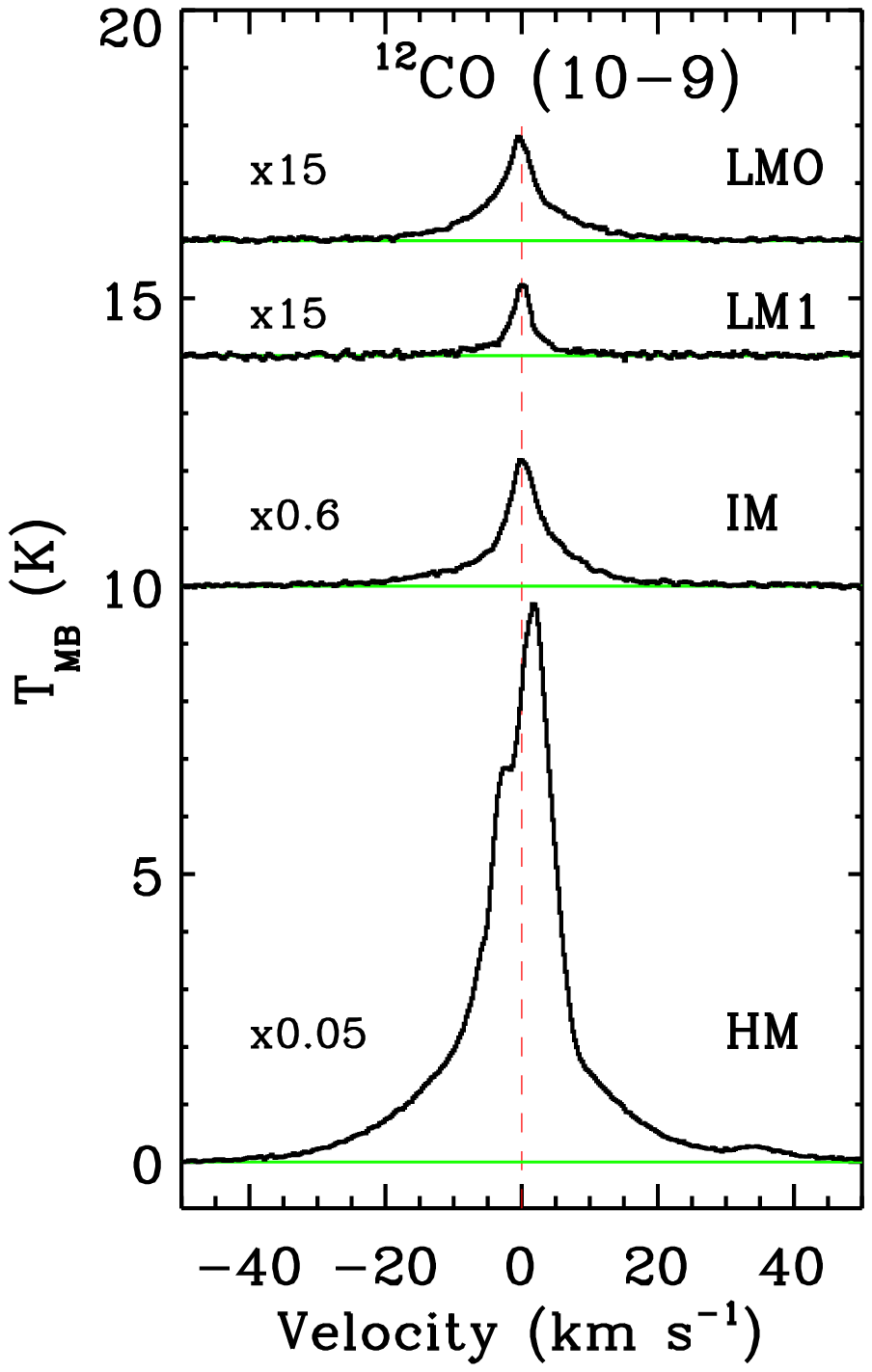}
    \includegraphics[scale=0.44, angle=0]{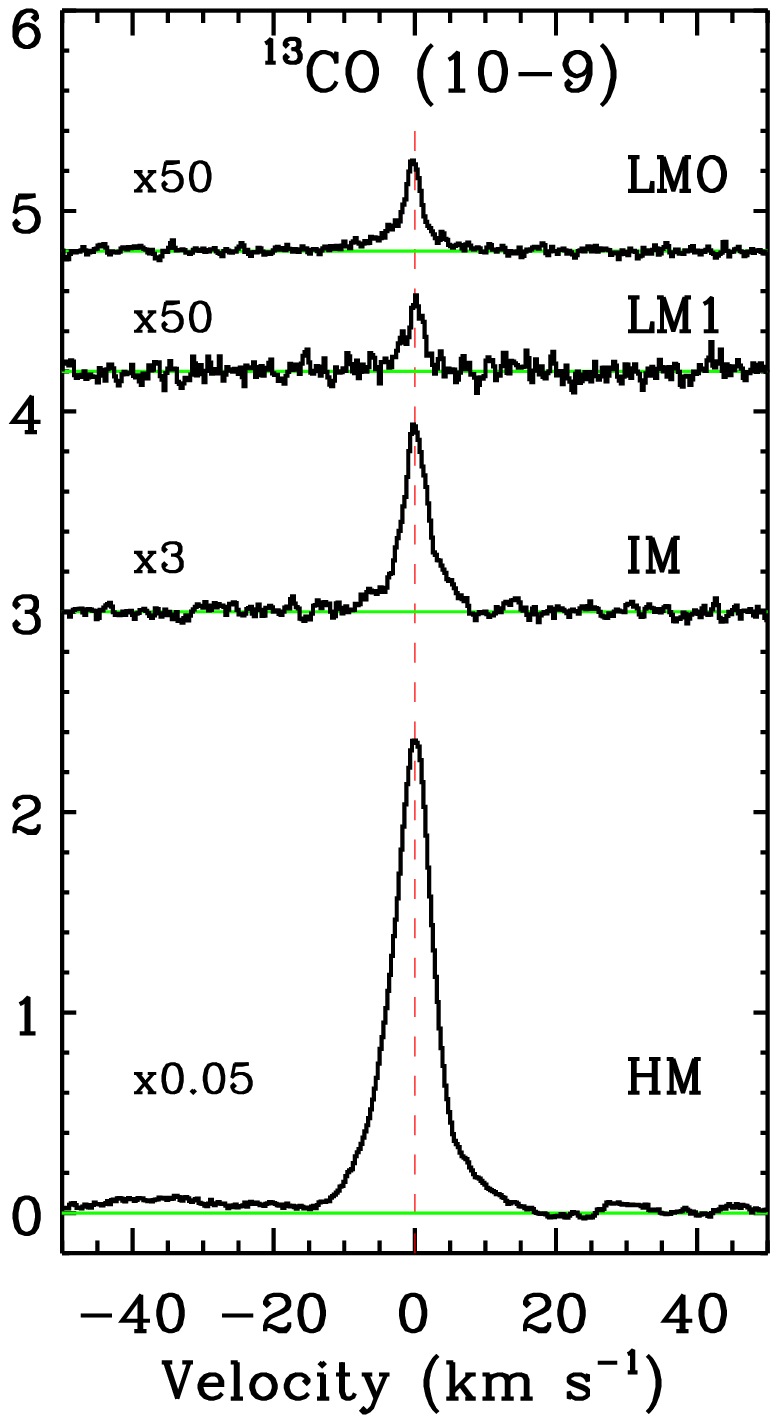}
    \caption{ $^{12}$CO~$J$=10--9 (left) and $^{13}$CO~$J$=10--9 (right) spectra of low-mass 
      class~0 (LM0), low-mass Class~I (LMI), intermediate-mass (IM) and 
      high-mass YSO (HM) averaged independently and compared. 
      All spectra were shifted to 0~km\,s$^{-1}$, re-binned to 0.27~km\,s$^{-1}$ 
      and the intensity of the emission line scaled to a common distance of 1 kpc before averaging. 
      The green line indicates the continuum level and the red dashed line the 0~km\,s$^{-1}$ value. 
      W43-MM1 was not included in the average of $^{13}$CO~$J$=10--9 high-mass spectra because of the strong
      absorption features caused by H$_2$O$^+$ (see Section~\ref{HIFIobservations}).}
    \label{fig:aver1213co}
  \end{figure}
}

\def\placeFigAveragediecico{
  \begin{figure}[!t]
    \centering
    \includegraphics[scale=0.32, angle=0]{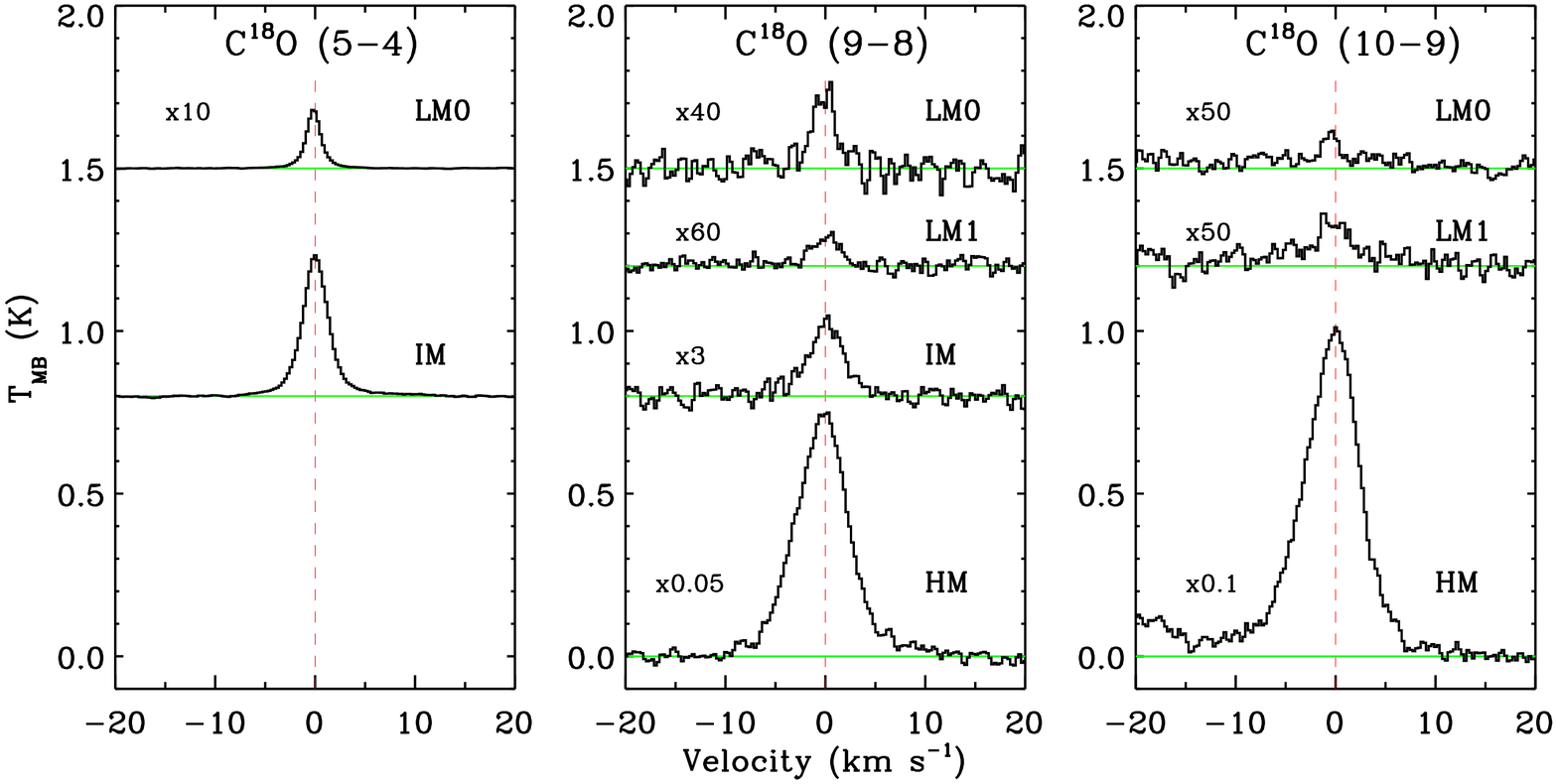}
    \caption{ Same as Fig. \ref{fig:aver1213co} but for the C$^{18}$O spectra from the 
      observed transitions:~$J$=5--4 (left), ~$J$=9--8 (centre) and ~$J$=10--9 (right). More
      details about these transitions in Appendix~\ref{Data}.}
    \label{fig:averc18o}
  \end{figure}
}


\def\placeFigPreLM{
  \begin{figure}[!t]
    \centering
    \includegraphics[scale=0.5, angle=0]{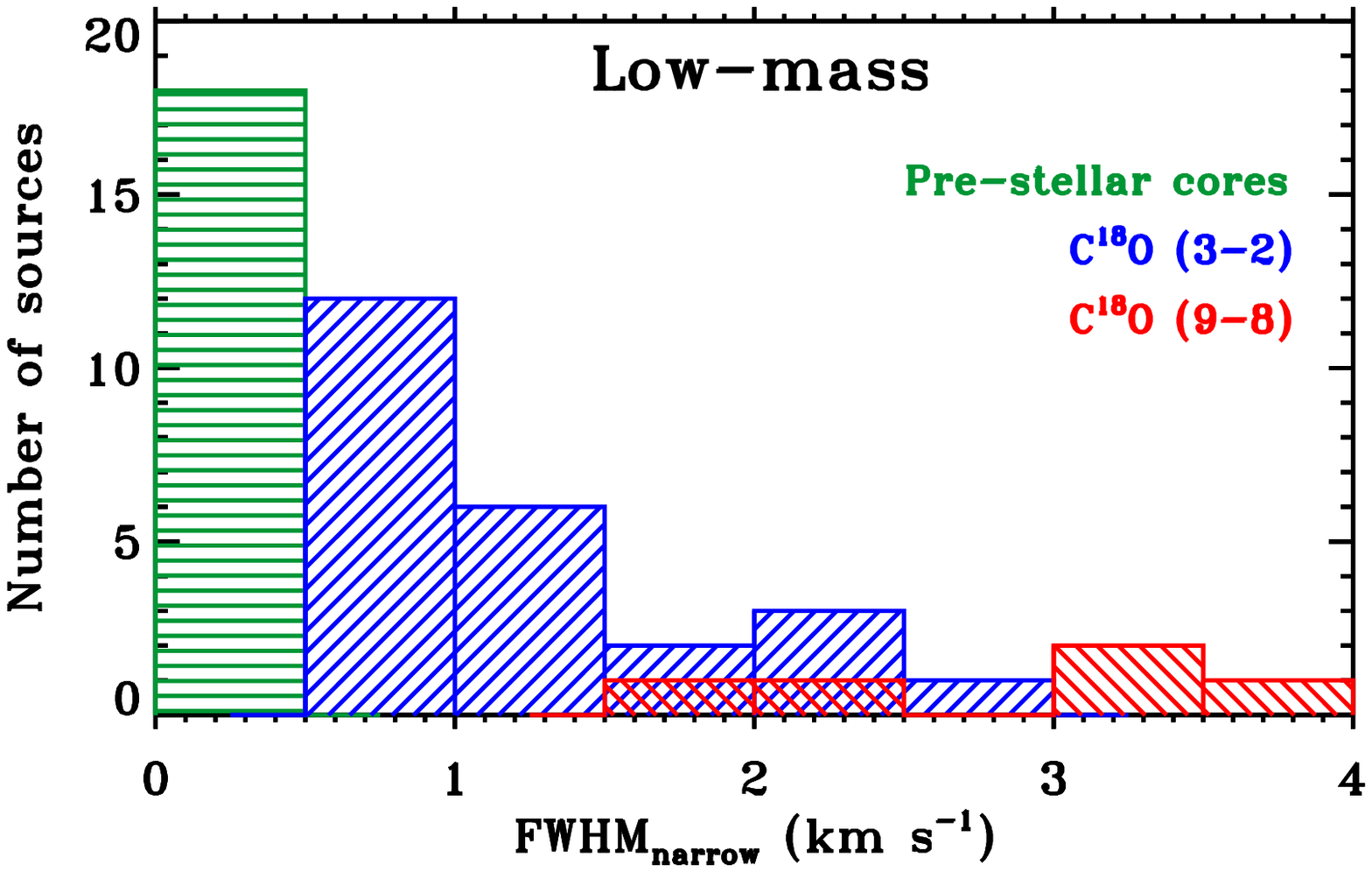}
    \includegraphics[scale=0.5, angle=0]{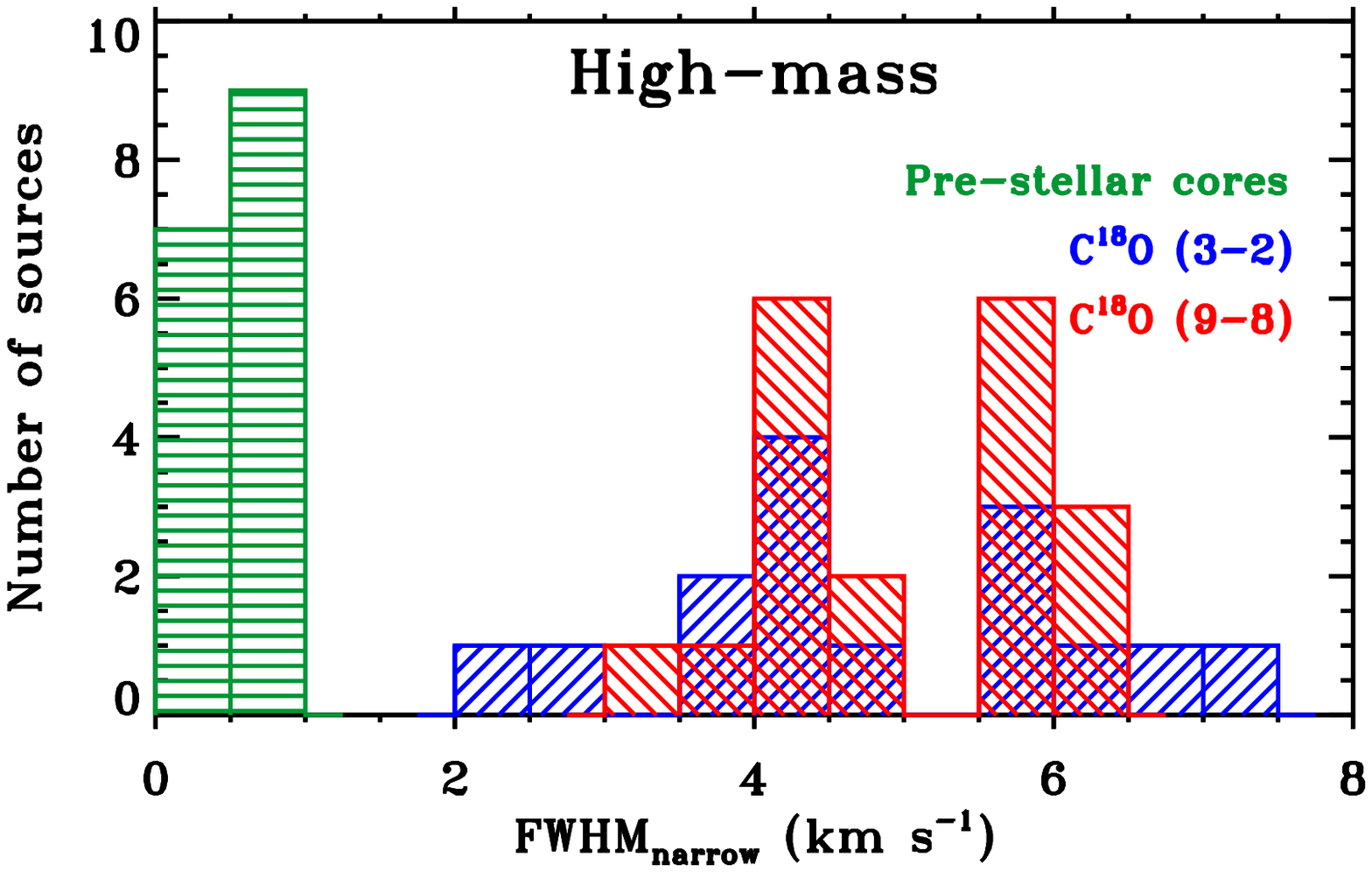}
    \caption{ Comparison of the observed width of the ammonia line for a sample of low-mass pre-stellar 
      cores collected by \cite{Jijina99} (green), width of the C$^{18}$O~$J$=3--2 spectra (blue)
      and C$^{18}$O~$J$=9--8 line widths (red) for the detected WISH sample of low-mass protostars (top).
      The histogram of the bottom presents the same values but for high-mass pre-stellar cores and
      the WISH high-mass sample of YSOs.}
    \label{fig:preLM}
  \end{figure}
}


\def\placeSpectraCO{
  \begin{figure*}[h]
    \centering
    \bigskip
    \includegraphics[scale=0.30, angle=0]{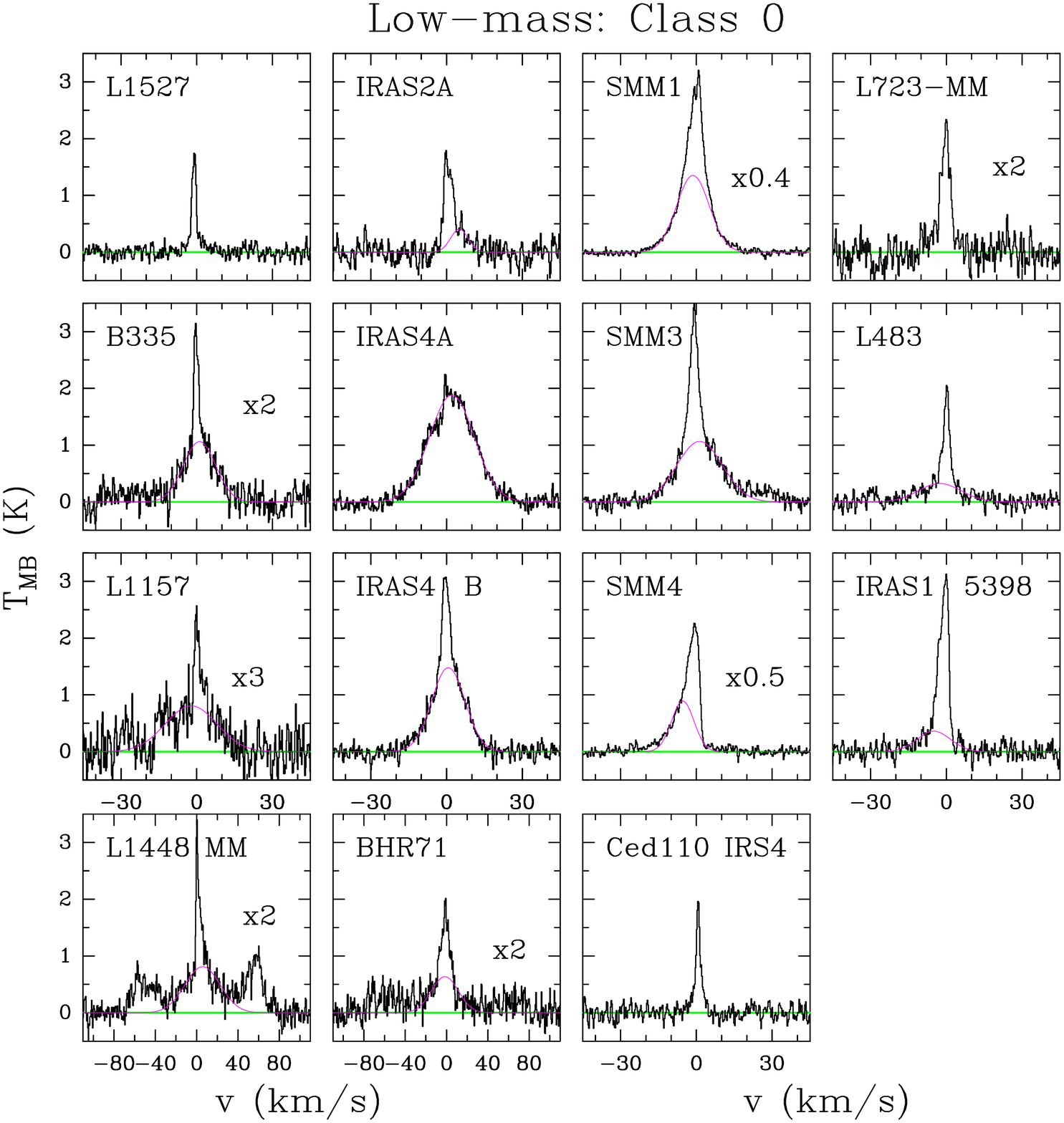}
    \bigskip
    \includegraphics[scale=0.30, angle=0]{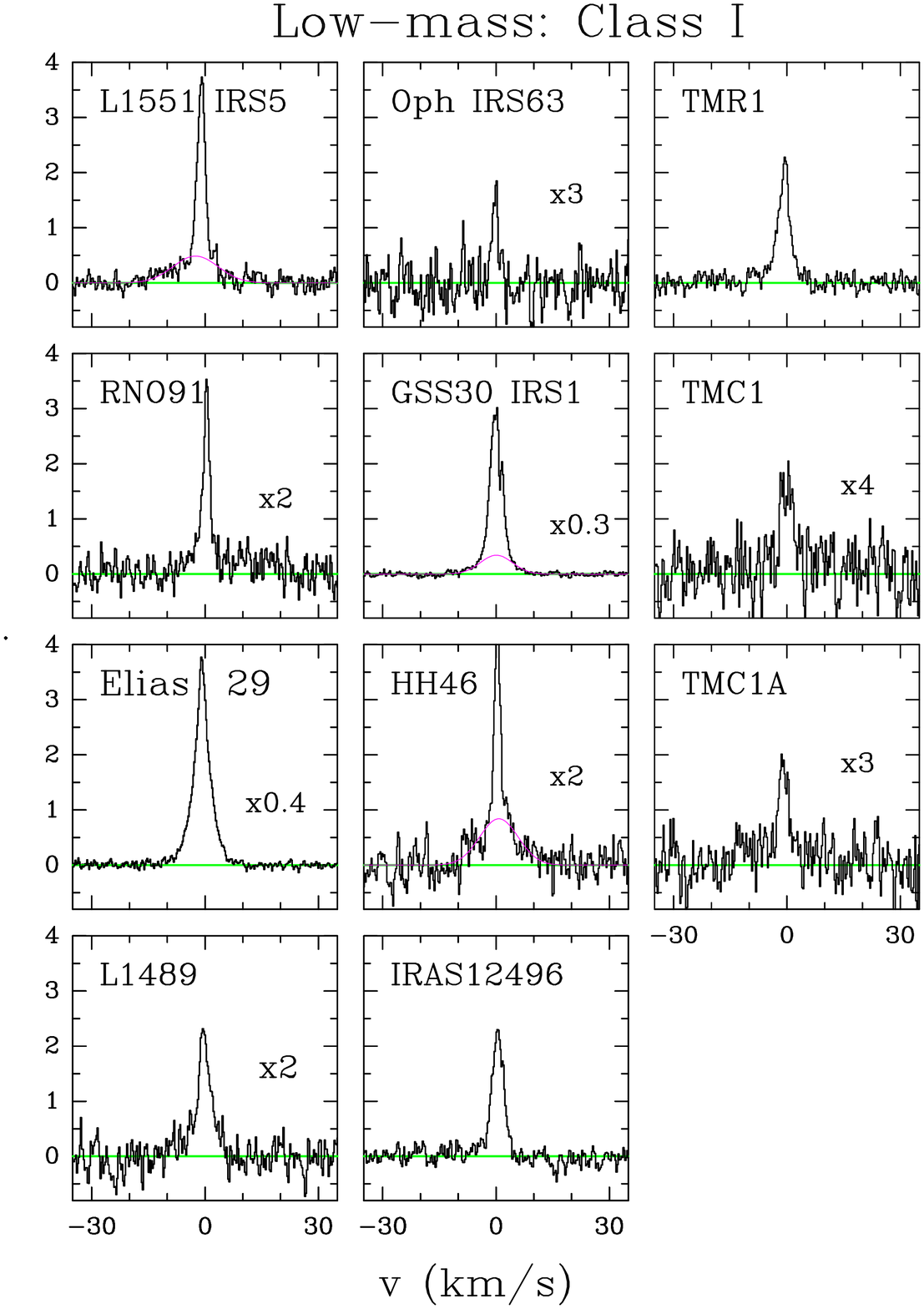}
    \smallskip
    \includegraphics[scale=0.30, angle=0]{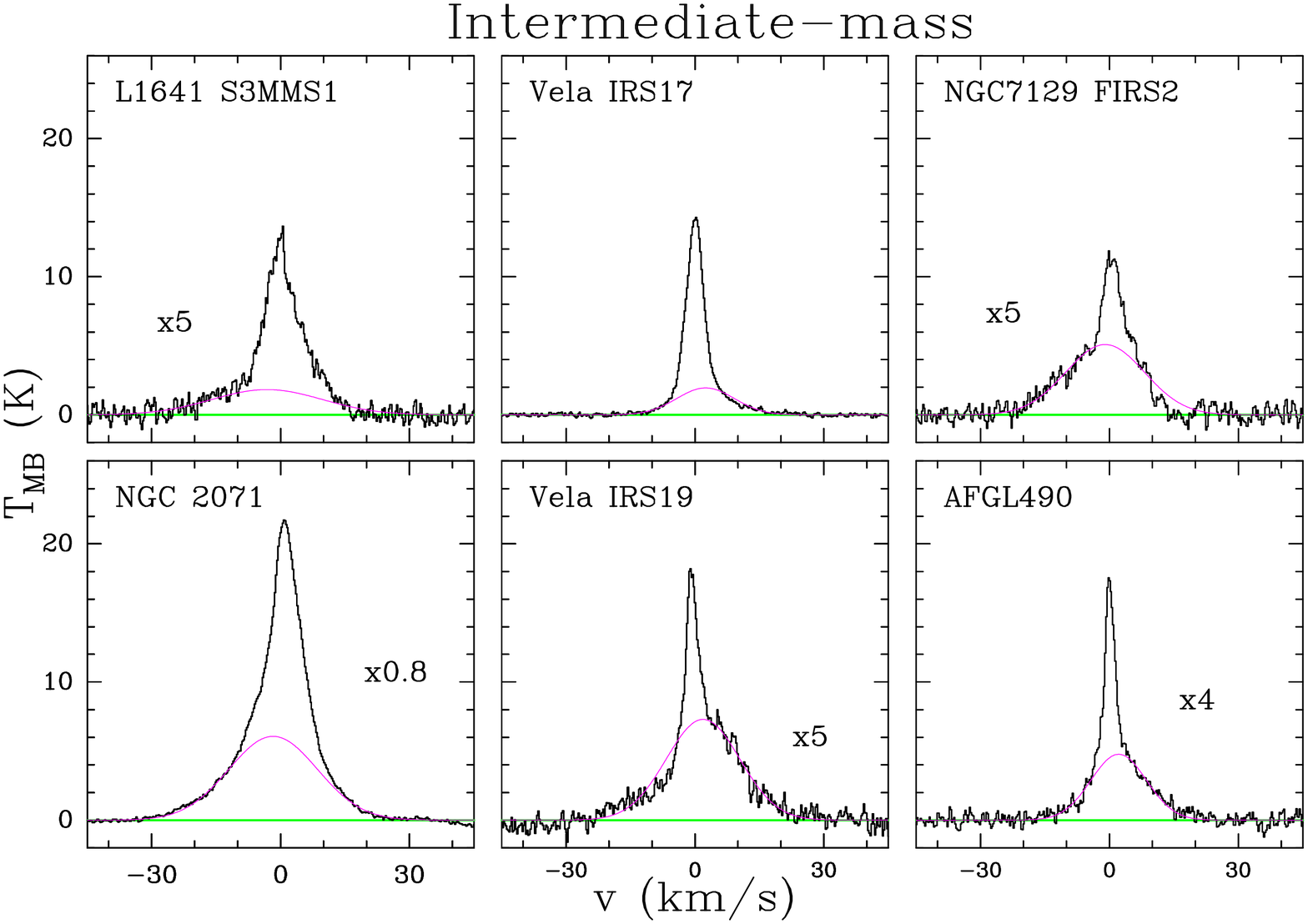}
    \caption{\label{fig:Spec12CO10-9} $^{12}$CO~$J$=10--9 spectra for low- and intermediate-mass YSOs. The green 
      line represents the baseline level and the pink Gaussian profile
      the broad velocity component for those sources for which a 2 Gaussian decomposition has been performed. 
      All the spectra have been shifted to zero velocity. The numbers indicate where the spectra have been 
      scaled for greater visibility.}
  \end{figure*}
}

\def\placeSpectratreceCO{
  \begin{figure*}[h]
    \centering
    \bigskip
    \includegraphics[scale=0.30, angle=0]{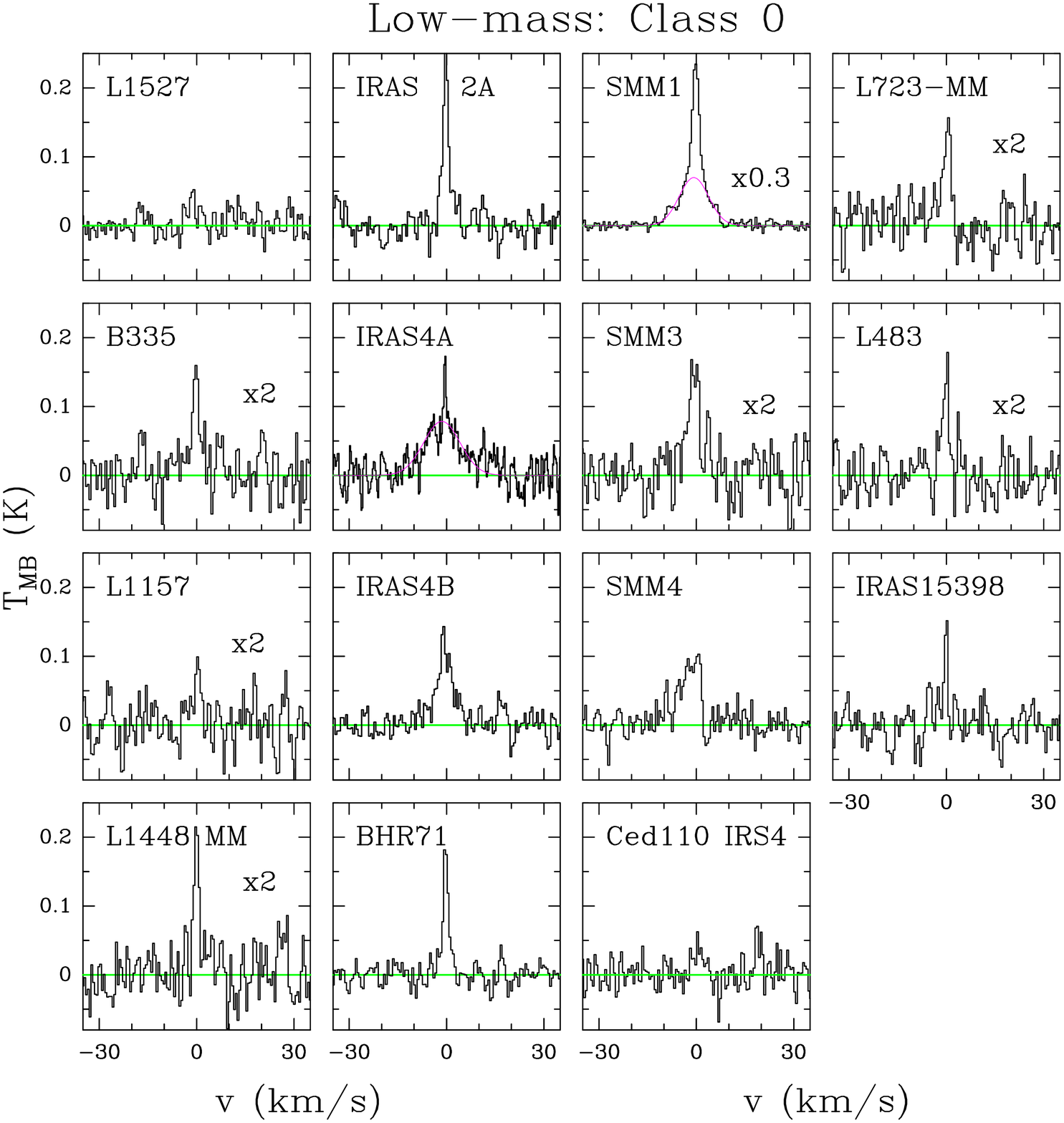}
    \bigskip
    \includegraphics[scale=0.30, angle=0]{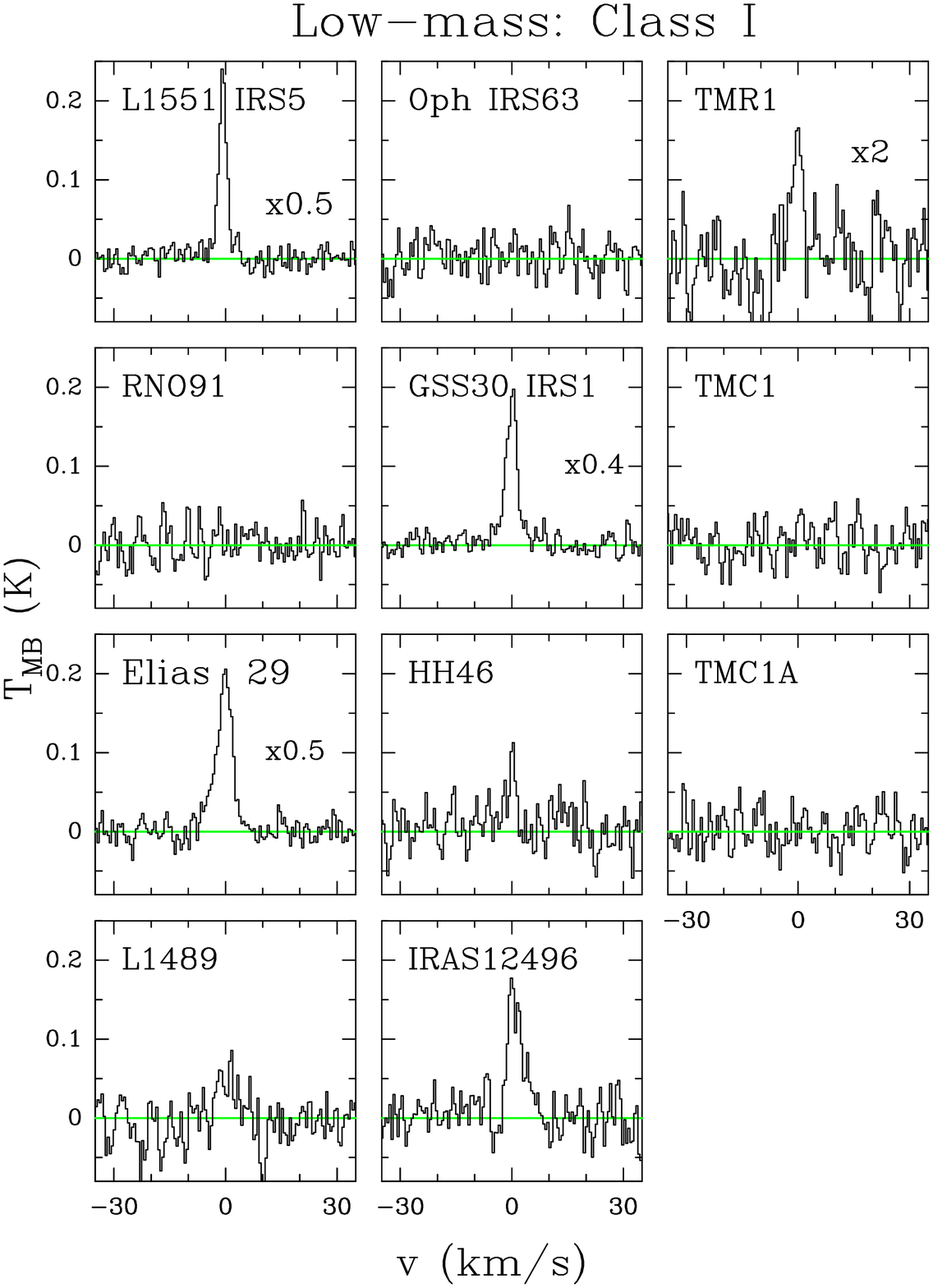}
    \bigskip
    \includegraphics[scale=0.30, angle=0]{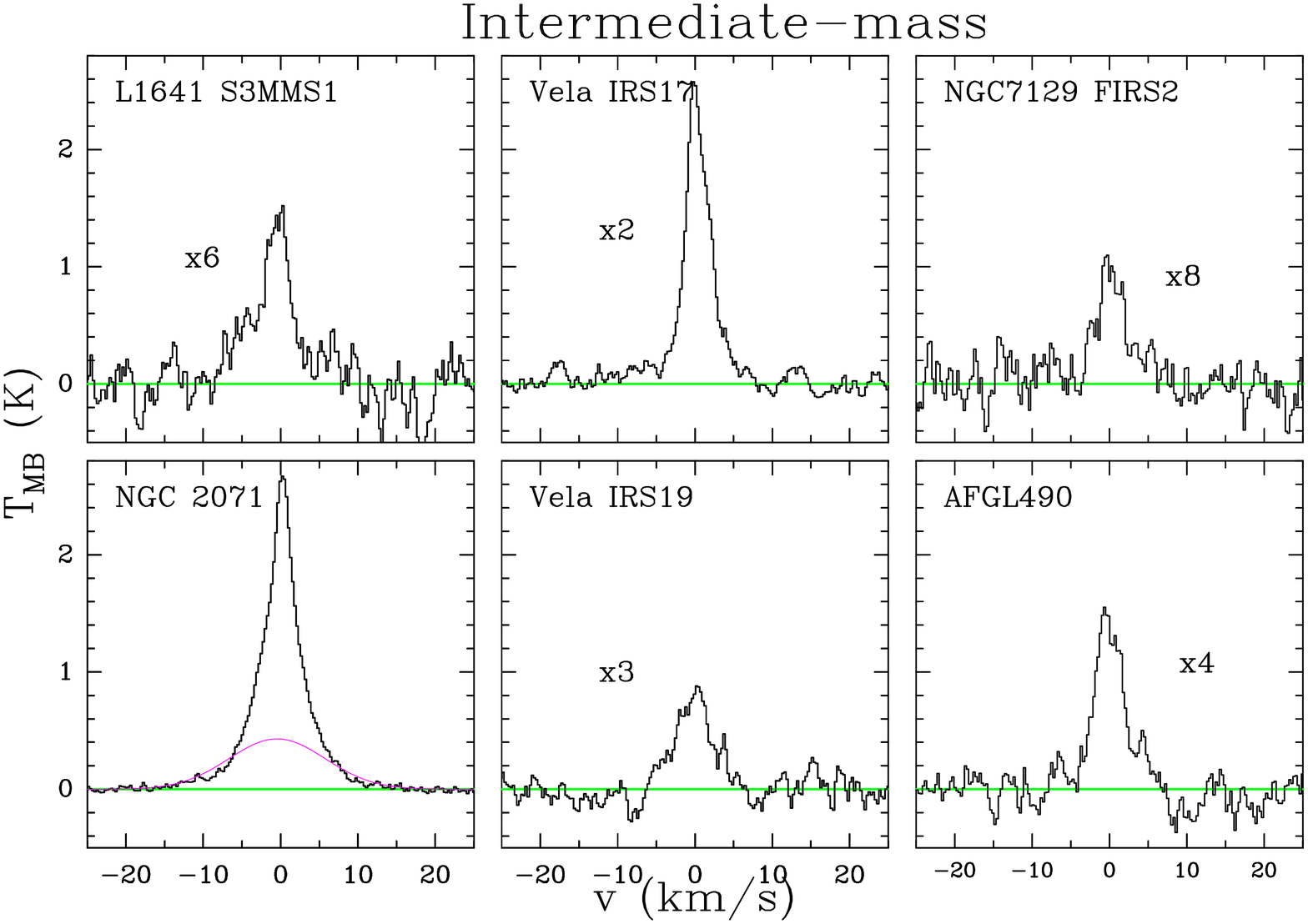}
    \smallskip
    \includegraphics[scale=0.30, angle=0]{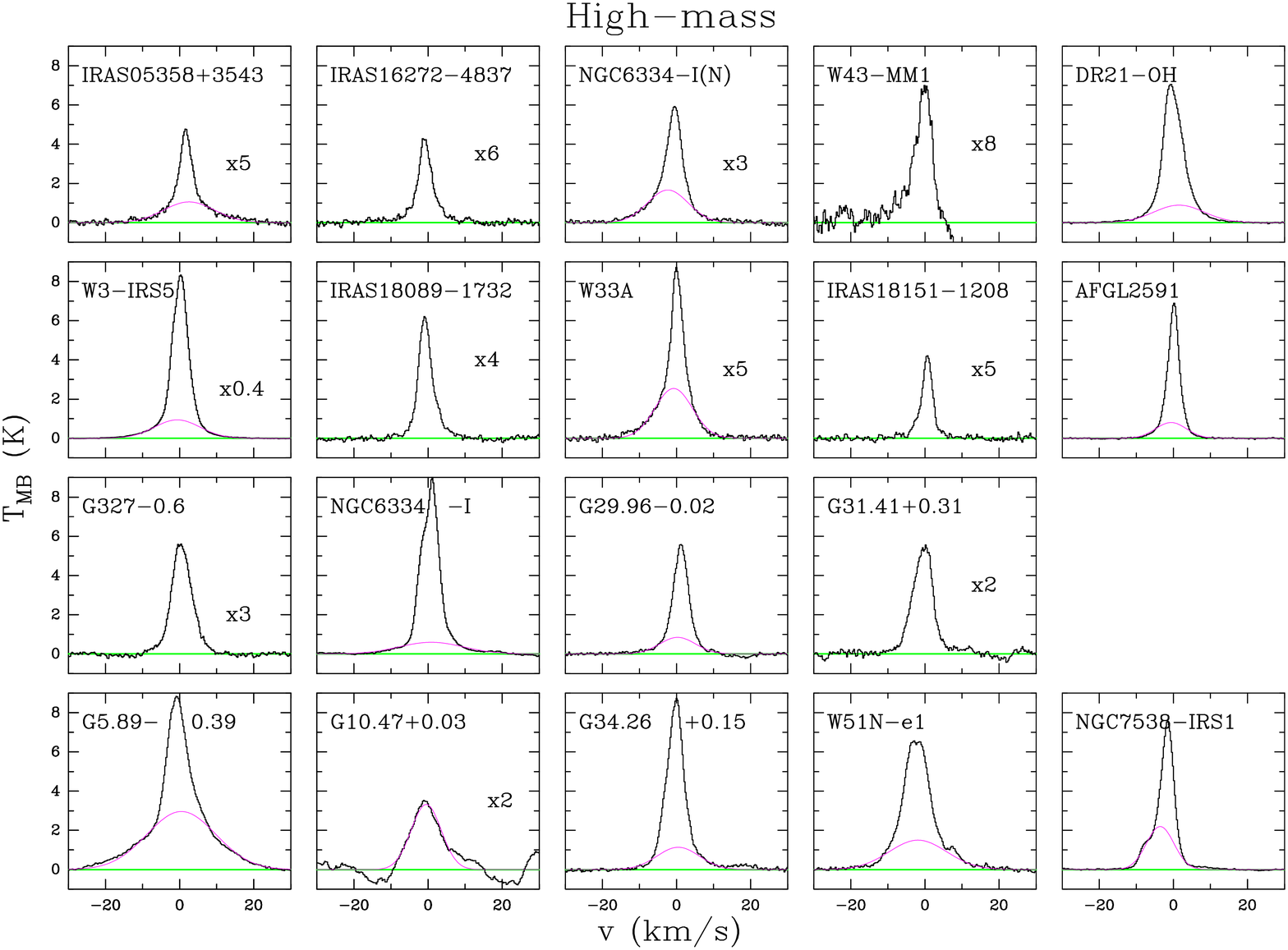}    
    \caption{\label{fig:Spec13CO10-9} Same as Fig.~\ref{fig:Spec12CO10-9} but for the $^{13}$CO~$J$=10--9 
      spectra from the observed low-, intermediate- and high-mass YSOs. In the figure with the high-mass 
      sample, the sources are presented according to their evolutionary stage. 
      In the first row we find the mid-IR-quiet high-mass protostellar objects (HMPOs), in the second
      row the mid-IR-bright HMPOs, in the third line hot molecular
      cores and in the last row the 
      ultra-compact H\small{II} regions.}
  \end{figure*}
}

\def\placeSpectraCochoOcinco{
  \begin{figure}
    \centering
    \bigskip
    \includegraphics[scale=0.30, angle=0]{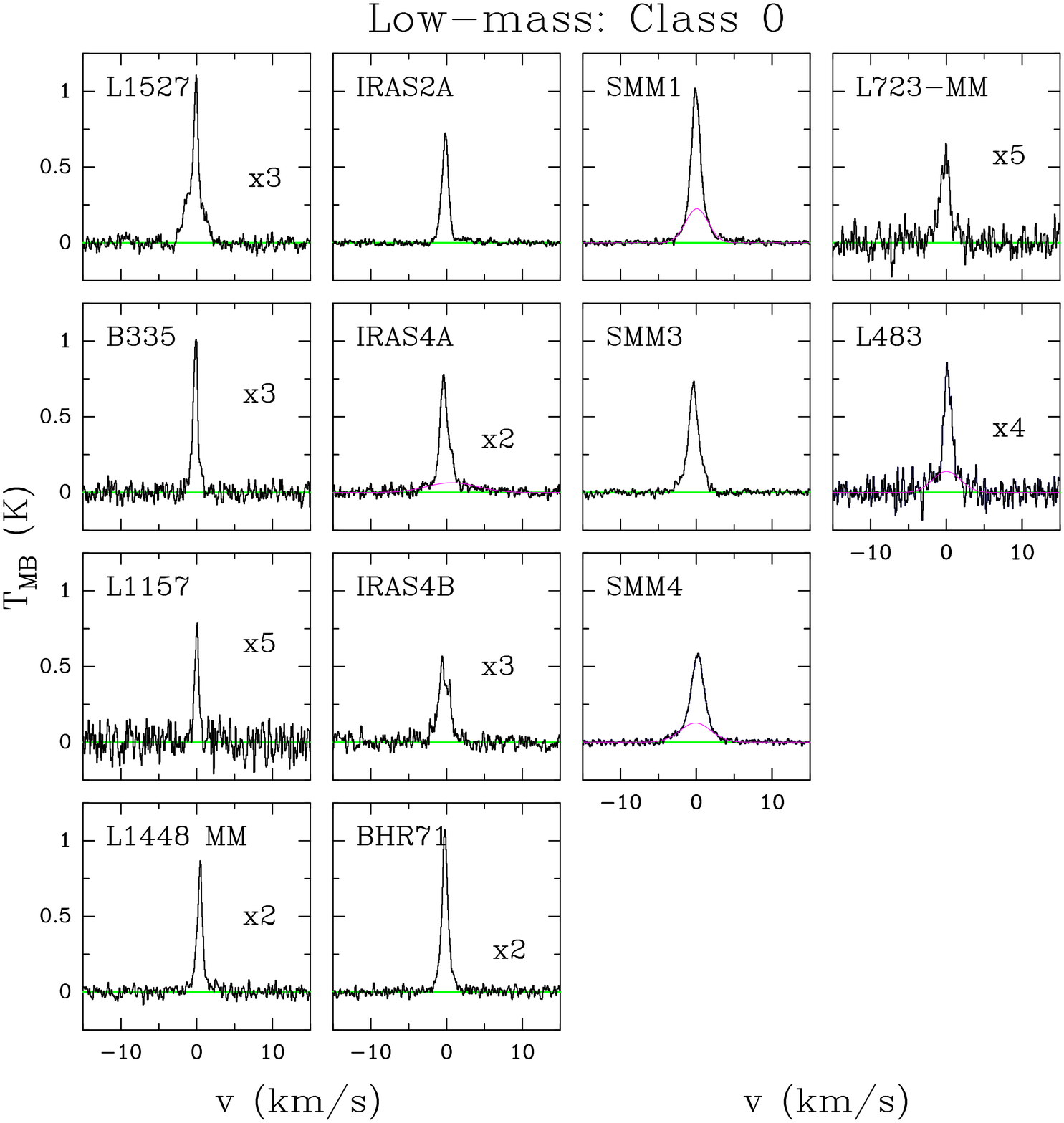}
    \bigskip
    \includegraphics[scale=0.30, angle=0]{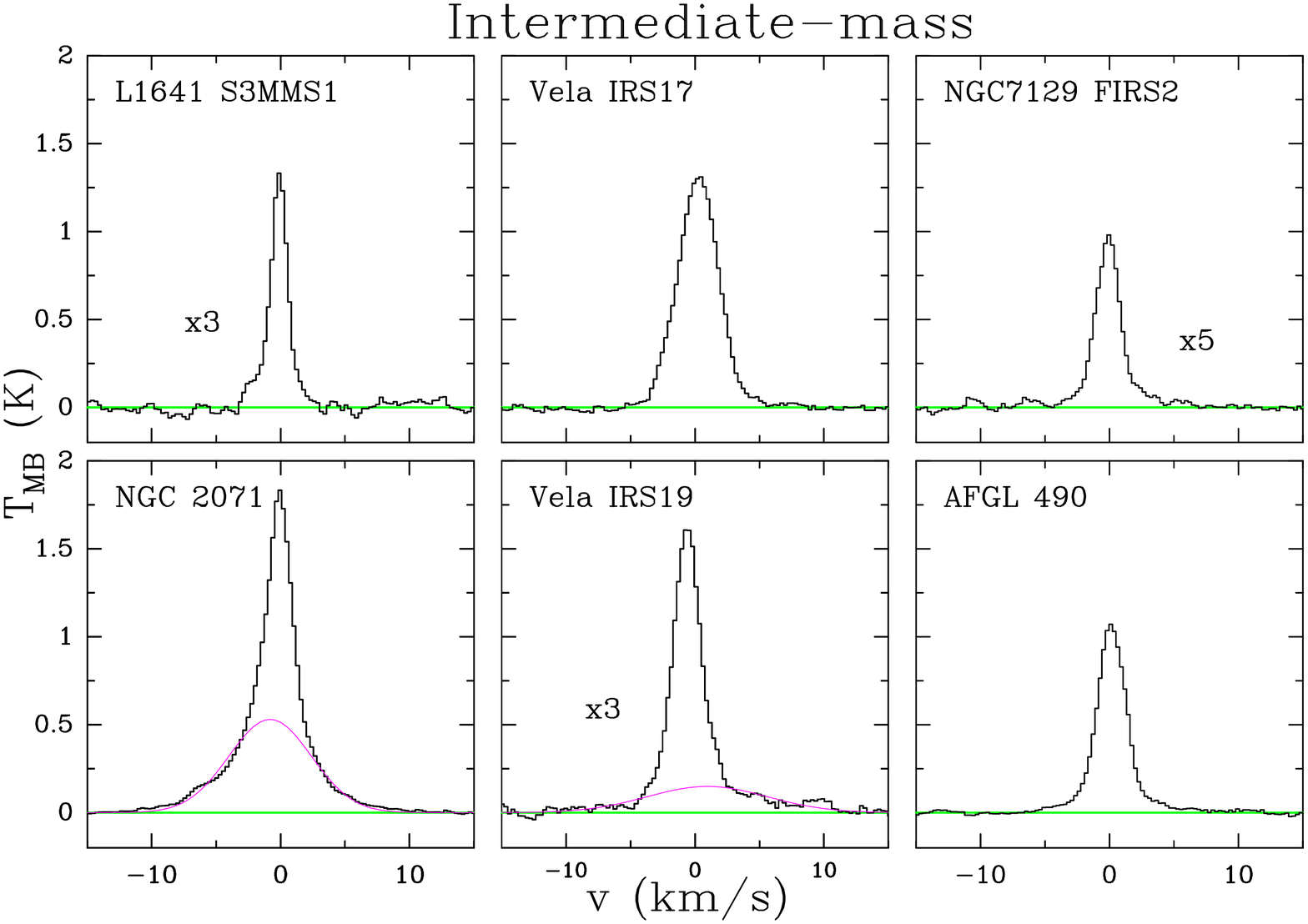}     
    \caption{\label{fig:SpecC18O5-4} Same as Fig.~\ref{fig:Spec12CO10-9} but for the C$^{18}$O~$J$=5--4 spectra 
      of the low-mass (Class~0) and intermediate-mass protostars. For the low-mass sample the HRS spectra are presented 
      while for the intermediate-mass objects the WBS data is used.}
  \end{figure}
}

\def\placeSpectraCochoOnueve{
  \begin{figure*}
    \centering
    \bigskip
    \includegraphics[scale=0.3, angle=0]{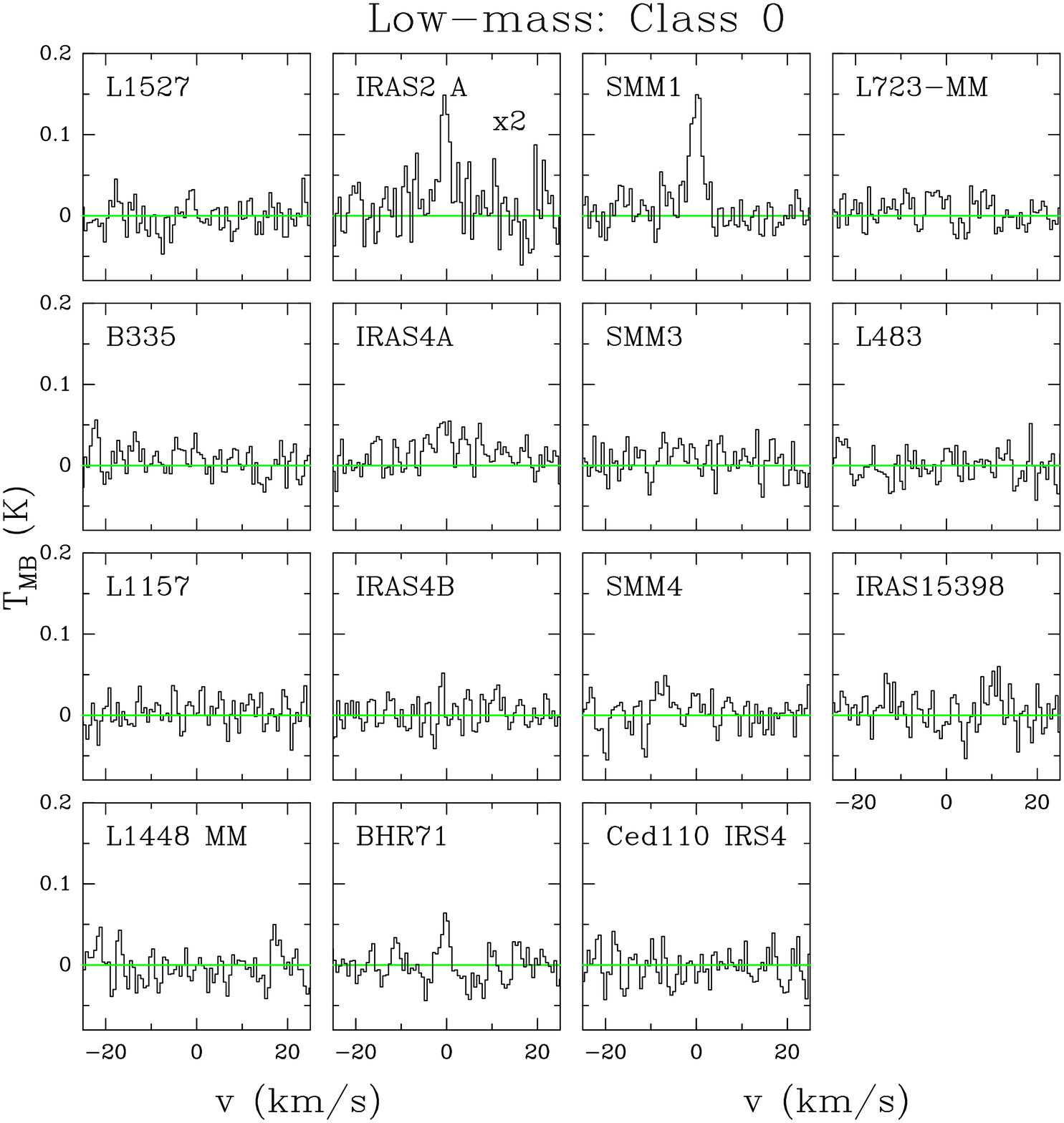}
    \bigskip
    \includegraphics[scale=0.3, angle=0]{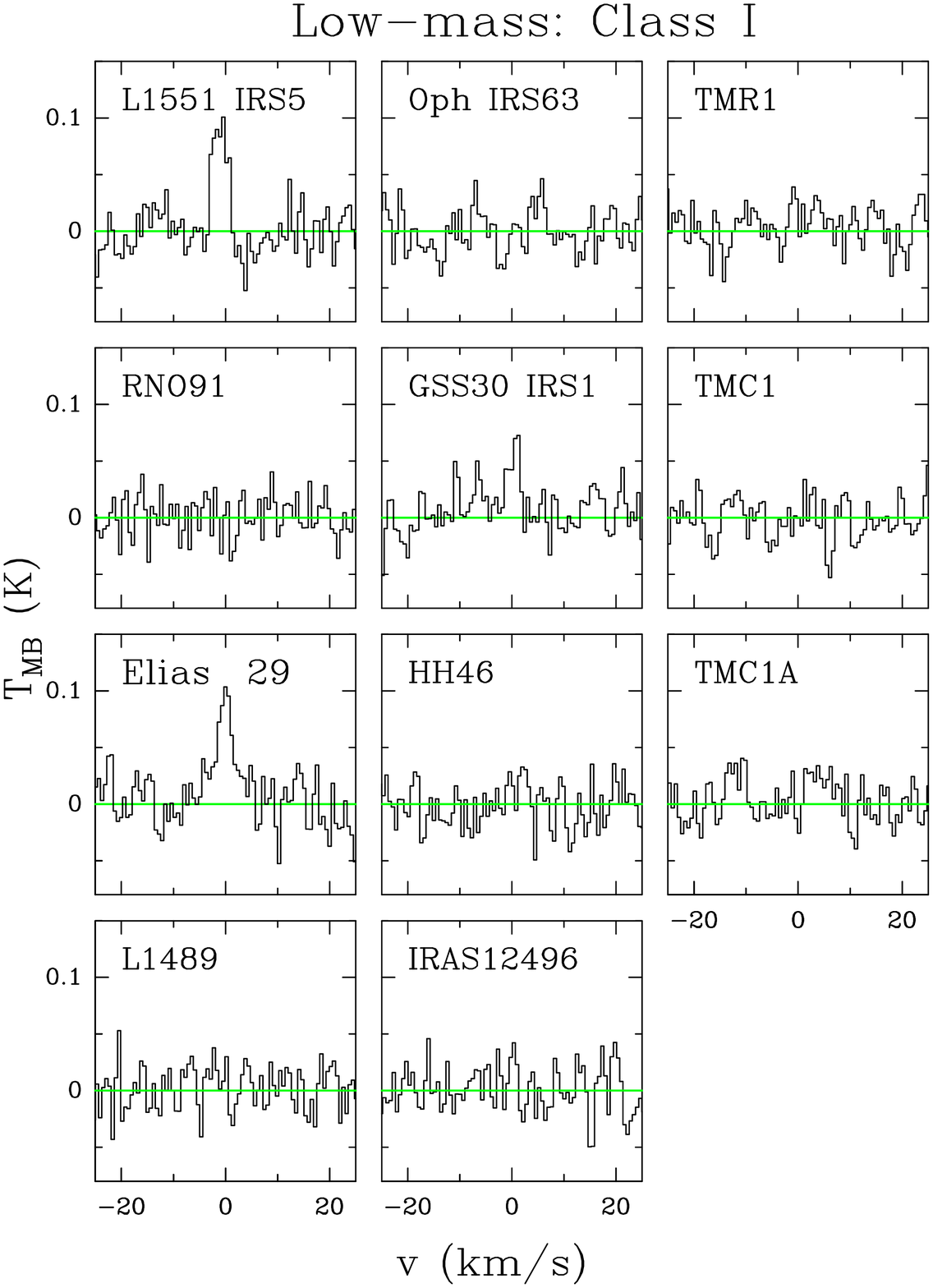}
    \bigskip
    \includegraphics[scale=0.30, angle=0]{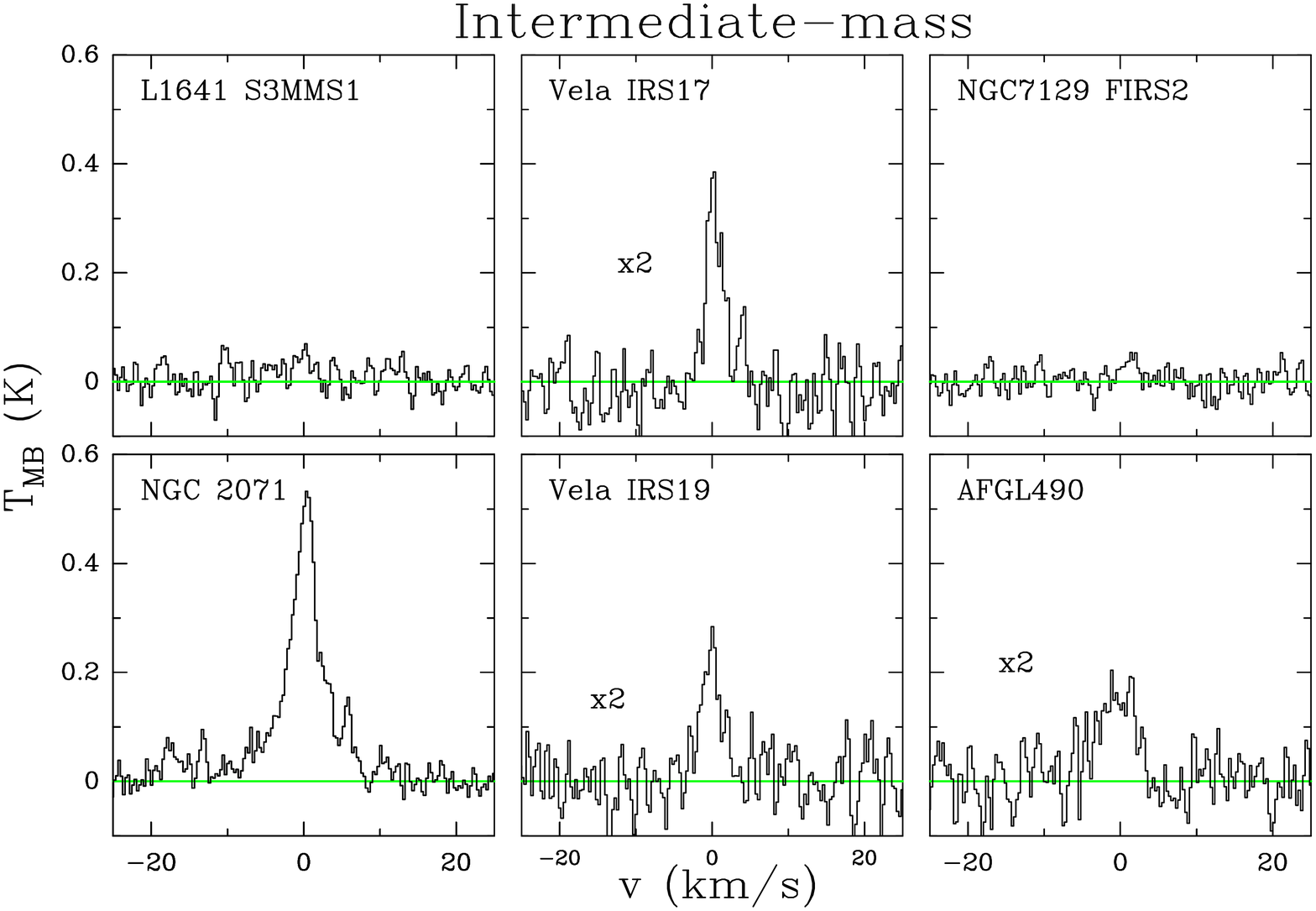}
    \bigskip
    \includegraphics[scale=0.3, angle=0]{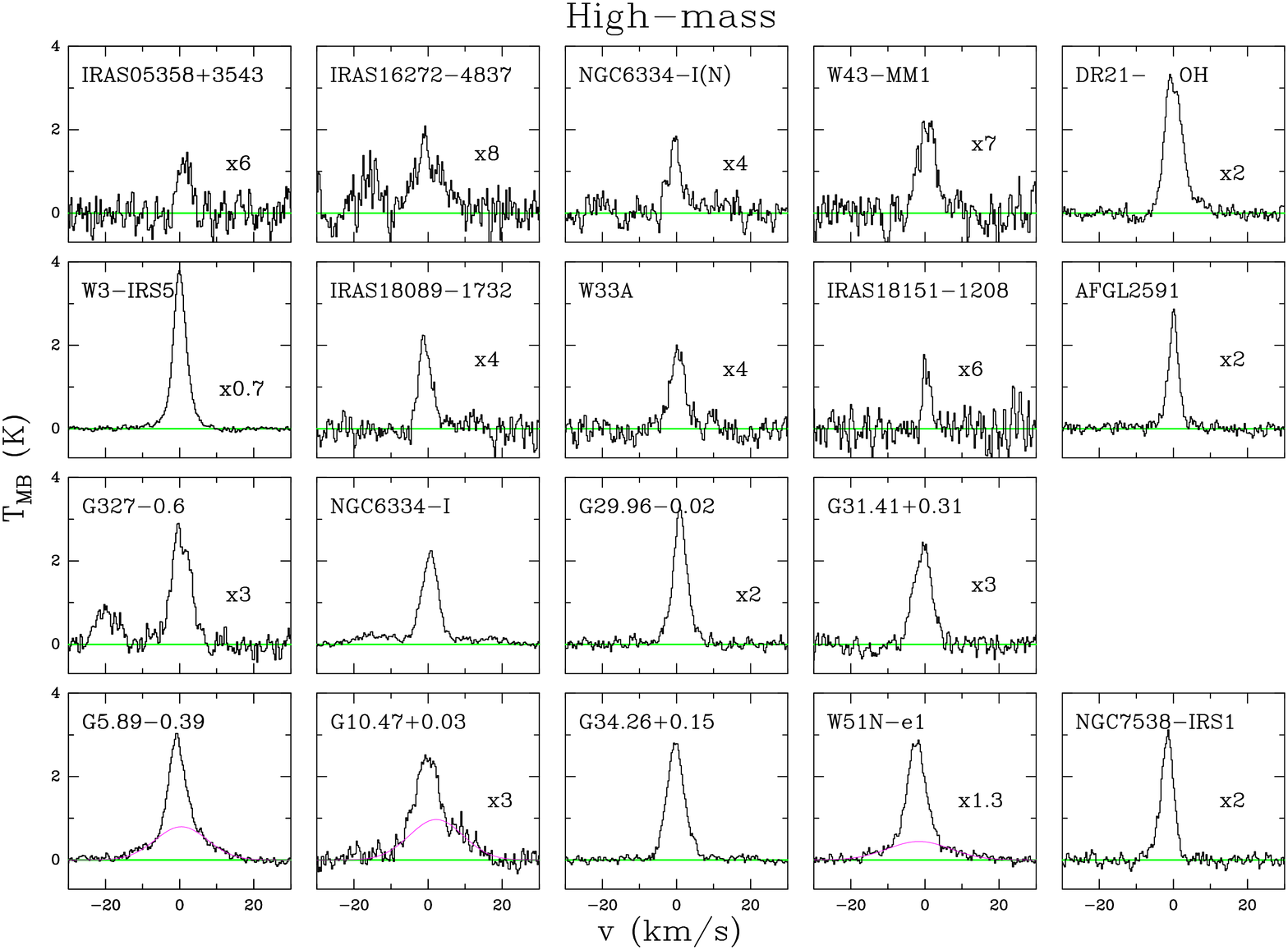}
    \caption{\label{fig:SpecC18O9-8} Same as Fig.~\ref{fig:Spec12CO10-9} but for the C$^{18}$O~$J$=9--8 spectra from
      the low-, intermediate- and high-mass YSOs. The high-mass objects are presented according to their 
      evolutionary stage, as is explained in Fig.~\ref{fig:Spec13CO10-9}.}
  \end{figure*}
}

\def\placeSpectraCochoOdiez{
  \begin{figure*}
    \centering
    \bigskip
    \includegraphics[scale=0.34, angle=0]{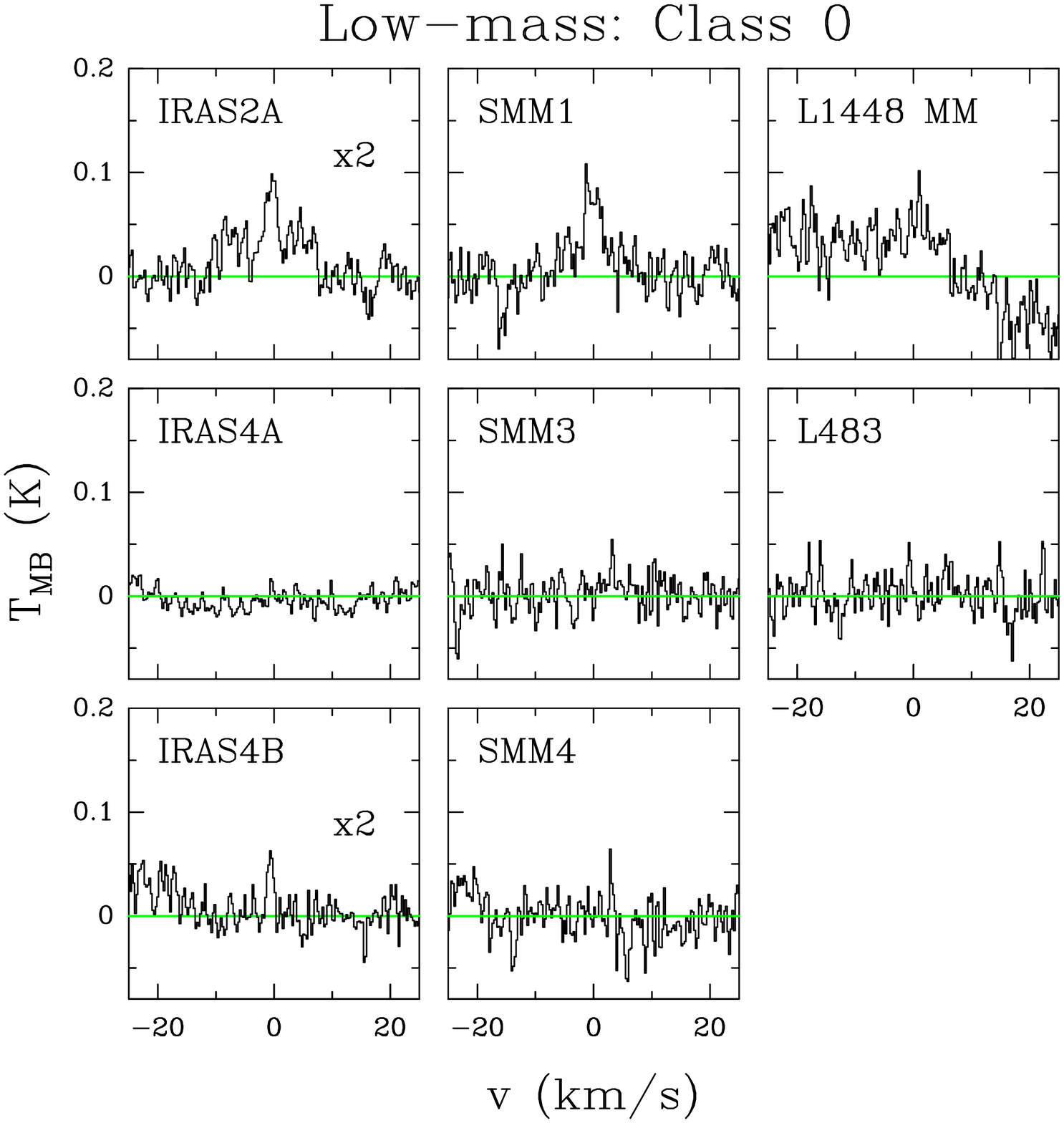}
    \bigskip
    \includegraphics[scale=0.34, angle=0]{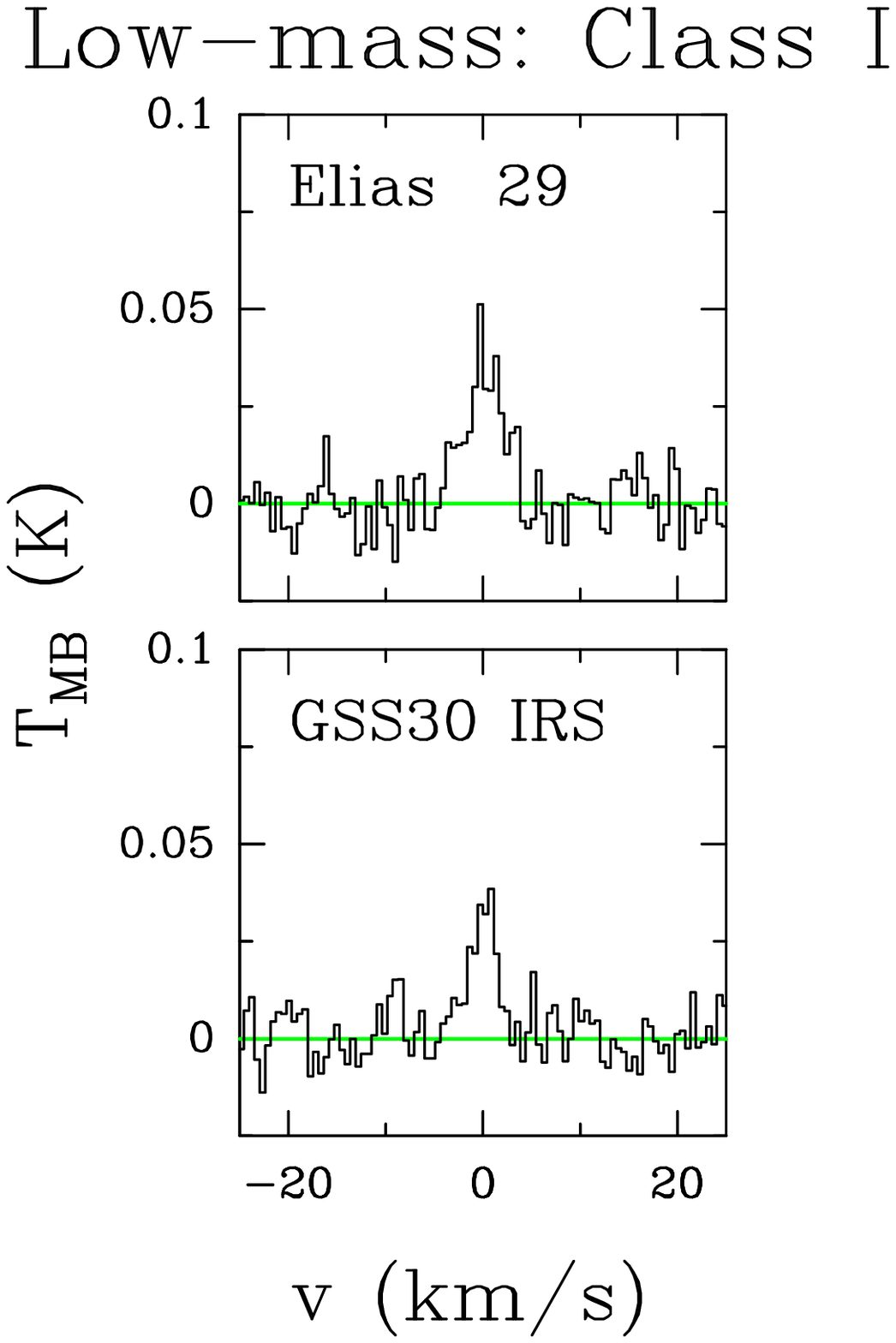}
    \bigskip
    \includegraphics[scale=0.30, angle=0]{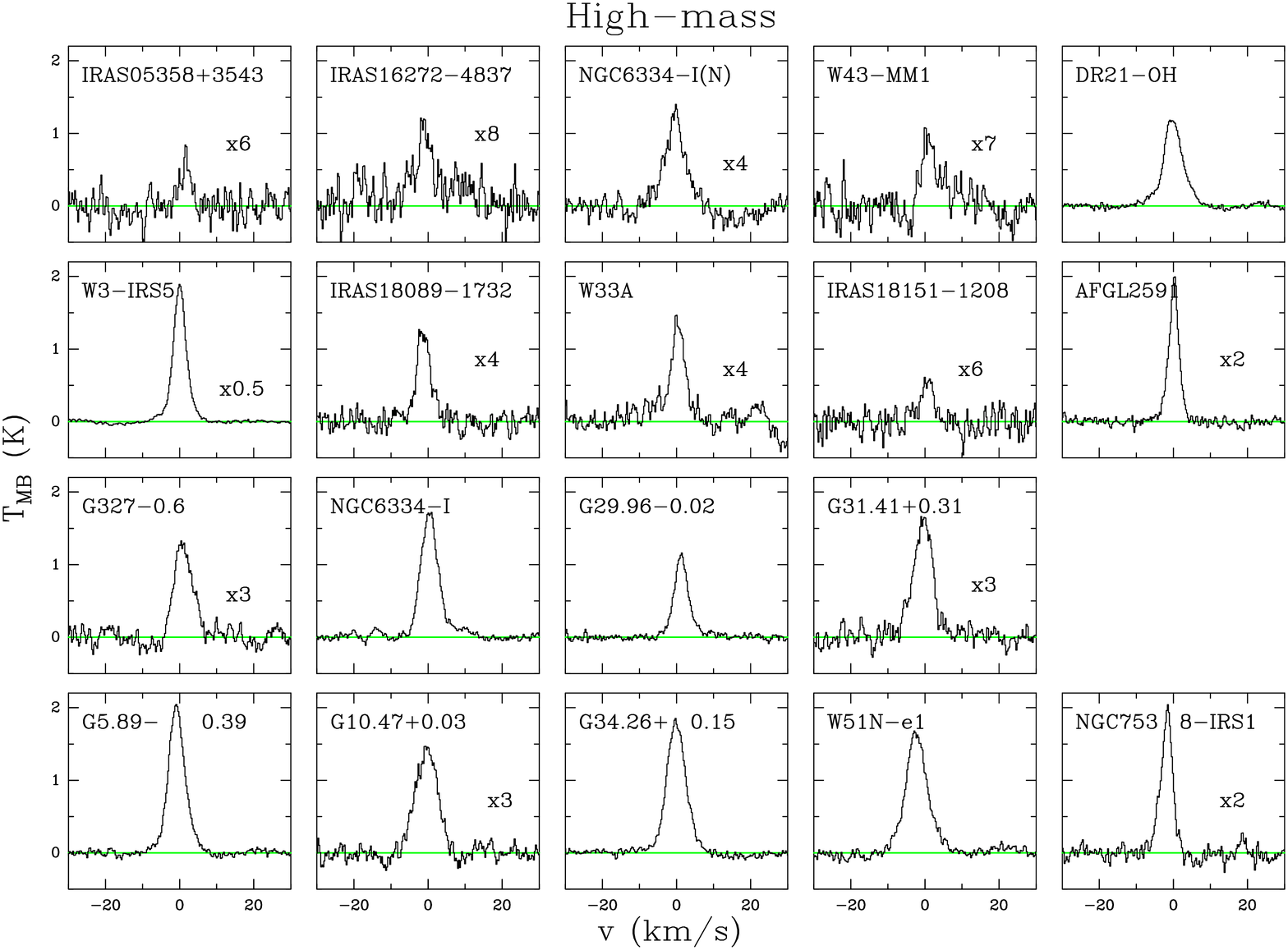}
    \caption{\label{fig:SpecC18O10-9} Same as Fig.~\ref{fig:Spec12CO10-9} but for the C$^{18}$O~$J$=10--9 spectra 
      of the observed low- and high-mass YSOs. In the case of the high-mass sources, the YSOs are disposed 
      as in Fig.~\ref{fig:Spec13CO10-9}. The line wings of the 3$_{12}-3_{03}$ water transition for the high-mass sources 
      have been fitted with a Gaussian profile, subtracted and the residuals plotted to isolate the C$^{18}$O~$J$=10--9 emission line.}
  \end{figure*}
}



\def\placeFigHIFIuvJCMTLM{
  \begin{figure}[!]
    \centering
    \includegraphics[scale=0.5, angle=0]{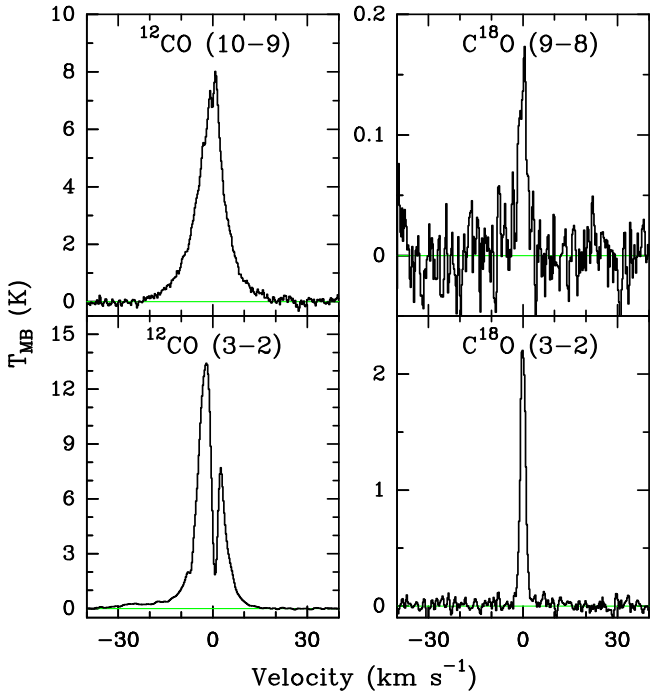}
    \caption{ Comparison between high-$J$ HIFI and low-$J$ JCMT spectra for the low-mass source Ser\,SMM1.
      $^{12}$CO~$J$=10--9 spectra (left-top) and 
      C$^{18}$O~$J$=9--8 line (right-top) observed with HIFI and for the $^{12}$CO~$J$=3--2 and 
      C$^{18}$O~$J$=3--2 lines (left-bottom and right-bottom respectively) observed with JCMT. 
      The spectra have been resampled to 0.27 km\,s$^{-1}$ and shifted
      to zero velocity. The green line indicates the baseline subtraction.}
    \label{fig:HIFIvsJCMTLM}
  \end{figure}
}

\def\placeFigHIFIuvJCMTIM{
  \begin{figure}[!h]
    \centering
    \includegraphics[scale=0.5, angle=0]{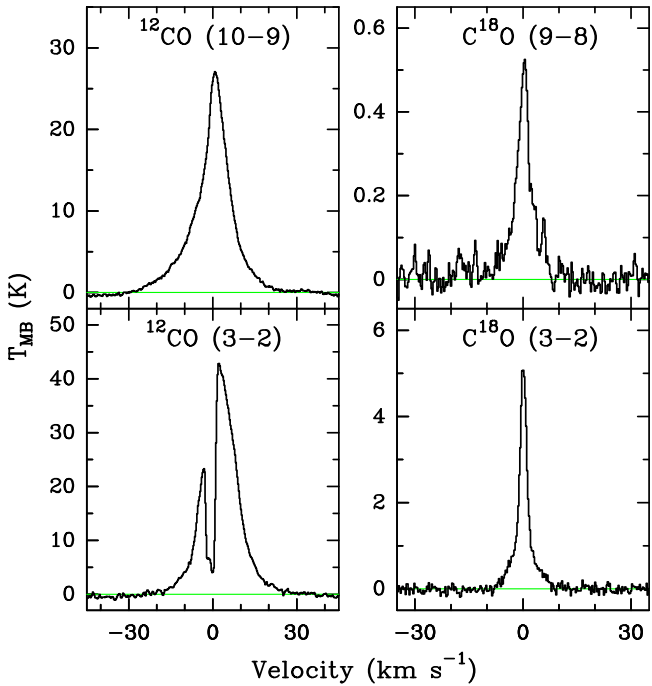}
    \caption{ Same as Fig.~\ref{fig:HIFIvsJCMTLM} but for the
      intermediate-mass source NGC2071.}
      \label{fig:HIFIvsJCMTIM}
  \end{figure}
} 

\def\placeFigHIFIuvJCMT{
  \begin{figure}[!h]
    \centering
    \includegraphics[scale=0.5, angle=0]{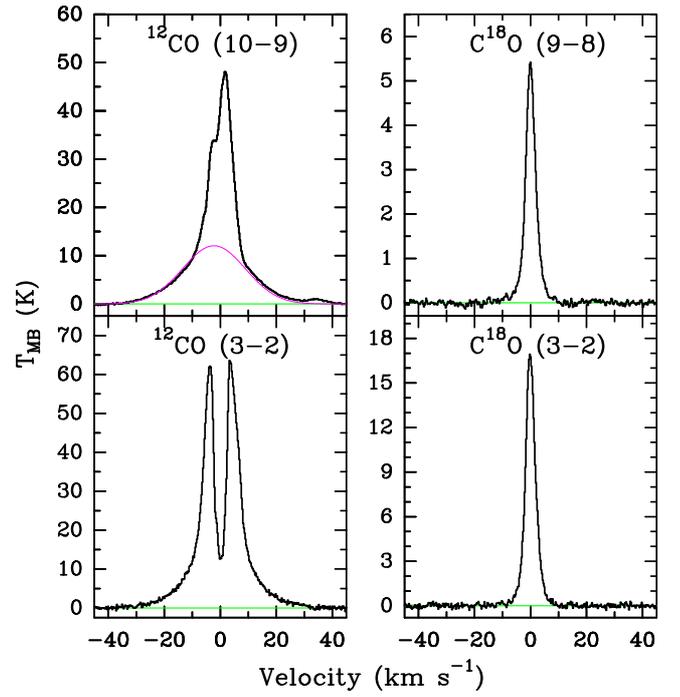}
    \caption{ Same as Fig.~\ref{fig:HIFIvsJCMTLM} but for the
      high-mass source W3-IRS5. The pink Gaussian profile represents the broad velocity component
      identified in the $^{12}$CO~$J$=10--9 line.}
    \label{fig:HIFIvsJCMT}
  \end{figure}
}

\def\placeSpectraCOJCMT{
  \begin{figure*}[h]
    \centering
    \bigskip
    \includegraphics[scale=0.30, angle=0]{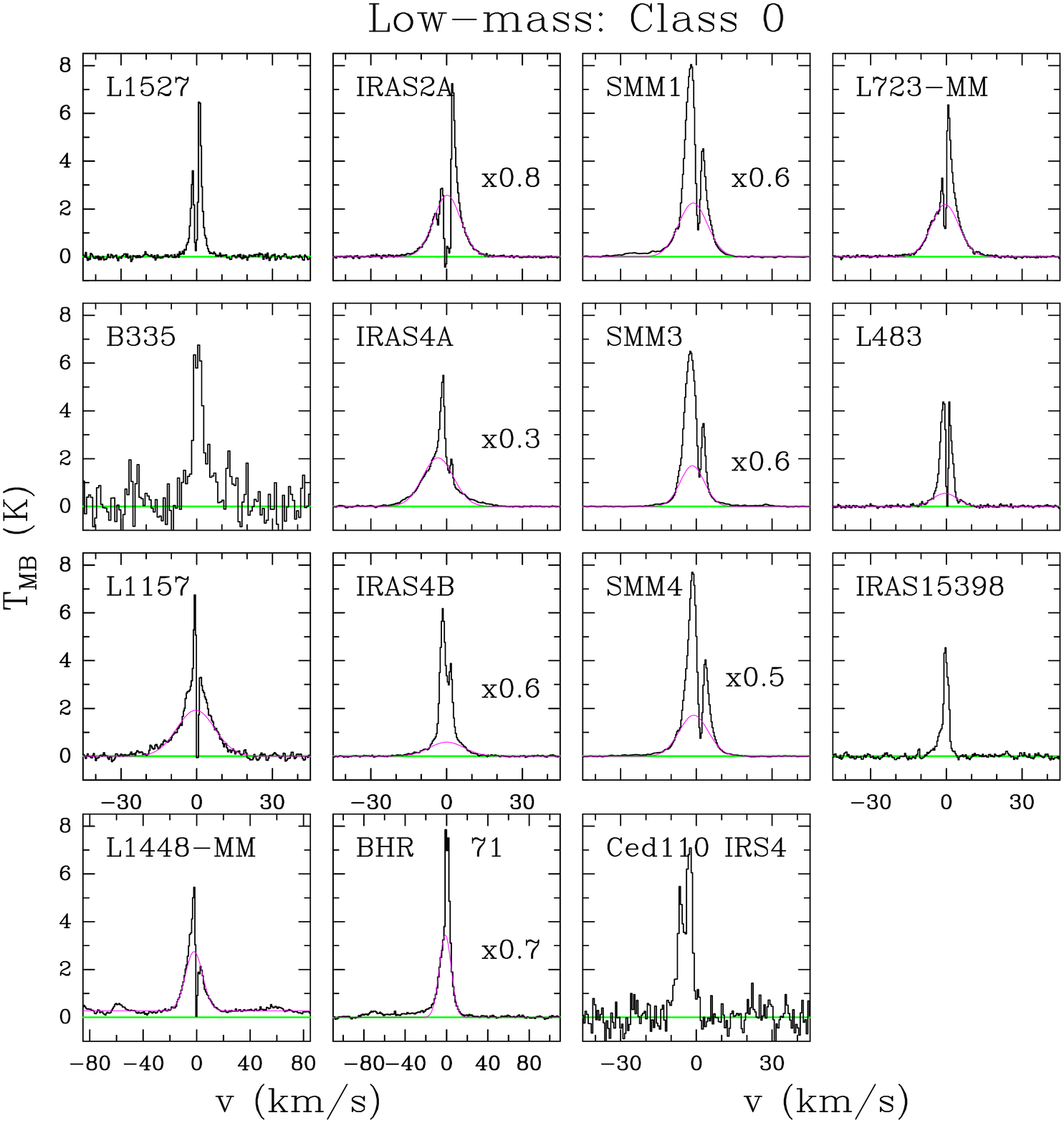}
    \bigskip
    \includegraphics[scale=0.30, angle=0]{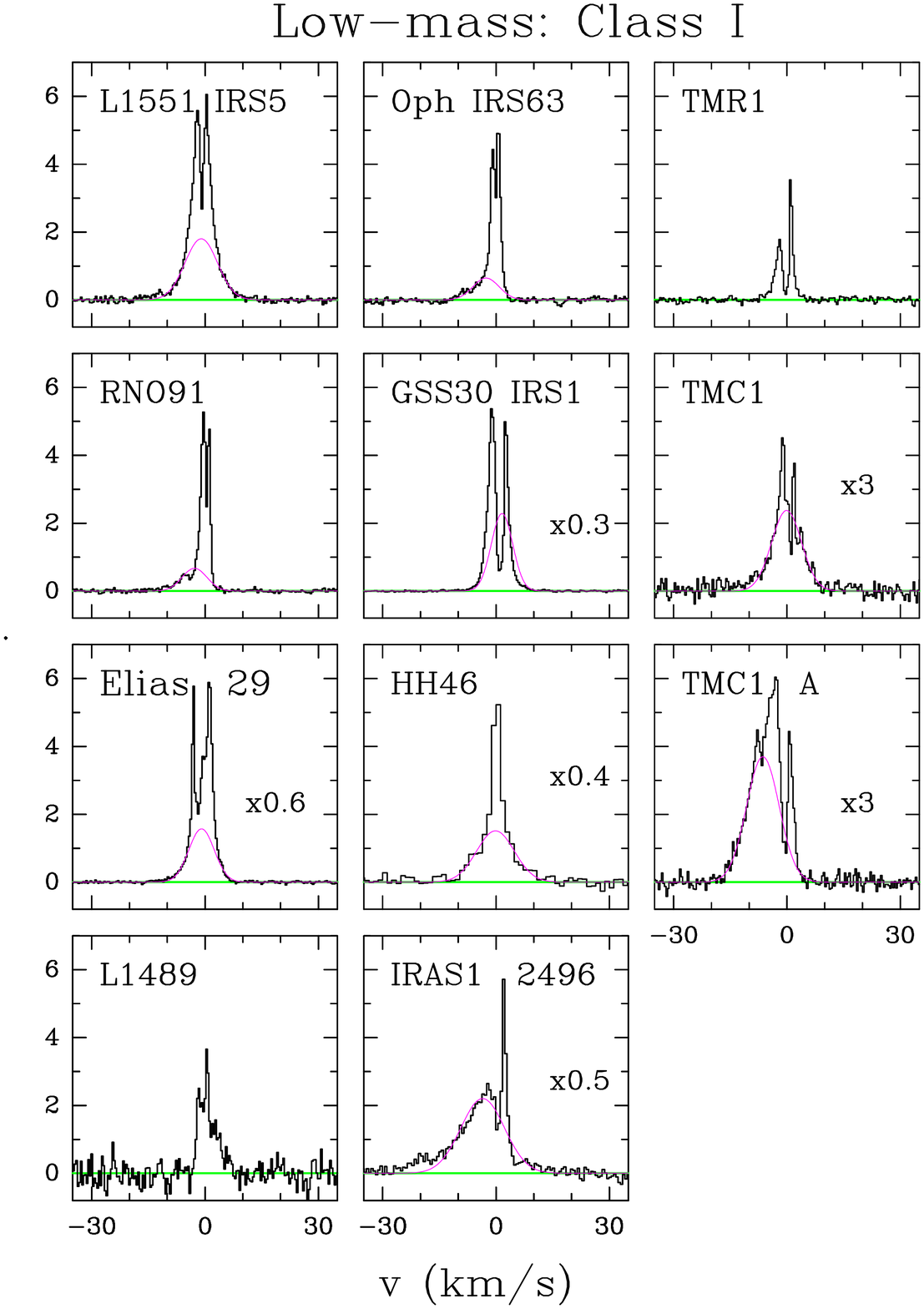}
    \smallskip
    \includegraphics[scale=0.30, angle=0]{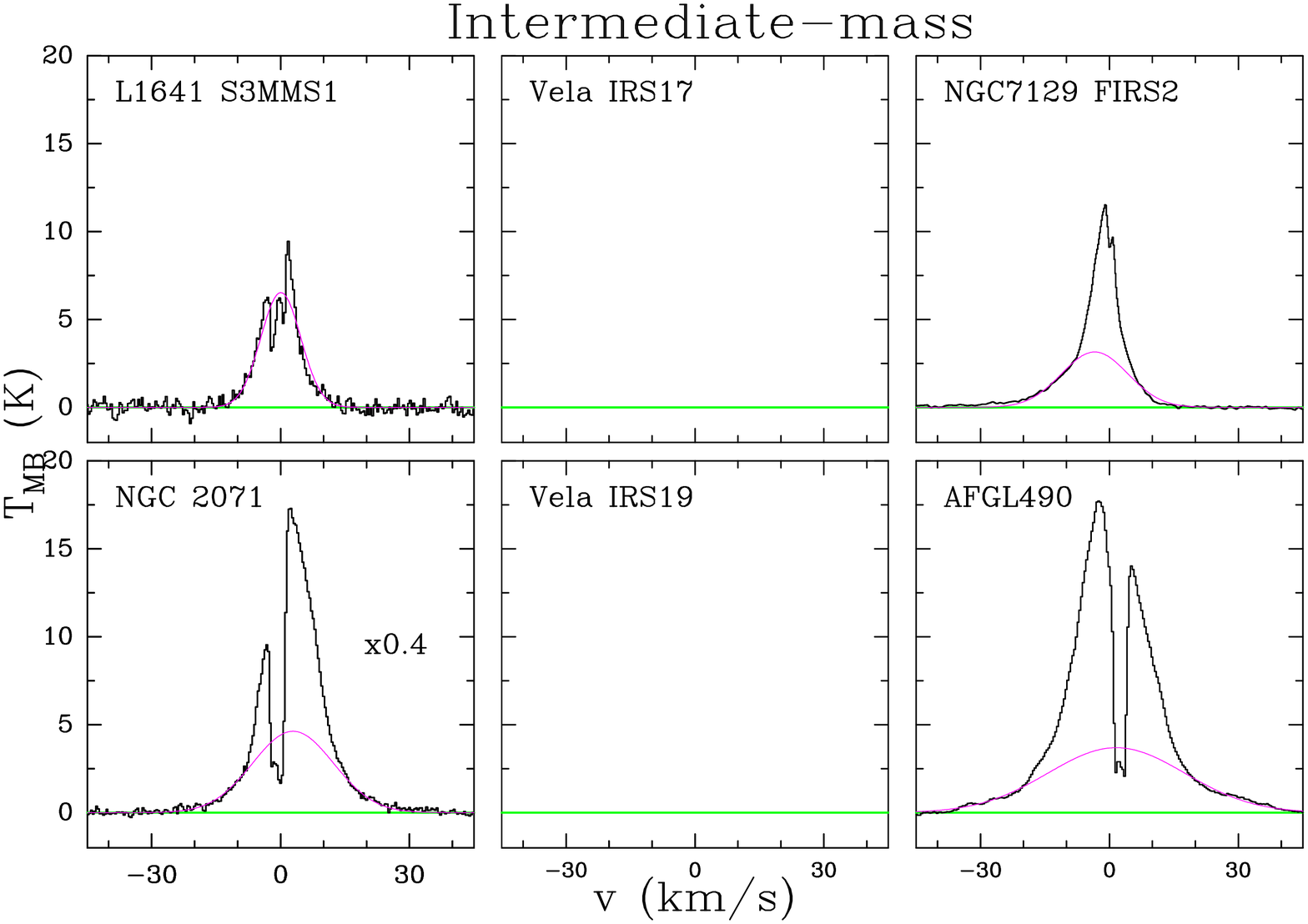}
    \smallskip
    \includegraphics[scale=0.30, angle=0]{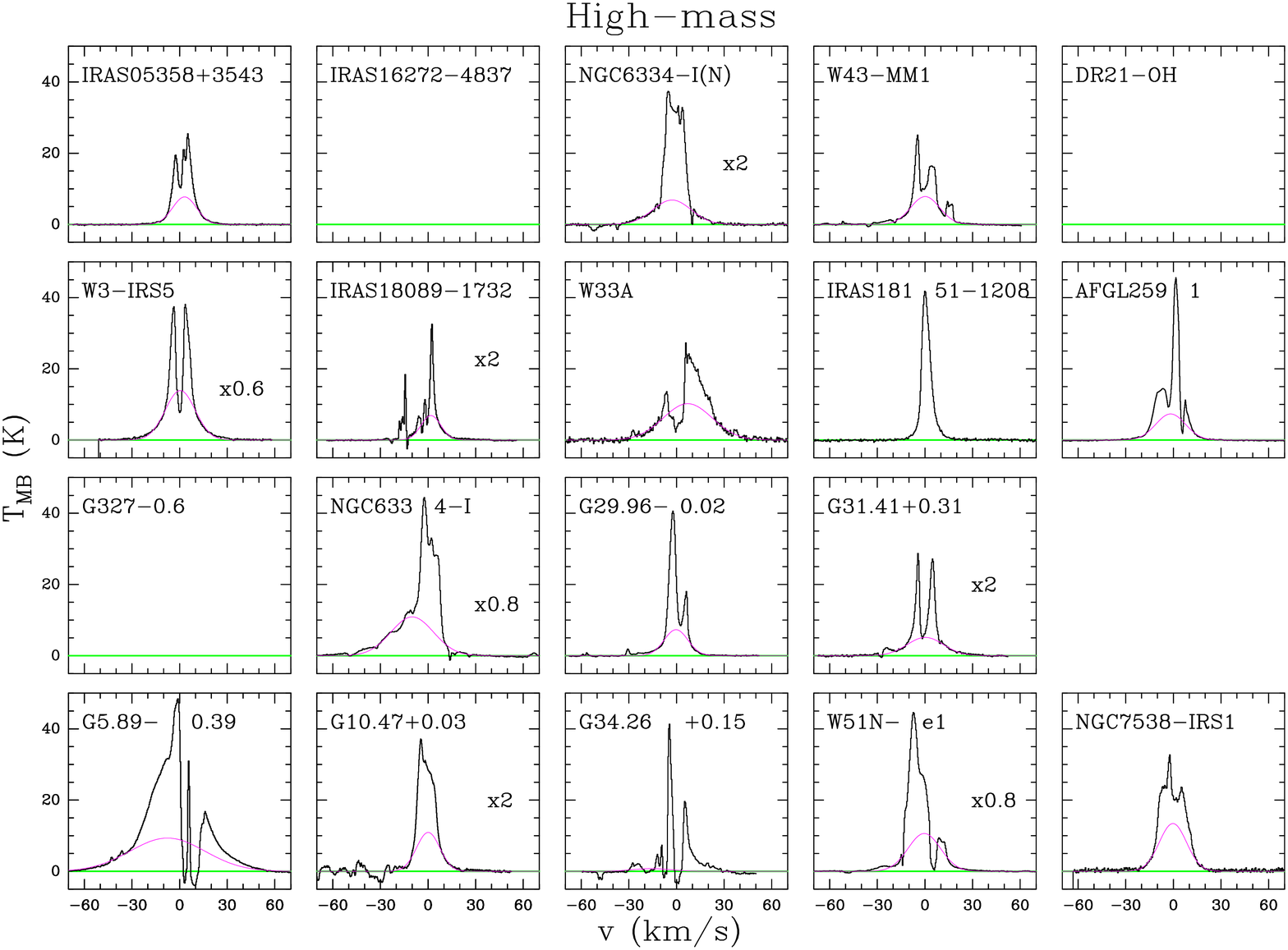}
    \caption{\label{fig:Spec12CO3-2} $^{12}$CO~$J$=3--2 spectra for low-, intermediate and high-mass YSOs.
      observed with the JCMT. The green 
      line represents the baseline level and the pink Gaussian profile
      the broad velocity component for those sources for which a 2 Gaussian decomposition has been performed. 
      All the spectra have been shifted to zero velocity. The numbers indicate where the spectra have been 
      scaled for greater visibility.}
  \end{figure*}
}

\def\placeSpectraCochoOnueveJCMT{
  \begin{figure*}
    \centering
    \bigskip
    \includegraphics[scale=0.3, angle=0]{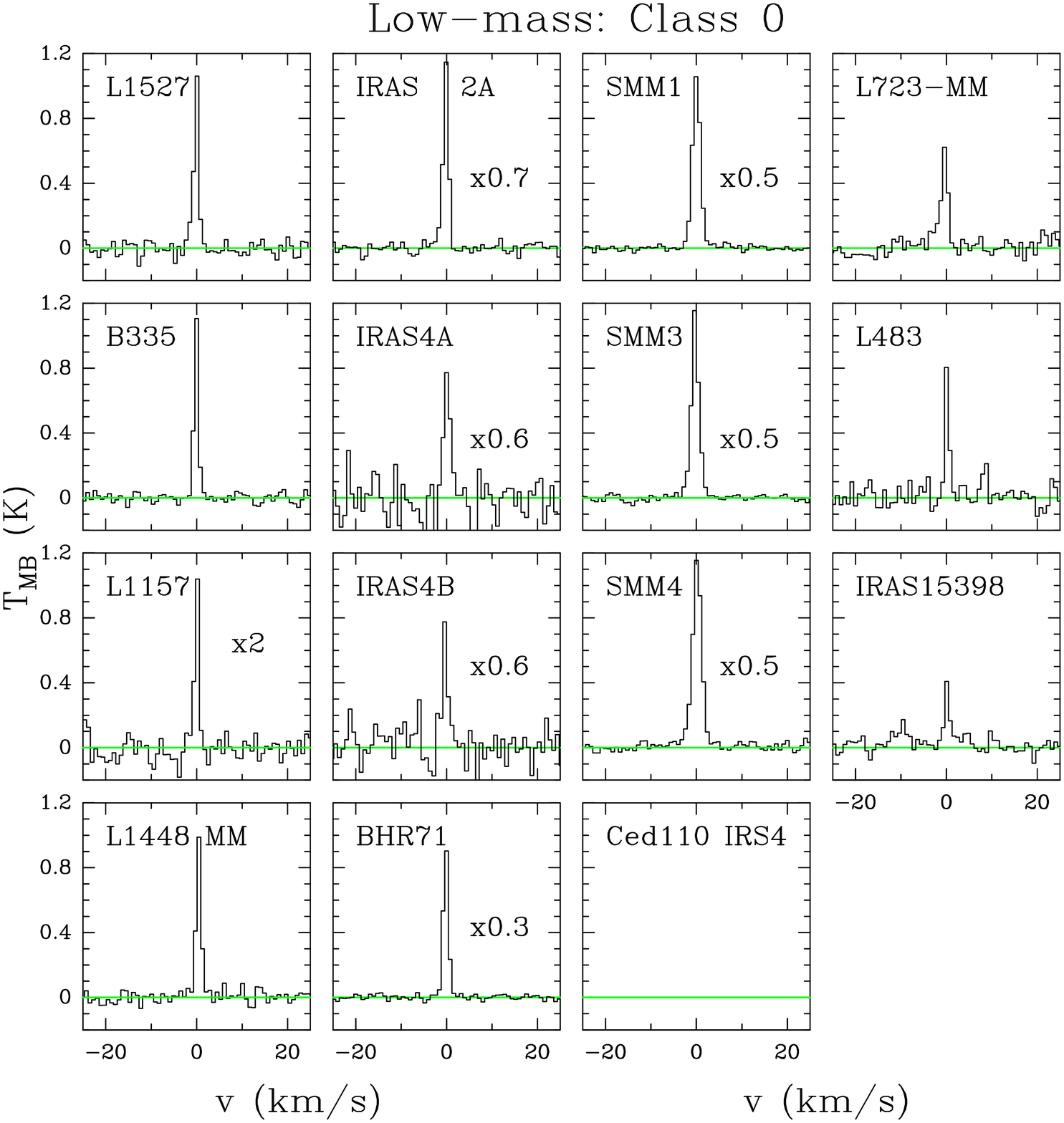}
    \bigskip
    \includegraphics[scale=0.3, angle=0]{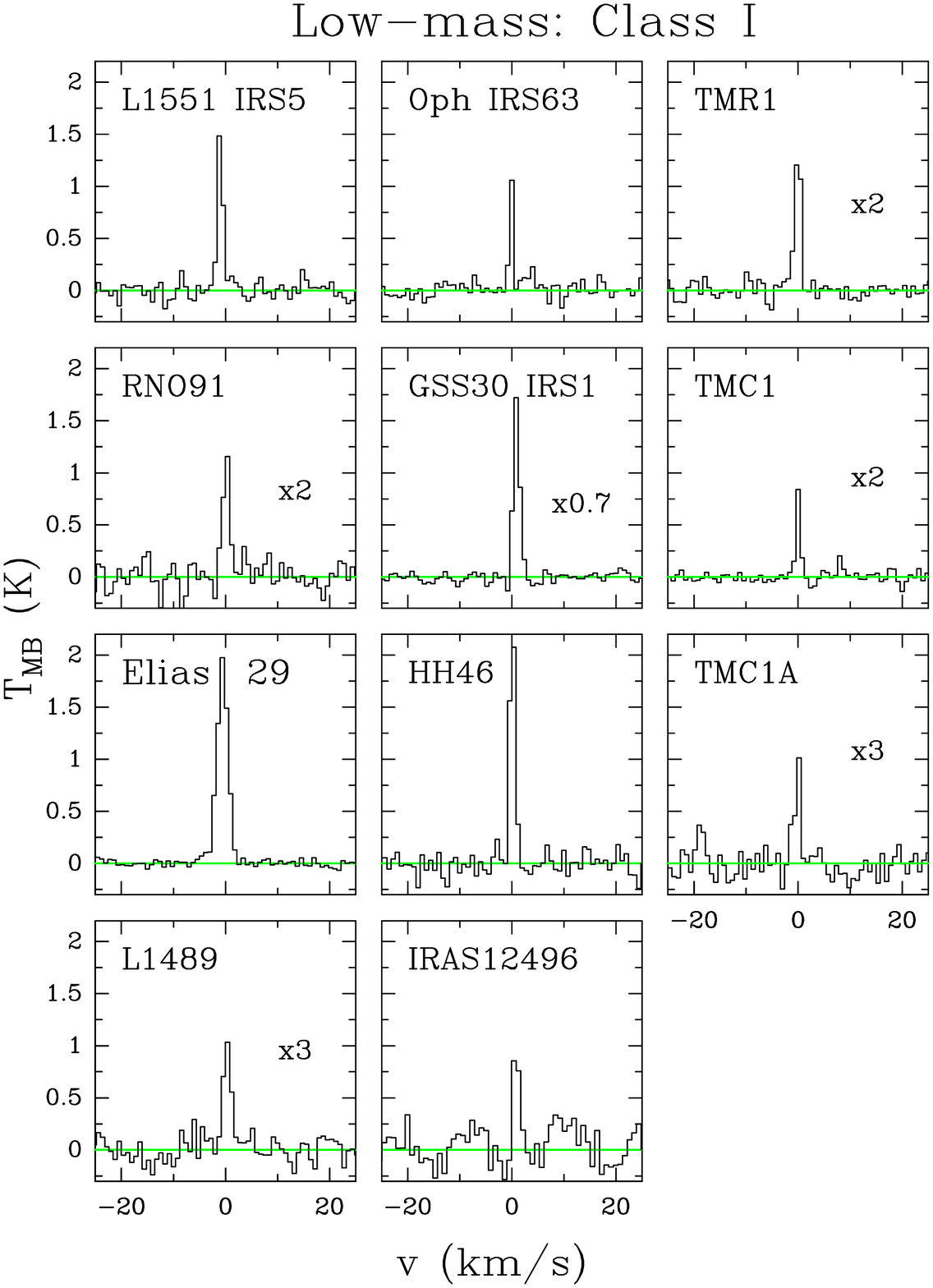}
    \bigskip
    \includegraphics[scale=0.30, angle=0]{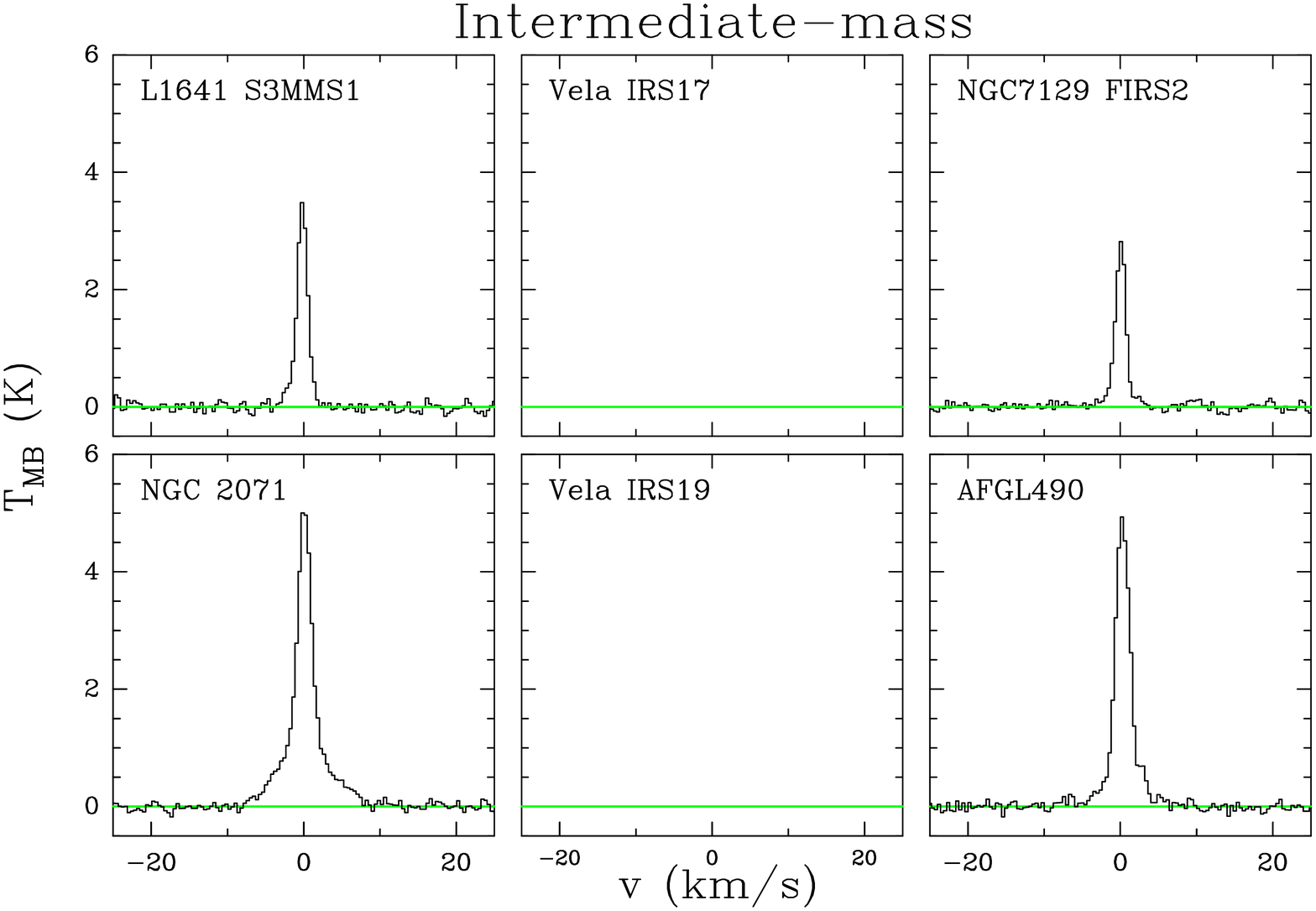}
    \bigskip
    \includegraphics[scale=0.3, angle=0]{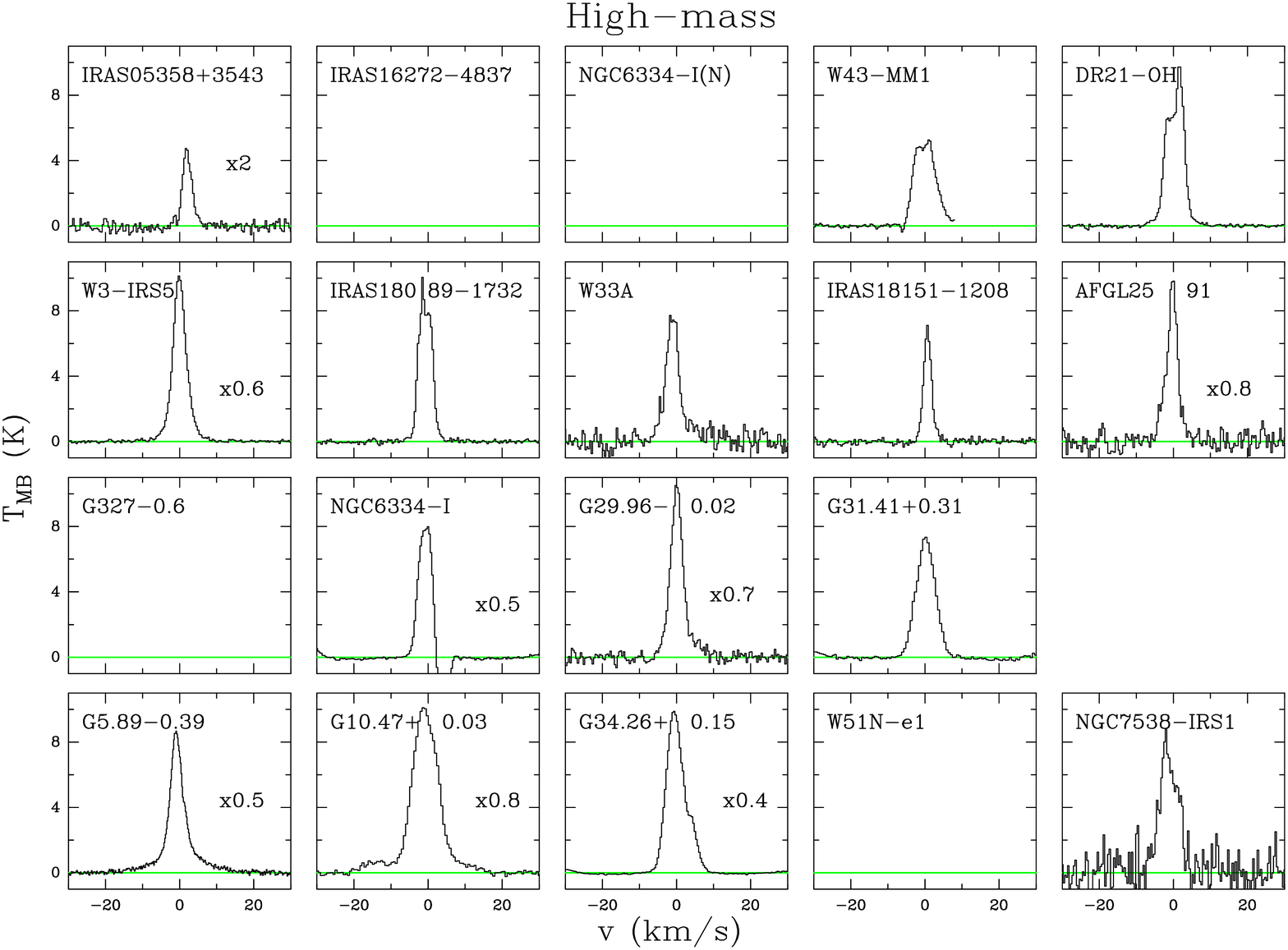}
    \caption{\label{fig:SpecC18O3-2} Same as Fig.~\ref{fig:Spec12CO3-2} but for the C$^{18}$O~$J$=3--2 spectra from
      the low-, intermediate- and high-mass YSOs.}
  \end{figure*}
}



\def\placeTableSources{
  \begin{table*}[!h]
    \begin{center}
      \caption{\label{tbl:sources} Source parameters.}
      \begin{tabular}{l c cc r c}
        \hline\hline
        Source     &  $\varv_{\rm LSR}$  & $L_{\rm bol}$ & $d$ & $M_{\rm env}$ & References\\
        & (km\,s$^{-1}$) &($L_\odot$) & (kpc)  & ($M_\odot$)& \\
        \hline
            {\bf Low-mass: Class\,0}&        &    &      &   & \\
            L\,1448-MM            & $+5.2$   & 9.0  & 0.235 &  3.9   & 1 \\
            NGC\,1333 IRAS\,2A    & $+7.7$   & 35.7 & 0.235 &  5.1   & 1 \\
            NGC\,1333 IRAS\,4A    & $+7.0$   & 9.1  & 0.235 &  5.6   & 1 \\
            NGC\,1333 IRAS\,4B    & $+7.4$   & 4.4  & 0.235 &  3.0   & 1 \\
            L\,1527               & $+5.9$   & 1.9  & 0.140 &  0.9   & 1 \\
            Ced110 IRS4           & $+4.2$   & 0.8  & 0.125 &  0.2   & 1 \\
            BHR\,71               & $-4.4$   & 14.8 & 0.200 &  2.7   & 1 \\
            IRAS\,15398           & $+5.1$   &  1.6 & 0.130 &  0.5   & 1 \\
            L\,483-MM             & $+5.2$   & 10.2 & 0.200 &  4.4   & 1 \\
            Ser SMM\,1            & $+8.5$   & 30.4 & 0.230 & 16.1   & 1 \\
            Ser SMM\,4            & $+8.0$   & 1.9  & 0.230 &  2.1   & 1 \\
            Ser SMM\,3            & $+7.6$   & 5.1  & 0.230 &  3.2   & 1 \\
            L\,723-MM             & $+11.2$  & 3.6  & 0.300 &  1.3   & 1 \\
            B\,335                & $+8.4$   & 3.3  & 0.250 &  1.2   & 1 \\
            L\,1157               & $+2.6$   & 4.7  & 0.325 &  1.5   & 1 \\
        \hline
            {\bf Low-mass: Class\,I}&        &    &     &   &\\
            L\,1489               & $+7.2$  &  3.8  & 0.140   &  0.2   & 1 \\
            L\,1551 IRS\,5        & $+6.2$  & 22.1  & 0.140   &  2.3   & 1 \\
            TMR\,1                & $+6.3$  &  3.8  & 0.140   &  0.2   & 1 \\
            TMC\,1A               & $+6.6$  &  2.7  & 0.140   &  0.2   & 1 \\
            TMC\,1                & $+5.2$  &  0.9  & 0.140   &  0.2   & 1 \\
            HH\,46                & $+5.2$  & 27.9  & 0.450   &  4.4   & 1 \\
            IRAS\,12496           & $+3.1$  & 35.4  & 0.178   &  0.8   & 1 \\
            Elias\,29             & $+4.3$  & 14.1  & 0.125   &  0.3   & 1 \\
            Oph\,IRS\,63          & $+2.8$  &  1.0  & 0.125   &  0.3   & 1 \\
            GSS\,30\,IRS1         & $+3.5$  & 13.9  & 0.125   &  0.6   & 1 \\
            RNO\,91               & $+0.5$  &  2.6  & 0.125   &  0.5   & 1 \\
         \hline
             {\bf Intermediate-mass} &        &    &     &   &\\
             NGC\,7129 FIRS\,2   &  $-$9.8  &   430  &  1.25 & 50.0   & 2 \\
             L1641\,S3\,MMS1     &     5.3  &    70  &  0.50 & 20.9   & 2 \\
             NGC\,2071           &     9.6  &   520  &  0.45 & 30.0   & 2 \\
             Vela\,IRS\,17       &     3.9  &   715  &  0.70 &  6.4   & 2 \\
             Vela\,IRS\,19       &    12.2  &   776  &  0.70 &  3.5   & 2 \\
             AFGL\,490           & $-$13.5  &  2000  &  1.00 & 45.0   & 2 \\
         \hline
             {\bf High-mass} &        &    &    &   &\\ 
             IRAS05358+3543      & $-$17.6     & $6.3\times10^3$     & 1.8 &   142   & 3 \\
             IRAS16272$-$4837    & $-$46.2     & $2.4\times10^4$     & 3.4 &  2170   & 3 \\
             NGC6334-I(N)        & $-$3.3      & $1.9\times10^3$     & 1.7 &  3826   & 3 \\
             W43-MM1             & $+$98.8     & $2.3\times10^4$     & 5.5 &  7550   & 3 \\
             DR21(OH)            & $-$3.1      & $1.3\times10^4$     & 1.5 &   472   & 3 \\
             W3-IRS5             & $-$38.4     & $1.7\times10^5$     & 2.0 &   424   & 3 \\
             IRAS18089$-$1732    & $+$33.8     & $1.3\times10^4$     & 2.3 &   172   & 3 \\
             W33A                & $+$37.5     & $1.1\times10^5$     & 3.8 &  1220   & 3 \\
             IRAS18151$-$1208    & $+$32.8     & $2.0\times10^4$     & 3.0 &   153   & 3 \\
             AFGL2591            & $-$5.5      & $2.2\times10^5$     & 3.3 &   320   & 3 \\
             G327$-$0.6          & $-$45.3     & $5.0\times10^4$     & 3.3 &  2044   & 3 \\
             NGC6334-I           & $-$7.4      & $2.6\times10^5$     & 1.7 &   500   & 3 \\
             G29.96$-$0.02       & $+$97.6     & $3.5\times10^5$     & 6.0 &   768   & 3 \\
             G31.41+0.31         & $+$97.4     & $2.3\times10^5$     & 7.9 &  2968   & 3 \\
             G5.89$-$0.39        & $+$10.0     & $5.1\times10^4$     & 1.3 &   140   & 3 \\
             G10.47+0.03         & $+$67.3     & $3.7\times10^5$     & 5.8 &  1168   & 3 \\
             G34.26+0.15         & $+$58.0     & $3.2\times10^5$     & 3.3 &  1792   & 3 \\
             W51N-e1             & $+$59.5     & $1.0\times10^5$     & 5.1 & 4530   & 3 \\
             NGC7538-IRS1        & $-$56.2     & $1.3\times10^5$     & 2.7 &   433   & 3 \\
             \hline
      \end{tabular}
      \end{center}
        {\bf Notes.} See \citet{vanDishoeck11} for the source coordinates.\\
        
      {\bf References.} 
      (1) Bolometric luminosities and envelope masses obtained from \cite{Kristensen12}. 
      (2) Envelope masses collected in \cite{Wampfler13}.
      (3) Bolometric luminosities (obtained from observations) and envelope masses calculated in
      \cite{vanderTakinprep}.\\
      
  \end{table*}
}

\def\placeTableOverviewCO{
  \begin{table*}
    \caption{\label{tbl:overview_co} Overview of the main properties of the observed lines.}
    \begin{center}
      \begin{tabular}{l r r r c c c c r r r r}
        \hline\hline
        Mol. & Trans.& $E_\mathrm{u}/k_{\mathrm{B}}$ & Frequency & Tel./Inst.-band&  $\eta_{\mathrm{MB}}$ & Beam & Spec. Resol. &
        \multicolumn{4}{c}{Exposure time (min)} \\
        \cline{9-12}
        \noalign{\smallskip}
        & & (K) & (GHz) &  & &size ($\arcsec$) &  (km\,s$^{-1}$) & LM0 & LMI & IM & HM\\
        \hline
        $^{12}$CO & 3--2  &  33.2 &   345.796 & JCMT    & 0.63 & 14 & 0.1/0.4  &  21 &  21 &  21 & 21 \\
                 & 10--9 & 304.2 &  1151.985 & HIFI-5a & 0.64 &20 & 0.13$^a$ &  10 &   7 &  10  & 20 \\
        \noalign{\smallskip}
        $^{13}$CO &  3--2 &  31.7 &   330.588 & JCMT    & 0.63 & 15 & 0.1/0.4  &  32 &  32 &  32 &  32\\
                 & 10--9 & 290.8 &  1101.350 & HIFI-4b & 0.74 & 21 & 0.14$^a$ &  40 &  30  &  40 &  42/59\\
        \noalign{\smallskip}
        C$^{18}$O &  3--2 & 31.6  &  329.331  & JCMT    & 0.63 & 15 & 0.1/0.4  &  39 &  39 &  24 &  24 \\
                 &  5--4 & 79.0  &  548.831  & HIFI-1a & 0.76 & 42 & 0.07$^b$ &  60 &  -- &  31 &  -- \\
                 &  9--8 & 237.0 &  987.560  & HIFI-4a & 0.74 & 23 & 0.15$^a$ &  20 &  20 &  20 &   7 \\
                 & 10--9 & 289.7 & 1097.163  & HIFI-4b & 0.74 & 21 & 0.14$^a$ &  30$^c$ &  300 &  30 &  30 \\
        \hline
      \end{tabular}
    \end{center}
        {\bf Notes.} LM0: low-mass Class~0 sources. LMI: low-mass Class~I objects.
        IM: intermediate-mass YSOs. HM: high-mass protostars.\\
        $^{(a)}$ WBS data. $^{(b)}$ HRS data. $^{(c)}$ NGC1333\,IRAS\,2A, 4A and 4B were observed for 300 min each.
  \end{table*}
}


\def\placeTableMedian{
  \begin{table}
    \caption{\label{tbl:median_values} Median values of the width of the broad velocity component of 
       $^{12}$CO~$J$=10--9 and $J$=3--2 spectra, and of the FWHM for the C$^{18}$O~$J$=3--2 
      and $J$=9--8 line profiles.}
    \begin{center}
      \begin{tabular}{lccc}
        \hline\hline
         & broad[$^{12}$CO]  &  C$^{18}$O~$J$=3--2 &  C$^{18}$O~$J$=9--8\\
        \noalign{\smallskip}
        & (km\,s$^{-1}$) & (km\,s$^{-1}$) & (km\,s$^{-1}$) \\
        \hline
        LM0 &  17.8 &  1.2  & 2.5\\
        LMI &  12.7 &  0.9  & 3.1\\
        IM  &  21.0 &  1.9  & 3.9\\
        HM  &  24.8 &  4.3  & 5.0\\
        \hline
      \end{tabular}
    \end{center}
        {\bf Notes.} LM0: low-mass Class~0 sources. LMI: low-mass Class~I protostars.
        IM: intermediate-mass YSOs. HM: high-mass objects.\\
  \end{table}
}


\def\placeTableCorrelationFits{
\begin{table}
  \begin{center}
    \caption{\label{tab:linear_fit} Slope, $b$, and intercept, $a$, of the calculated power-law 
      fit for each CO and isotopologue line versus bolometric luminosity, 
      together with their errors and the Pearson correlation coefficient, $r$.
      The fitted equation is: \mbox{$\log$\,($L_{\rm{CO}}$) $=a+b\,\cdot$\,$\log$\,($L_{\rm{bol}}$)}.}
    \begin{tabular}{l ccc}
      \hline\hline
      Line                        &        $a$       &        $b$       & r\\
      \hline
      \noalign{\smallskip}
      ${^{12}}$CO~$J$=10--9        & $-$2.9 $\pm$ 0.2 &  0.84 $\pm$ 0.06 &  0.92\\
      ${^{13}}$CO~$J$=10--9        & $-$4.4 $\pm$ 0.2 &  0.97 $\pm$ 0.03 &  0.98\\
      C${^{18}}$O~$J$=3--2         & $-$4.1 $\pm$ 0.1 &  0.93 $\pm$ 0.03 &  0.98\\
      C${^{18}}$O~$J$=5--4         & $-$3.5 $\pm$ 0.2 &  0.78 $\pm$ 0.08 &  0.93\\
      C${^{18}}$O~$J$=9--8         & $-$5.2 $\pm$ 0.2 &  1.03 $\pm$ 0.05 &  0.97\\
      C${^{18}}$O~$J$=10--9        & $-$5.2 $\pm$ 0.3 &  0.96 $\pm$ 0.06 &  0.96\\
      \hline
    \end{tabular}
  \end{center}  

\end{table}
}

\def\placeTableCOWTpeak{
  \begin{table*}
    \begin{center}
      \caption{\label{table:12CO10-9-W-Tpeak} Observed and fitted properties of the $^{12}$CO~$J$=10--9 line profiles.
        The subscript $B$ refers to the broad velocity component and the $N$ to the narrow component.}
      \begin{tabular}{l rr rr rr rr}
        \hline\hline
        Source    & rms$^a$ &${\int{T_{\rm{MB}}{\rm{d}}\varv^{b}}}$ &  $T_{\rm{MB  B}}^{\rm{peak}}$ &  $T_{\rm{MB  N}}^{\rm{peak}}$ & 
        $V_{\rm{B}}^{\rm{peak}}$  & $V_{\rm{N}}^{\rm{peak}}$ & $FWHM_{\rm{B}}$  & $FWHM_{\rm{N}}$ \\
        & (mK) &   (K km\,s$^{-1}$) & (K) & (K) & (km\,s$^{-1}$)  & (km\,s$^{-1}$)  &(km\,s$^{-1}$)  &(km\,s$^{-1}$) \\
        \hline
            {\bf Low-mass: Class\,0}&    &    &   &  &&&&\\ 
            L\,1448-MM$^d$         &   91 &   21.5 &   0.4  &    0.9  &   10.3 &   6.0 &    41.0  &  4.8 \\
            NGC\,1333 IRAS\,2A     &  104 &   20.3 &   0.4  &    1.5  &   12.8 &   8.2 &     8.3  &  3.9 \\
            NGC\,1333 IRAS\,4A$^c$ &  105 &   45.5 &   1.8  &    $-$  &   10.3 &   $-$ &    24.8  &  $-$ \\
            NGC\,1333 IRAS\,4B     &  104 &   32.4 &   1.4  &    1.5  &    8.1 &   7.1 &    16.5  &  3.3 \\
            L\,1527$^c$            &   93 &    4.8 &   $-$  &    1.6  &    $-$ &   4.9 &     $-$  &  2.4 \\
            Ced110\,IRS4$^c$       &  127 &    4.9 &   $-$  &    1.6  &    $-$ &   4.2 &     $-$  &  2.5 \\
            IRAS\,15398            &  132 &   16.5 &   0.5  &    2.5  &  $-$1.0 &   4.1 &     15.0  &  4.2 \\
            BHR\,71$^d$$^e$         &  111 &    9.9 &   0.2  &    0.6  &  $-$8.5 &  $-$5.4 &    30.8  &  9.3 \\
            L\,483-MM              &  108 &   10.9 &   0.3  &    1.5  &    2.8 &   5.3 &    19.1  &  2.9 \\
            Ser SMM\,1$^e$         &   98 &   81.3 &   3.4  &    4.1  &    7.1 &   8.6 &    15.6  &  5.9 \\
            Ser SMM\,3             &  102 &   33.7 &   1.0  &    2.0  &    8.8 &   7.0 &    20.8  &  4.1 \\
            Ser SMM\,4             &   97 &   39.0 &   1.6  &    3.3  &    2.1 &   6.8 &    10.7  &  4.3 \\
            L\,723-MM$^c$          &  110 &    6.8 &   $-$  &    1.1  &    $-$ &  10.9 &     $-$  &  4.9 \\
            B335                   &  120 &   11.1 &   0.5  &    1.0  &    8.9 &   8.3 &    16.1  &  2.3 \\
            L\,1157                &  103 &    9.5 &   0.3  &    0.5  &    0.4 &   2.9 &    26.8  &  2.7 \\
       \hline
           {\bf Low-mass: Class\,I}&    &    &        &   &&&&   \\
           L\,1489$^c$        &   123  &    5.8 &    $-$  &    0.9  &    $-$ &    7.0 &    $-$  &   4.9 \\
           L\,1551 IRS\,5     &   113  &   15.9 &    0.4  &    3.1  &    4.2 &    6.2 &   16.1  &   2.5 \\
           TMR\,1 $^c$        &   113  &    9.3 &    $-$  &    1.8  &    $-$ &    5.5 &    $-$  &   3.9 \\
           TMC\,1A$^c$        &   137  &    4.2 &    $-$  &    0.5  &    $-$ &     5.7 &   $-$  &   3.6 \\
           TMC\,1$^c$         &   119  &    2.7 &    $-$  &    0.4  &    $-$ &     5.2 &   $-$  &   4.4 \\
           HH\,46             &   127  &    9.5 &    0.4  &    1.7  &    6.0 &     5.6 &  12.7  &   1.8 \\
           IRAS\,12496$^c$    &   100  &    10.0 &    $-$  &    2.1  &    $-$ &     2.8 &   $-$  &   3.8 \\
           GSS\,30\,IRS1$^e$  &   127  &   44.6 &    1.1  &    8.0  &    2.9 &     2.7 &  10.0  &   3.7 \\
           Elias\,29$^c$      &   120  &   47.0 &    $-$  &    7.8  &    $-$ &     4.2 &   $-$  &   5.1 \\
           Oph\,IRS\,63$^c$   &   124  &    1.0 &    $-$  &    0.5  &    $-$ &     2.6 &   $-$  &   2.1 \\
           RNO\,91$^c$        &   114  &    7.6 &    $-$  &    1.5  &    $-$ &     0.4 &   $-$  &   2.4 \\
       \hline
           {\bf Intermediate-mass}&    &    &        &     &&&& \\
           L1641\,S3\,MMS1    &   100  &   31.4 &    0.8  &    1.4  &    5.5 &   5.3 &   21.1  &   7.0 \\
           Vela\,IRS\,19      &   110  &   42.2 &    1.3  &    2.1  &   14.9 &  11.5 &   22.0  &   3.3 \\
           Vela\,IRS\,17      &   107  &   94.0 &    1.9  &   11.3  &    5.6 &   3.9 &   15.8  &   4.6 \\
           NGC\,7129 FIRS\,2$^e$  &   100  &   29.8 &    0.9  &    1.2  &  $-$11.5 &  $-$8.9 &   20.9  &   5.1 \\
           NGC\,2071          &   162  &  421.6 &    7.0  &   13.9  &    8.2 &  10.9 &   24.3  &   7.6 \\
           AFGL\,490          &   129  &   29.6 &    1.2  &    3.0  &  $-$11.5 & $-$13.4 &   15.6  &   2.7 \\
       \hline
           {\bf High-mass}&    &    &        &     &&&& \\
           W3$-$IRS5$^e$       &   102  &  674.8 &   10.8  &   32.0  &  $-$40.7 &  $-$37.5 &  28.6  &   8.4 \\
           
           \hline
      \end{tabular}  
    \end{center}
        {\bf Notes.} $^{(a)}$ In 0.27 km\,s$^{-1}$ bins.
        $^{(b)}$ Integrated over the entire line, not including ``bullet'' emission.
        $^{(c)}$ Single Gaussian fit. $^{(d)}$ EHV emission features
        removed from the spectra by using two additional Gaussian fit
        profiles. $^{(e)}$ Self-absorption features detected.\\
  \end{table*}
}

\def\placeTabletreceCOWTpeak{
  \begin{table*}
    \begin{center}
      \caption{\label{table:13CO10-9-W-Tpeak} Observed and fitted properties of the $^{13}$CO~$J$=10--9 line profiles for 
        the detected sources. The subscript $B$ refers to the broad velocity component and the $N$ to 
        the narrow component. }
      \begin{tabular}{l rr rr rr rr}
        \hline\hline
        Source    & rms$^a$ &${\int{T_{\rm{MB}}{\rm{d}}\varv^{\rm{b}}}}$ &  $T_{\rm{MB  B}}^{\rm{peak}}$ &  $T_{\rm{MB  N}}^{\rm{peak}}$ & 
        $V_{\rm{B}}^{\rm{peak}}$  & $V_{\rm{N}}^{\rm{peak}}$ & $FWHM_{\rm{B}}$  & $FWHM_{\rm{N}}$ \\
        & (mK) &   (K km\,s$^{-1}$) & (K) & (K) & (km\,s$^{-1}$)  & (km\,s$^{-1}$)  &(km\,s$^{-1}$)  &(km\,s$^{-1}$) \\
        \hline
            {\bf Low-mass: Class\,0}&    &    &   &  &&&& \\ 
            L\,1448-MM         &    20  &    0.2 &    $-$  &   0.1  &   $-$ &    4.7 &   $-$ &   1.6 \\
            NGC\,1333 IRAS\,2A &    15  &    0.8 &    $-$  &   0.2  &   $-$ &    7.5 &   $-$ &   2.1 \\
            NGC\,1333 IRAS\,4A &    23  &    1.1 &    0.07 &   0.09 &   5.7 &    6.7 &  13.2 &   0.7 \\
            NGC\,1333 IRAS\,4B &    17  &    0.8 &    $-$  &   0.1  &   $-$ &    6.9 &   $-$ &   6.8 \\
            IRAS 15398         &    24  &    0.2 &    $-$  &   0.16  &   $-$ &    5.0 &   $-$ &   1.1 \\
            BHR\,71            &    17  &    0.4 &    $-$  &   0.2  &   $-$ &   --4.6 &   $-$ &   1.8 \\
            L\,483-MM          &    16  &    0.3 &    $-$  &   0.1  &   $-$ &    5.0 &   $-$ &   2.4 \\
            Ser SMM\,1         &    22  &    4.0 &    0.2  &   0.5  &   7.6 &    8.3 &  10.6 &   2.4 \\
            Ser SMM\,3         &    20  &    0.3 &    $-$  &   0.1  &   $-$ &    7.2 &   $-$ &   4.0 \\
            Ser SMM\,4         &    25  &    0.6 &    $-$  &   0.1  &   $-$ &    5.7 &   $-$ &   6.0 \\
            L\,723-MM          &    16  &    0.3 &    $-$  &   0.1  &   $-$ &   11.5 &   $-$ &   2.7 \\
            B335               &    20  &    0.3 &    $-$  &   0.1  &   $-$ &    8.1 &   $-$ &   2.3 \\
            L\,1157            &    20  &    0.1 &    $-$  &   0.05 &   $-$ &    3.0 &   $-$ &   2.2 \\
        \hline
            {\bf Low-mass: Class\,I}&    &    &        &    &&&& \\
            L\,1489            &    32  &    0.4 &    $-$  &   0.05 &  $-$ &    7.8 &   $-$ &   7.3 \\
            L\,1551 IRS\,5     &    29  &    1.3 &    $-$  &   0.47  &   $-$ &    6.5 &   $-$ &   2.4 \\
            TMR\,1             &    29  &    0.3 &    $-$  &   0.08  &   $-$ &    6.0 &   $-$ &   3.6 \\
            HH\,46             &    30  &    0.1 &    $-$  &   0.11  &   $-$ &    5.4 &   $-$ &   1.5 \\
            IRAS\,12496        &    28  &    0.8 &    $-$  &   0.15  &   $-$ &    3.1 &   $-$ &   4.2 \\
            GSS\,30\,IRS1      &    33  &    2.0 &    $-$  &   0.46  &   $-$ &    2.8 &   $-$ &   3.4 \\
            Elias\,29          &    28  &    2.1 &    $-$  &   0.38  &   $-$ &    4.7 &   $-$ &   5.1 \\
        \hline
            {\bf Intermediate-mass}&    &    &        &    &&&& \\
            L1641\,S3\,MMS1    &    38  &    1.8 &    $-$  &   0.2  &   $-$ &    4.7 &   $-$  &  6.1 \\
            Vela\,IRS\,19      &    41  &    1.1 &    $-$  &   0.2  &   $-$ &   12.1 &   $-$  &  5.9 \\
            Vela\,IRS\,17      &    43  &    4.9 &    $-$  &   0.8  &   $-$ &    0.4 &   $-$  &  4.4 \\
            NGC\,7129 FIRS\,2  &    20  &    0.9 &    $-$  &   0.1  &   $-$ &  --9.6 &   $-$  &  4.3\\
            NGC\,2071          &    17  &   16.6 &    0.4  &   2.1  &   9.1 &    9.8 &  14.7  &  4.7 \\
            AFGL\,490          &    47  &    2.6 &    $-$  &   0.3  &   $-$ & --13.5 &   $-$  &  4.8 \\
        \hline
            {\bf High-mass}&    &    &        &     &&&& \\
            IRAS05358+3543      &    18  &   6.2 &    0.2  &   0.7  &  --15.1 & --15.9 &  15.4  &  3.7 \\
            IRAS16272$-$4837    &    18  &   3.9 &    $-$  &   0.7  &   $-$ & --47.0 &  $-$   &  4.8 \\
            NGC6334-I(N)        &    23  &  13.8 &    0.5  &   1.3  &  --5.7 &  --3.8 &  13.3  &  4.4 \\
            W43-MM1             &    38  &   4.5 &    $-$  &   0.8  &   $-$  &  98.4 &   $-$  &  5.8 \\
            DR21(OH)            &    18  &  52.2 &    0.9  &   5.7  &   --1.9 &  --3.3 &  16.0  &  5.8 \\
            W3-IRS5             &    21  & 129.6 &    2.1  &  17.0  &  --38.7 & --38.3 &  14.7  &  4.7 \\
            IRAS18089$-$1732    &    27  &   8.3 &    $-$  &   1.4  &   $-$  &  33.0 &   $-$  &  4.6 \\
            W33A                &    19  &  11.4 &    0.5  &   1.1  &   36.8 &  37.7 &  11.7  &  3.5 \\
            IRAS18151$-$1208    &    19  &   3.4 &    $-$  &   0.8  &   $-$  &  33.4 &   $-$  &  3.6 \\
            AFGL2591            &    19  &  29.5 &    0.8  &   5.4  &   --5.8 &  --5.4 &   9.8  &  3.3 \\
            G327$-$0.6          &    25  &  12.8 &    $-$  &   1.8  &   $-$  & --44.8 & $-$   &  6.6 \\
            NGC6334-I           &    23  &  57.7 &    0.7  &   7.2  & --6.7  & --6.8 &  17.8  &  5.4 \\
            G29.96$-$0.02       &    46  &  31.3 &    0.6  &   4.6  &   96.9 &  98.8 &  13.6  &  4.6 \\
            G31.41+0.31         &    52  &  17.4 &    $-$  &   2.5  &   $-$  &  96.8 &   $-$  &  6.1 \\
            G5.89$-$0.39        &    34  & 133.7 &    2.6  &   6.2  &   10.5 &   9.4 &  21.9  &  5.9 \\
            G10.47+0.03         &   200  &  26.2 &    1.8  &   $-$  &   66.8 &  $-$  &   10.0 &  $-$ \\
            G34.26+0.15         &    59  &  58.8 &    1.3  &   6.7  &   58.0 &  57.7 &  12.0  &  5.0 \\
            W51N-e1             &    50  &  70.2 &    1.3  &   5.0  &   58.6 &  57.1 &  18.6  &  7.2 \\
            NGC7538-IRS1        &    24  &  41.0 &    2.0  &   5.3  &  --59.8 & --57.6 &   8.7  &  3.5 \\
         \hline
      \end{tabular} 
    \end{center}
    {\bf Notes.} $^{(a)}$ In 0.27 km\,s$^{-1}$ bins.
        $^{(b)}$ Integrated over the entire line.\\
        
  \end{table*}
}

\def\placeTableCochoOcincoWTpeak{
  \begin{table*}
    \begin{center}
      \caption{\label{table:C18O5-4-W-Tpeak} Observed and fitted properties of the C$^{18}$O~$J$=5--4 line profiles for 
        the observed sources. The subscript $B$ refers to the broad velocity component and the $N$ to the 
        narrow component.}
      \begin{tabular}{l rr rr rr rr}
        \hline\hline
        Source    & rms$^a$ &${\int{T_{\rm{MB}}{\rm{d}}\varv^{b}}}$ &  $T_{\rm{MB  B}}^{\rm{peak}}$ &  $T_{\rm{MB  N}}^{\rm{peak}}$ & 
        $V_{\rm{B}}^{\rm{peak}}$  & $V_{\rm{N}}^{\rm{peak}}$ & $FWHM_{\rm{B}}$  & $FWHM_{\rm{N}}$ \\
        & (mK) &   (K km\,s$^{-1}$) & (K) & (K) & (km\,s$^{-1}$)  & (km\,s$^{-1}$)  &(km\,s$^{-1}$)  &(km\,s$^{-1}$) \\
        \hline
            {\bf Low-mass: Class\,0}&    &    &   &  &&&& \\ 
            L\,1448-MM         &      4 &     0.47 &      $-$  &	0.37  &	$-$ &	5.1 &	$-$ &	1.3  \\
            NGC\,1333 IRAS\,2A &      2 &     0.87 &      $-$  &	0.52  &	$-$ &	7.5 &	$-$ &	1.4  \\
            NGC\,1333 IRAS\,4A &      4 &     0.68 &      0.03 &	0.32  &	7.9 &	6.9 &	7.8 &	1.3  \\
            NGC\,1333 IRAS\,4B &      3 &     0.32 &      $-$  &	0.14  &	$-$ &	7.0 &	$-$ &	1.9  \\
            L\,1527            &      3 &     0.55 &      $-$  &	0.23  &	$-$ &	5.7 &	$-$ &	2.0  \\
            BHR\,71            &      3 &     0.63 &      $-$  &	0.37  &	$-$ & --4.7 &   $-$ &	1.4  \\
            L\,483-MM          &      3 &     0.31 &      0.03 &	0.16  &	5.2 &	5.4 &   4.3 &	1.1  \\
            Ser SMM\,1         &      4 &     1.93 &      0.13 &	0.84  & 8.7 &	8.4 &	4.2 &	1.4  \\
            Ser SMM\,3         &      5 &     1.40 &      $-$  &	0.55  &	$-$ &	7.6 &	$-$ &	2.0  \\
            Ser SMM\,4         &      3 &     1.55 &      0.13 &	0.44  &	7.6 &   7.7 &	4.7 &	1.8  \\
            L\,723-MM          &      5 &     0.19 &      $-$  &	0.09  & $-$ &	10.0 &	$-$ &	1.9  \\
            B335               &      4 &     0.26 &      $-$  &	0.22  &	$-$ &	 8.2 &	$-$ &	1.2  \\
            L\,1157            &      4 &     0.15 &      $-$  &	0.09  & $-$ &	 2.6 &	$-$ &	1.3  \\
        \hline
             {\bf Intermediate-mass}&    &    &        &  &&&&  \\
             L1641\,S3\,MMS1    &     20 &     0.73 &     $-$  &	0.40  & $-$ &	 5.2 &	 $-$ &	1.7  \\
             Vela\,IRS\,19      &     10 &     1.67 &     0.05 &	0.46  &	12.5 &	11.6 &	 9.2 &	2.4  \\
             Vela\,IRS\,17      &     25 &     5.32 &     $-$  &	 1.25 &	$-$ &	 4.1 &	 $-$ &	3.7  \\
             NGC\,7129 FIRS\,2  &      4 &     0.55 &     $-$  &	 0.18 & $-$ &  --9.9 & $-$ &	2.3  \\
             NGC\,2071          &      5 &     7.29 &     0.55 &	1.33  &	8.9 &	 9.6 &	 7.4 &  2.0  \\
             AFGL\,490          &     14 &     3.55 &     $-$  &	1.04  & $-$ &  --13.3 & $-$ &	2.7  \\
             \hline
      \end{tabular}
      \end{center}
    {\bf Notes.} $^{(a)}$ In 0.27 km\,s$^{-1}$ bins.
        $^{(b)}$ Integrated over the entire line.\\
  \end{table*}
}

\def\placeTableCochoOnueveWTpeak{
  \begin{table*}
    \begin{center}
      \caption{\label{table:C18O9-8-W-Tpeak} Observed and fitted properties of the C$^{18}$O~$J$=9--8 
        narrow line profiles for the detected sources.}
      \begin{tabular}{l rc cc c}
        \hline\hline
        Source    & ${\int{T_{\rm{MB}}{\rm{d}}\varv^{b}}}$ &  $T_{\rm{MB}}^{\rm{peak}}$ & $FWHM$  & rms$^a$ & $N_{\rm{H_2}}$$^c$\\
        &   (K km\,s$^{-1}$) & (K) & (km\,s$^{-1}$)  & (mK) & (cm$^{-2}$)\\
        \hline
            {\bf Low-mass: Class\,0}&    &    &   &  &\\ 
            NGC\,1333 IRAS\,2A &     0.21 &      0.07  &       2.0  &     20  &  1.2$\times$10$^{21}$\\
            Ser SMM\,1         &     0.70 &      0.13  &       3.0  &     23  &  4.1$\times$10$^{21}$\\
            \hline
            {\bf Low-mass: Class\,I}&    &    &        &    \\
            L\,1551 IRS\,5     &     0.22 &      0.10  &       3.1  &     24  &  1.3$\times$10$^{21}$\\
            GSS\,30\,IRS1      &     0.23 &      0.06  &       2.2  &     25  &  1.4$\times$10$^{21}$\\
            Elias\,29          &     0.40 &      0.08  &       3.9  &     30  &  2.3$\times$10$^{21}$\\
         \hline
             {\bf Intermediate-mass}&    &    &        &    &\\
             Vela\,IRS\,19      &     0.37 &      0.11  &       3.2  &     24  &  2.2$\times$10$^{21}$\\
             Vela\,IRS\,17      &     0.69 &      0.16  &       2.8  &     25  &  4.0$\times$10$^{21}$\\
             NGC\,2071          &     2.97 &      0.41  &       4.6  &     25  &  1.7$\times$10$^{22}$\\
             AFGL\,490          &     0.45 &      0.08  &       5.4  &     20  &  2.6$\times$10$^{21}$\\
             \hline
             {\bf High-mass}&    &    &        &    & \\
             IRAS05358+3543      &     0.93 &      0.22  &       4.0  &     56  &  5.4$\times$10$^{21}$\\
             IRAS16272$-$4837    &     1.18 &      0.14  &       5.9  &     55  &  6.9$\times$10$^{21}$\\
             NGC6334-I(N)        &     1.82 &      0.37  &       3.9  &     55  &  1.1$\times$10$^{22}$\\
             W43-MM1             &     1.81 &      0.29  &       5.9  &     49  &  1.0$\times$10$^{22}$\\
             DR21(OH)            &     9.11 &      1.51  &       5.5  &     57  &  5.3$\times$10$^{22}$\\
             W3-IRS5             &    25.43 &      4.80  &       4.2  &     66  &  1.5$\times$10$^{23}$\\
             IRAS18089$-$1732    &     3.26 &      0.52  &       4.1  &     53  &  1.9$\times$10$^{22}$\\
             W33A                &     3.17 &      0.43  &       4.8  &     46  &  1.9$\times$10$^{22}$\\
             IRAS18151$-$1208    &     0.60 &      0.24  &       4.2  &     54  &  3.5$\times$10$^{21}$\\
             AFGL2591            &     4.56 &      1.21  &       3.1  &     39  &  2.7$\times$10$^{22}$\\
             G327$-$0.6          &     6.16 &      0.83  &       6.1  &     57  &  3.6$\times$10$^{22}$\\
             NGC6334-I           &    10.97 &      2.00  &       5.0  &     54  &  6.4$\times$10$^{22}$\\
             G29.96$-$0.02       &     7.50 &      1.41  &       4.3  &     41  &  4.4$\times$10$^{22}$\\
             G31.41+0.31         &     5.26 &      0.73  &       6.0  &     36  &  3.1$\times$10$^{22}$\\
             G5.89$-$0.39$^d$    &    31.35 &     2.13  &       5.9  &     51  &  1.8$\times$10$^{23}$\\
             G10.47+0.03$^d$     &     5.71 &      0.53  &       6.4  &     79  &  1.1$\times$10$^{23}$\\
             G34.26+0.15         &    16.26 &      2.53  &       5.5  &     49  &  9.5$\times$10$^{22}$\\
             W51N-e1$^d$         &    15.22 &      1.83  &       5.7  &     48  &  8.9$\times$10$^{22}$\\
             NGC7538-IRS1        &     6.32 &      1.37  &       4.0  &     48  &  3.7$\times$10$^{22}$\\
             \hline
      \end{tabular} 
      \end{center}
    {\bf Notes.} $^{(a)}$ In 0.27 km\,s$^{-1}$ bins.
    $^{(b)}$ Integrated over the entire line.
    $^{(c)}$ Column density of H$_2$ calculated for an excitation temperature of 75~K and a C$^{18}$O/H$_2$
    abundance ratio of 5$\times$10$^{-7}$. 
    $^{(d)}$ Two Gaussian fit. Only the values from the narrow velocity component are presented. \\
  \end{table*}
}

\def\placeTableCochoOdiezWTpeak{
  \begin{table}
    \begin{center}
      \caption{\label{table:C18O10-9-W-Tpeak} Observed and fitted properties of the C$^{18}$O~$J$=10--9 line profiles for 
        the detected sources.}
      \begin{tabular}{l rr rr}
        \hline\hline
        Source    & ${\int{T_{\rm{MB}}{\rm{d}}\varv^{b}}}$ &  $T_{\rm{MB}}^{\rm{peak}}$ & $FWHM$  & rms$^a$ \\
        &   (K km\,s$^{-1}$) & (K) & (km\,s$^{-1}$)  & (mK) \\
        \hline
            {\bf Low-mass: Class\,0}&    &    &   &  \\ 
            NGC\,1333 IRAS\,2A &     0.34 &      0.04  &      10.7  &      8 \\
	    NGC\,1333 IRAS\,4B &     0.05 &      0.03  &      1.4  &      10 \\
            Ser SMM\,1         &     0.61 &      0.07  &      4.8  &     17 \\
            \hline
            {\bf Low-mass: Class\,I}&    &    &        & \\
	    GSS\,30\,IRS1      &     0.12 &      0.04  &       3.1  &     8  \\
	    Elias\,29          &     0.22 &      0.04  &       4.7  &     7  \\
             \hline
             {\bf High-mass}&    &    &        &     \\
             IRAS05358+3543     &     0.42 &      0.10  &       3.8  &     26 \\
             IRAS16272$-$4837    &     0.49 &      0.11  &       4.8  &     26 \\
             NGC6334-I(N)        &     2.31 &      0.28  &       7.5  &     20 \\
             W43-MM1             &     0.92 &      0.12  &       5.3  &     23 \\
             DR21(OH)            &     7.94 &      1.08  &       6.0  &     33 \\
             W3-IRS5             &    17.04 &      3.39  &       4.1  &     22 \\
             IRAS18089$-$1732    &     1.54 &      0.30  &       4.5  &     28 \\
             W33A                &     1.72 &      0.30  &       4.3  &     22 \\
             IRAS18151$-$1208    &     0.48 &      0.09  &       4.2  &     22 \\
             AFGL2591            &     3.80 &      0.88  &       3.4  &     23 \\
             G327$-$0.6          &     3.03 &      0.42  &       5.8  &     34 \\
             NGC6334-I           &    10.09 &      1.55  &       5.3  &     28 \\
             G29.96$-$0.02       &     5.79 &      1.02  &       4.5  &     30 \\
             G31.41+0.31         &     3.98 &      0.52  &       6.1  &     38 \\
             G5.89$-$0.39        &    11.59 &      2.08  &       7.0  &     27 \\
             G10.47+0.03         &     3.44 &      0.47  &       7.1  &     26 \\
             G34.26+0.15         &    11.80 &      1.73  &       6.0  &     29 \\
             W51N-e1             &     10.49 &      1.48  &       6.8  &     52 \\
             NGC7538-IRS1        &     3.95 &      0.88  &       3.9  &     37 \\
             \hline
      \end{tabular}
      \end{center}
        {\bf Notes.} $^{(a)}$ In 0.27 km\,s$^{-1}$ bins.
        $^{(b)}$ Integrated over the entire line.\\
  \end{table}
}

\def\placeTableJCMT{
  \begin{table}
    \caption{\label{tbl:jcmt_data} Values of the FWHM for the $^{12}$CO~$J$=3-2 broad velocity 
      component and the FWHM and integrated intensity for the C$^{18}$O~$J$=3-2 spectra.}
      \begin{center}
      \begin{tabular}{l cc c} 
        \noalign{\smallskip}
        \hline\hline
        Source & $^{12}$CO~$J$=3-2  &  \multicolumn{2}{c}{C$^{18}$O~$J$=3-2}\\
        \cline{3-4} 
        & FWHM$_B$ & FWHM  & ${\int{T_{\rm{MB}}{\rm{d}}\varv}}$\\
        & (km\,s$^{-1}$) & (km\,s$^{-1}$) & (K\,km\,s$^{-1}$)\\ 
        \hline
        {\bf Low-mass: Class\,0}&    &   \\
	L\,1448-MM	&  18.5 	&  1.1 & 1.5\\
	NGC\,1333\,IRAS2A	&  13.8 	&  1.2 & 2.8\\
	NGC\,1333\,IRAS4A	&  20.1		&  1.5 & 2.5\\
	NGC\,1333\,IRAS4B	&  16.2		&  0.8 & 1.6\\        
	L\,1527$^a$	&   $-$ 	&  0.7 & 1.5\\
	Ced110\,IRS4$^a$&   $-$ 	 &  $-$  &$-$\\
        BHR\,71 	&  16.4 	&  1.2 & 4.1\\
	IRAS\,15398$^a$	&   $-$ 	&  0.7 & 0.8\\
	L\,483-MM	&  10.2 	&  0.6 & 1.2\\
	Ser\,SMM1	&  14.1 	&  2.1 & 4.8\\
	Ser\,SMM4	&  14.2 	&  2.3 & 6.0 \\
	Ser\,SMM3	&  11.6 	&  1.7 & 4.9\\
	L\,723-MM	&  13.2 	&  1.5 & 1.0\\
	B\,335$^a$	&  $-$		&  0.8 & 1.4\\
	L\,1157		&  18.9 	&  0.9 & 0.6\\
        \hline
            {\bf Low-mass: Class\,I}&    &   \\
	L\,1489$^a$	&   $-$ 	&  2.0 & 0.7\\
	L\,1551-IRS5	&  10.4 	&  0.9 & 2.3\\
  	TMR1$^a$	&  $-$  	&  1.2 & 1.1\\  
 	TMC1		&  13.7 	&  0.7 & 0.9\\ 
 	TMC1A		&  10.2		&  1.0 & 0.9\\
	HH46		&  12.8 	&  0.9 & 3.2\\
	IRAS\,12496	&  19.5 	&  1.2 & 1.4\\
	Elias\,29 	&   7.8 	&  2.5 & 5.3\\
	Oph\,IRS\,63	&   8.5 	&  0.6 & 1.1\\
	GSS\,30-IRS1	&   7.4 	&  1.2 & 3.7\\
	RNO\,91		&   9.3 	&  0.6 & 1.1\\
        \hline
            {\bf Intermediate-mass}&    &  \\
	NGC\,7129\,FIRS2& 18.2		&  1.6 & 5.1\\
	L1641\,S3\,MMS1	& 13.3  	&  1.6 & 6.4\\
        NGC2071 	& 25.5  	&  2.8 & 14.1\\
        AFGL\,490 	& 37.4  	&  2.3 & 13.2\\
        \hline
        {\bf High-mass}& & \\
        IRAS05358+3543 	& 20.0  	&  3.0 & 7.7\\
        NGC6334-I(N) 	& 30.1		&  $-$ & $-$\\
        W43-MM1		& 27.6		&  6.7 & 36.9\\
        DR21-OH 	&  $-$ 		&  5.5 & 49.8\\
        W3-IRS5 	& 24.7  	&  4.2 & 76.1\\
        IRAS18089-1732	& 15.7		&  4.1  & 39.7 \\
        W33A 		& 36.7  	&  4.3 & 41.7\\
	IRAS18151-1208 $^a$  &  $-$	&  2.5  & 19.3\\
        AFGL2591 	& 22.0 		&  3.6 & 45.9\\
        NGC6334-I 	& 35.7  	&  3.7 & 50.3\\
        G29.96-0.02 	& 18.0  	&  4.0 & 65.9\\
        G31.41+0.31 	& 24.9 	        &  7.3 & 44.5 \\
        G5.89-0.39	& 53.5		&  4.8  & 95.0 \\
        G10.47+0.03 	& 14.1  	&  7.0 & 82.2\\
        G34.26+0.15 	& 23.9  	&  6.0 & 153.2\\
	W51-e1		& 30.3		&  $-$ &  $-$\\
        NGC7538-IRS1 	& 20.8  	&  5.9 & 50.3\\
        \hline
      \end{tabular}
      \end{center} 
          {\bf Notes.} $^{(a)}$ No broad velocity component identified in the 
          $^{12}$CO~$J$=3-2 convolved central spectrum.
  \end{table}
}


\begin{abstract}{
  \textit{Context.} 
  Our understanding of the star formation process has traditionally been confined to certain 
  mass or luminosity boundaries because most studies focus only on
  low-, intermediate-
  or high-mass star-forming regions. Therefore, the processes that regulate the formation of these
  different objects have not been effectively linked.  
  As part of the ``Water In Star-forming regions with \textit{Herschel}'' (WISH) key program, water and other important
  molecules, such as CO and OH, have been observed in 51 embedded young stellar objects (YSOs). 
  The studied sample covers a range of luminosities from $<1$ to $> 10^5$ L$_{\odot}$. \\
  \textit{Aims.} We analyse the CO line emission towards a large sample of embedded protostars in terms of both
  line intensities and profiles. 
  This analysis covers a wide luminosity range in order to achieve a better understanding of 
  star formation without imposing luminosity boundaries. In particular, this paper aims to 
  constrain the dynamics of the environment in which YSOs form.\\
  \textit{Methods.} \textit{Herschel}-HIFI spectra of the $^{12}$CO~$J$=10--9, $^{13}$CO~$J$=10--9 and 
  C$^{18}$O~$J$=5--4, $J$=9--8 and 
  $J$=10--9 lines are analysed for a sample of 51 embedded protostars. In addition, JCMT spectra of 
  $^{12}$CO~$J$=3--2 and C$^{18}$O~$J$=3--2 extend this analysis to
  cooler gas components.
  We focus on characterising the shape and intensity of the CO emission line profiles 
  by fitting the lines with one or two Gaussian profiles. 
  We compare the values and results of these fits across the entire luminosity range  
  covered by WISH observations. The effects of
  different physical parameters as a function of luminosity and the
  dynamics of the envelope-outflow
  system are investigated.\\
  \textit{Results.} All observed CO and isotopologue spectra show a strong linear correlation between 
  the logarithms of the line and bolometric luminosities across six orders of magnitude on both axes. 
  This suggests that the high-$J$ CO lines primarily trace the amount of dense gas associated with YSOs and
  that this relation can be extended to larger (extragalactic) scales. 
  The majority of the detected $^{12}$CO line profiles can be decomposed into a broad and a narrow Gaussian component, 
  while the C$^{18}$O spectra are mainly fitted with a single Gaussian.
  For low- and intermediate-mass protostars, the width of the 
  C$^{18}$O~$J$=9--8 line is roughly twice that of the C$^{18}$O~$J$=3--2 line, suggesting increased 
  turbulence/infall in the warmer inner envelope. For high-mass protostars, the line widths are comparable 
  for lower- and higher-$J$ lines. 
  A broadening of the line profile is also observed from pre-stellar cores to embedded protostars, which is due 
  mostly to non-thermal motions (turbulence/infall).
  The widths of the broad $^{12}$CO~$J$=3--2 and $J$=10--9 velocity components 
  correlate with those of the narrow 
  C$^{18}$O~$J$=9--8 profiles, suggesting that the entrained outflowing gas
  and envelope motions are related independent of the mass of the protostar.
  These results indicate that physical processes in protostellar envelopes have similar 
  characteristics across the studied luminosity range. 
}

\end{abstract}

   \keywords{Astrochemistry –-- Stars: formation –-- Stars: protostars –-- ISM: molecules 
     –-- ISM: kinematics and dynamics –-- line: profiles}
   \maketitle



\section{Introduction}
The evolution of a protostar is closely related to the initial mass of the molecular core
from which it forms and to the specific physical and chemical properties of the original
molecular cloud (e.g., \citealt{Shu93}; \citealt{vanDishoeckBlake98}; \citealt{McKee07}). 
During the early stages of their formation, young stellar objects (YSOs) are embedded in large, cold and 
dusty envelopes which will be accreted or removed by the forming star.
Depending on the mass of the star-forming region, the parameters and mechanisms that rule several processes of the 
star formation, such as driving agent of the molecular outflow and accretion rates, will vary. 

Molecular outflows are crucial for removing angular momentum and mass from the protostellar system (see review
by \citealt{Lada99}). They have been extensively studied for low-mass YSOs (e.g., \citealt{Cabrit92}; 
\citealt{Bachiller99})
where they are better characterised than for massive protostars (e.g., \citealt{Shepherd&Churchwell96}; 
\citealt{Beuther&Shepherd05}).
The reason is related to the short life-time \citep{Mottram11} and the 
large distances (few kpc) associated with massive YSOs. This means that 
outflows from massive stars are less well resolved than their low-mass counterparts.
The agent that drives the molecular outflow, either jets or winds from the disk and/or stars
(e.g., \citealt{Churchwell99}; \citealt{Arce07}), might be different 
depending on the mass of the star-forming region. Therefore, the interaction of the outflow
with the surrounding material and especially the resulting chemistry may differ across the mass range. 

The accretion rates are also different depending on the mass of the forming star.
Typical values for low-mass star formation are 
10$^{-7}$--10$^{-5}$~M$_\odot$\,yr$^{-1}$ (\citealt{Shu77}; \citealt{Bontemps96}) 
whereas higher values are necessary in order to overcome radiation pressure
and form massive stars within a free-fall time (e.g., \citealt{Jijina96}). These values range from
10$^{-4}$ to 10$^{-3}$~M$_\odot$\,yr$^{-1}$ for sources with $>10^{4}$~L$_\odot$ (e.g., \citealt{Beuther02}).
In addition, for the low-mass sources, the accretion episode finishes before the protostar reaches the
main sequence, while massive YSOs still accrete circumstellar material after reaching the hydrogen burning 
phase (\citealt{Palla&Stahler93}; \citealt{Cesaroni05}). 

Another difference is that the ionising radiation created by main-sequence OB stars is much more powerful than 
that generated by a single low-mass protostar. Therefore, photon dominated regions (PDR) 
and H{\sc{ii}} regions are formed in areas of massive star formation, affecting 
the kinematics, temperature and chemistry of the surrounding material \citep{Hollenbach&Tielens99}. 
In addition, the strength of stellar winds and their interaction with the 
envelope material is different depending on the stellar spectral type of the YSOs.

Because of these differences, the study of star formation 
has traditionally been restricted to mass boundaries, 
focused either on low-mass ($M<$ 3~M$_\odot$) or high-mass ($M>$ 8~M$_\odot$) YSOs. 
One of the goals of the ``Water In Star-forming regions 
with \textit{Herschel}'' (WISH) key program \citep{vanDishoeck11} 
is to offer a complete description of the interaction of young stars with their surroundings
as a function of mass.
For this purpose, and in order to constrain the physical and chemical processes that determine 
star formation, water and other key molecules like CO have been observed for a large sample of 
embedded YSOs (51 sources). The targeted objects cover a vast range of luminosities (from $<$ 
1~L$_\odot$ to $>$ 10$^5$~L$_\odot$) and different evolutionary stages (more details in Section~\ref{Sample}).
With the Heterodyne Instrument for the Far-Infrared (HIFI; \citealt{deGraauw10}) on board the \textit{Herschel
Space Observatory} (\citealt{Pilbratt10}), high spectral resolution data of high frequency 
molecular lines have been obtained.
These can be used to probe the physical conditions, chemical composition and dynamics of protostellar 
systems (e.g., \citealt{Evans99}; \citealt{Jorgensen02}; \citealt{vanderWiel13sub}).
 
Due to its high, stable abundance and strong lines, CO is one of the most important and often used 
molecules to probe the different physical components of the YSO environment (envelope, outflow, disk).
In particular, molecular outflows are traced by $^{12}$CO emission through maps in the line
wings (e.g, \citealt{Curtis10}). Its isotopologue C$^{18}$O 
is generally thought to probe quiescent gas in the denser part of the protostellar envelope, 
whereas $^{13}$CO lines 
originate in the extended envelope and the outflow cavity walls (e.g., \citealt{Spaans95};
\citealt{Graves10}; \citealt{Yildiz12}). 
In addition, CO has a relatively low critical densities, due to its  
small permanent dipole moment ($\sim$0.1~Debye), 
and relatively low rotational energy levels, so this molecule is easily 
excited and thermalised by collisions with H$_2$ in a typical star
formation environment. For this reason, measurements of 
CO excitation provide a trustworthy estimate of the gas kinetic temperature. 
Moreover, integrated intensity measurements can be used to
obtain column densities of warm gas, 
providing a reference to determine the abundances of other species, such as water and H$_2$. 

Most CO observations from ground-based sub-millimetre telescopes have been limited to low-$J$ 
rotational transitions (up to upper transition $J_{\rm{u}}$=3, i.e., upper-level energy $E_{\rm{u}}/k_{\rm{B}}\sim$35~K), 
or mid-$J$ transitions ($J_{\rm{u}}$=6, with a $E_{\rm{u}}/k_{\rm{B}}$ of $\sim$100~K).
Thanks to HIFI, spectrally resolved data for high-$J$ CO transitions ($J_{\rm{u}}$ up to 16,  
$E_{\rm{u}}/k_{\rm{B}}\sim$600~K) are observable for the first time, so warm gas directly associated with the 
forming star is probed (e.g., \citealt{Yildiz10}, 2012; \citealt{Plume12}; \citealt{vanderWiel13sub}). 
Therefore, a uniform probe of 
the YSOs over the entire relevant range of $E_{\rm{u}}/k_{\rm{B}}$ (from 10--600~K) is achieved by combining HIFI data
with complementary spectra from single-dish ground-based telescopes.
These observations are indispensable in order to ensure a self-consistent data set for analysis.
Finally, the study of these lines in our Galaxy is crucial in order to compare them
with the equivalent lines targeted in high-redshift galaxies which are often used
to determine star-formation rates on larger scales. 

In this paper we present $^{12}$CO~$J$=10--9, $^{13}$CO~$J$=10--9, C$^{18}$O~$J$=5--4, $J$=9--8 
and $J$=10--9 HIFI spectra of 51 YSOs. Complementing these data, $^{12}$CO and C$^{18}$O~$J$=3--2 spectra 
observed with the James Clerk Maxwell Telescope (JCMT) are included in the analysis in order to use CO to its full 
diagnostic potential and extend the analysis to different regions of the protostellar environment with
different physical conditions.
Section~\ref{Observations} describes the sample, the observed CO data and
the method developed to analyse the line profiles.
A description of the morphology of the spectra, an estimation of the kinetic temperatures 
and correlations regarding the line luminosities of each isotopologue transition  
are presented in Section~\ref{Results}.
These results are also compared to other YSO parameters such as luminosity and envelope mass.
In Section~\ref{Discussion} we discuss the results to constrain 
the dynamics of individual velocity components of protostellar
envelopes, characterise the turbulence in the envelope-outflow system and consider high-$J$ CO 
as a dense gas tracer. Our conclusions are 
summarised in Section~\ref{Conclusions}.


\section{Observations}\label{Observations}

\subsection{Sample}\label{Sample}
The sample discussed in this paper is drawn from the WISH survey 
and covers a wide range of luminosities and different evolutionary stages.
A total of 51 sources are included in this study, which can be classified into three groups according to 
their bolometric luminosities, $L_{\rm{bol}}$.
The sub-sample of low-mass YSOs, characterized by $L_{\rm{bol}} <$ 50~L$_\odot$, is composed of 15 Class~0 
and 11 Class~I protostars (see \citealt{Evans09} for details of the classification). 
Six intermediate-mass sources were observed with 
70~L$_\odot$ $< L_{\rm{bol}} <$ 2$\times$10$^3$~L$_\odot$. Finally, 19 high-mass YSOs with 
$L_{\rm{bol}} >2\times$10$^3$ L$_\odot$
complete the sample. The bolometric luminosity of the sample members, together with their 
envelope masses ($M_{\rm{env}}$), distances ($d$) and source velocities ($\varv_{\mathrm{LSR}}$) 
are summarised in Table~\ref{tbl:sources}. For more information
about the sample studied in WISH, see \citet{vanDishoeck11}. 
Focusing on the evolutionary stages, the sub-sample of low-mass YSOs ranges from Class~0 to Class~I,
the intermediate-mass objects from Class~0 to Class~I as well,
and in the case of the high-mass sources, from (mid-IR-quiet/mid-IR-bright) 
massive young stellar objects (MYSOs) to ultra-compact \ion{H}{ii} regions (UC\ion{H}{ii}).
\placeTableSources

\subsection{HIFI observations}\label{HIFIobservations}
The sources were observed with the Heterodyne
Instrument for the Far Infrared (HIFI) on the \textit{Herschel Space 
Observatory}. The HIFI CO and isotopologue lines studied in this paper are: 
$^{12}$CO~$J$=10--9, $^{13}$CO~$J$=10--9,
C$^{18}$O~$J$=5--4, $J$=9--8 and $J$=10--9. 
The upper-level energies and frequencies of these lines together with the HIFI bands, 
main beam efficiencies ($\eta_{\mathrm{MB}}$), 
beam sizes, spectral resolution and integration times are presented in Table \ref{tbl:overview_co}.
With the exception of the $^{12}$CO~$J$=10--9 line, all
isotopologue line observations were obtained together with H$_2$O lines. 
The $^{12}$CO~$J$=10--9 line was targeted for the low- and intermediate-mass
sample but only for one high-mass object (W3-IRS5). The $^{13}$CO~$J$=10--9 and C$^{18}$O~$J$=9--8
lines were observed for the entire sample, while C$^{18}$O~$J$=5--4 only for the Class~0 and
intermediate-mass protostars. C$^{18}$O~$J$=10--9 was observed for all
low-mass Class~0 sources, two low-mass Class~I (Elias\,29 and
GSS\,30\,IRS1), one intermediate-mass YSO (NGC\,7129) and the entire high-mass sub-sample.

Single-pointing observations were performed for all targets in dual-beam-switch (DBS) mode, 
chopping to a reference position 3$\arcmin$ from the target. 
There is no contamination due to emission at the off position except for 
the $^{12}$CO~$J$=10--9 spectrum of NGC1333\,IRAS2A and IRAS4A (see \citealt{Yildiz10} for more details). 
These spectra have been corrected and presented in this paper without contamination. 
In the case of W43-MM1, the absorption features found in the $^{13}$CO~$J$=10--9 spectrum are caused by
H$_2$O$^+$ \citep{Wyrowski10}.

HIFI has two backends: the Wide Band Spectrometer (WBS) and the High
Resolution Spectrometer (HRS). Both spectrometers simultaneously measure two polarizations, horizontal (H) and vertical
(V). For more details, see \citet{Roelfsema12}.
The WBS has a constant spectral resolution of 1.1~MHz, whereas the 
HRS has different configuration modes with four possible spectral resolutions: 0.125, 0.25, 0.5 and 1.0~MHz.
The spectral resolution for each of the studied HIFI lines is listed in Table~\ref{tbl:overview_co}. 
The WBS data present lower noise than the HRS data (factor of $\sqrt{2}$) and provide 
a good compromise between noise and resolution.
Therefore, the WBS data are the primary focus of this paper.
HRS observations are only used for analysing the C$^{18}$O~$J$=5--4 line for the low-mass sources
because their narrow line profiles require the higher spectral resolution provided by these data.

The data reduction was performed using the standard HIFI pipeline in the \textit{Herschel} Interactive
Processing Environment (HIPE\footnote{HIPE is a joint development by the Herschel Science Ground Segment 
Consortium, consisting of ESA, the NASA Herschel Science Centre, and the HIFI, PACS and SPIRE consortia.})
ver. 8.2 \citep{OttS10}, resulting in absolute calibration on the corrected antenna temperature $T_{A}^{*}$  
scale, and velocity calibration with a $\varv_{\mathrm{LSR}}$ precision of a few m\,s$^{-1}$.  
The version of the calibration files used is 8.0, released in February 2012.  
The flux scale accuracy was estimated to be 10\% for bands 1, 4 and 5. 
Subsequently, the data were exported to 
{\small{GILDAS-\verb1CLASS1}}\footnote{{http://www.iram.fr/IRAMFR/GILDAS/}} 
for further analysis. The H and V polarizations were observed simultaneously 
and the spectra averaged to improve the signal-to-noise ratio.
In order to avoid possible discrepancies between both signals, the two polarisations were inspected 
for all the spectra presented in this paper with no differences $>$ 20\% found.
Afterwards, line intensities were converted to main-beam brightness temperatures through the 
relation \mbox{$T_{\mathrm{MB}} = T_{A}^{*}/ \eta_{\mathrm{MB}}$} (see \citealt{Wilson&Rohlfs09} 
for further information about radio-astronomy terminology). 
The main beam efficiency, $\eta_{\mathrm{MB}}$, for each HIFI band was taken from \citet{Roelfsema12}
and listed in Table~\ref{tbl:overview_co}. 
The final step of the basic reduction was the subtraction of a
constant or linear baseline.

\subsection{JCMT ground-based observations}\label{JCMTobservations}
Complementary data from the 15-m James Clerk Maxwell
Telescope (JCMT) on Mauna Kea, Hawaii are also included in this paper,
in particular for the high-mass sources for which $^{12}$CO~$J$=10--9 data are not available.
Jiggle map observations of $^{12}$CO~$J$=3--2 and
C$^{18}$O~$J$=3--2 for a sub-sample of YSOs were obtained with 
the Heterodyne Array Receiver Program (HARP, \citealt{Buckle09}) in August 2011 and summer 2012 (proposal M11BN07 and M12BN06). 
For the sources and transitions not included in the proposal, comparable data were obtained from 
the JCMT public archive. Four low-mass sources were observed with the 
12-m Atacama Pathfinder Experiment Telescope, APEX, 
because these protostars are not visible from the JCMT (see Appendix~\ref{JCMT_Data}).   
Further information about the low-mass YSOs and data can be found in \citet{Yildiz13sub}. 

The HARP instrument is a $4\times4$ pixel receiver array, although during the observation period
one of the receivers (H14) was not operational. The lines were observed in position-switching mode, 
with the off-positions carefully chosen in order to avoid contamination.
For the most massive and crowded regions, test observations of the off-position were taken for this purpose. 
The spatial resolution of the JCMT at the observed frequencies is
$\sim$14$\arcsec$, with a main beam efficiency of 
0.63\footnote{{http://www.jach.hawaii.edu/JCMT/spectral$\_$line/General/status.html}}.
This same value of \mbox{$\eta_{\mathrm{MB}}$} was used for the data obtained from the JCMT archive because the
small variations of this parameter ($<$~10$\%$) recorded over time
are negligible compared to the calibration uncertainties of the JCMT ($\sim$20$\%$, \citealt{Buckle09}). 
Some of the spectra collected from the JCMT archive were observed in a lower spectral resolution setting. Therefore, for
these data the spectral resolution is 0.4~km\,s$^{-1}$ instead of 0.1~km\,s$^{-1}$ (indicated in Table~\ref{tbl:overview_co}).

In the first step of the reduction process, the raw ACSIS data downloaded from the JCMT archive were transformed 
from \textit{sdf} format to \textit{fits} format using the 
Starlink\footnote{{http://starlink.jach.hawaii.edu/starlink}} package for each and every pixel.
Next, the data were converted to {\small{\verb1CLASS1}} format and the central spectrum was extracted after 
convolving the map 
to the same beam size as the $^{12}$CO~$J$=10--9 HIFI observations (20$\arcsec$).  
Line intensities were then converted to the main-beam brightness temperature scale and linear 
baselines subtracted. Since this manuscript focuses on analysing and comparing the central spectrum 
of the studied YSOs, the full JCMT spectral maps will be presented and discussed in a forthcoming paper.

\placeTableOverviewCO

\subsection{ Decomposition method}\label{Decomposition} 
 
In order to quantify the parameters that fit each spectrum, the following procedure was applied to
all spectra. First, the data were re-sampled to 0.27~km\,s$^{-1}$  
so that the results could be compared in a systematic manner. Then, the spectra were
fitted with a single Gaussian profile using the {\small{IDL}} function \textit{mpfitfun}, 
after which we plotted the residuals obtained from the fit to confirm whether the line profile 
hid an additional Gaussian component. 
For those sources whose profiles showed clear sub-structure, i.e., the residuals were larger than  
3 sigma rms, a two Gaussian component fit was used instead. Examples of the decomposition
procedure are shown in Fig.~\ref{fig:decomposition}. 
The results of this process  
together with the rms and integrated intensity for all lines 
are presented in Tables \ref{table:12CO10-9-W-Tpeak} to \ref{table:C18O10-9-W-Tpeak} in Appendix \ref{Data}.

All HIFI lines are observed in emission and none of the HIFI spectra present clear infall signatures.
Moreover, some CO lines show weak self-absorption features, which are of 
marginal significance and will not be discussed further.
In addition, extremely high velocity (EHV) emission features have been identified. The EHV components 
are knotty structures spaced regularly and 
associated with shocked material moving at velocities of hundreds of km\,s$^{-1}$ 
(e.g. \citealt{Bachiller90}). These structures have been detected in
the $^{12}$CO~$J$=10--9 spectra for the low-mass Class~0 sources L1448-MM
and BHR\,71 (\citealt{Kristensen11}; \citealt{Yildiz13sub}).  
These EHV components were not included in the study of the line profiles, 
so the residuals were analysed after fitting each of these features with a Gaussian function
and subtracting them from the initial profile. 

The method used for examining the data is similar to that introduced by \citet{Kristensen10} for several water lines 
in some of the WISH low-mass YSOs, applied to high-$J$ CO in \citet{Yildiz10} 
and extended in \citet{Kristensen12} for the 557 GHz 1$_{10}$--1$_{01}$ water 
line profiles of the entire low-mass sample. 
The emission lines are classified as narrow or broad if the full width half maximum (FWHM) of the 
Gaussian component is lower or higher than 7.5~km\,s$^{-1}$, respectively.
This distinction is made because 7.5~km\,s$^{-1}$ is the maximum width obtained in the single Gaussian fit of 
the HIFI C$^{18}$O lines, which is considered as narrow and traces the dense warm 
quiescent envelope material (see Section \ref{Discussion} for further analysis and discussion). 

The narrow component identified in the high-$J$ CO isotopologue
lines is always seen in emission, unlike a component of similar
width seen in the 1$_{10}$-1$_{01}$
line of water (\citealt{Kristensen12}). The narrow components in
high-$J$ CO and low-$J$ water probe entirely different parts of the
protostar: the former traces the quiescent warm envelope material,
the latter traces the cold outer envelope and ambient cloud. 
The broad emission in low-$J$ CO is typically narrower than in water
and traces entrained outflow material.
Only the highest-$J$ lines observed by HIFI
trace the same warm shocked gas as seen in the water lines \citep{Yildiz13sub}. 
To summarise, the components identified in the CO and
isotopologue data cannot be directly compared to those observed in
the H$_2$O 1$_{10}$-1$_{01}$ lines because the physical and chemical conditions
probed by water are different to those probed by CO (see \citealt{Santangelo12};
\citealt{Vasta12}).

In the analysis of the JCMT data, the FWHM of the broad velocity component for the complex
$^{12}$CO~$J$=3--2 line profiles was disentangled by masking in each spectrum the narrower 
emission and self-absorption features. The width of the narrow C$^{18}$O~$J$=3--2 lines 
was constrained by fitting these profiles with a single Gaussian. 
The results of these fits are presented in Table~\ref{tbl:jcmt_data} in Appendix~\ref{JCMT_Data}.

\placeFigDecomposition



\placeFigRepresentadoceco
\placeFigRepresentacDieciochoo
\hyphenation{sy-mme-tric}

\section{Results}\label{Results}

One of the aims of this paper is to characterise how the observed 
emission lines compare as a function of source luminosity.
In order to simplify the comparison across the studied mass range, 
the main properties and parameters of the HIFI and JCMT lines, such as line morphology, total intensity 
and kinetic temperature, are presented in this section. 
A more detailed description of the line profiles is reserved for Appendices~\ref{Data} and \ref{JCMT_Data}
(Figs.~\ref{fig:Spec12CO10-9} to \ref{fig:SpecC18O10-9} show the HIFI spectra and 
Figs.~\ref{fig:Spec12CO3-2} and \ref{fig:SpecC18O3-2} the JCMT data). 

Further study and analysis of each sub-sample will be presented in several forthcoming papers.
The CO lines for the low-mass sources and their excitation will be discussed by \citet{Yildiz13sub}. 
A review of the intermediate-mass sources focused on the water lines will be performed by
\cite{McCoeyinprep}.
In the case of high-mass YSOs, low-$J$ H$_2$O line profiles will be studied in detail by 
\cite{vanderTakinprep}.

In this manuscript a summary with the main characteristics of the studied emission line profiles 
is presented in Section~\ref{Line-profiles}. 
Section~\ref{Correlations} describes the calculation of the line luminosity, $L_{\rm{CO}}$, for each observed isotopologue  
together with its correlation with $L_{\rm{bol}}$. Finally, in Section~\ref{Tkin}, an estimation of the
kinetic temperature is obtained for two sources, an intermediate-mass
and a high-mass YSO, and compared with values obtained for low-mass sources.

\subsection{ Characterisation of the line profiles}\label{Line-profiles}

Figures \ref{fig:1213co} and \ref{fig:c18o} show characteristic profiles of 
each transition and YSO sub-type, so the line profiles can be compared across the luminosity range. 
$^{12}$CO~$J$=10--9 spectra present more intense emission lines than the other observed isotopologues and
more complex line profiles. Two velocity components are identified and most of the $^{12}$CO~$J$=10--9 
spectra can be 
well fitted by two Gaussian profiles (Fig.~\ref{fig:decomposition}a). 
Weak self-absorption features are also observed in some sources, such as Ser\,SMM1.
$^{13}$CO~$J$=10--9 profiles are weaker and narrower than $^{12}$CO~$J$=10--9 spectra.  
Some of the detected lines, especially for the high-mass sample, are fitted using two Gaussian components
(Fig.~\ref{fig:decomposition}a). 
In the case of the C$^{18}$O~$J$=5--4 spectra, a weak broad velocity component is identified 
in 6 sources (indicated in Table~\ref{table:C18O5-4-W-Tpeak}), 
due to the long exposure time and the high $S/N$ reached for this transition.
The width of this broad component is narrower by a factor of 2--3 
than that detected for the $^{12}$CO and $^{13}$CO~$J$=10--9 lines.
Similarly, two velocity components have been identified for the C$^{18}$O~$J$=9--8 line in three high-mass 
sources: G10.47+0.03, W51N-e1 and G5.89-0.39 (see Fig.~\ref{fig:SpecC18O10-9}). These massive objects 
present the widest broad velocity components for both $^{13}$CO and C$^{18}$O~$J$=10--9 spectra.
The width of the broad C$^{18}$O~$J$=9--8 component is slightly smaller than that identified in the 
$^{13}$CO~$J$=10--9 emission for each of these YSOs. 

Two velocity components were previously identified in approximately half of the 20
deeply embedded young stars in the Taurus molecular cloud studied by \citet{Fuller&Ladd02}
using lower-$J$ C$^{18}$O observations.
They found typical FWHM line widths of $\sim$0.6 and $\sim$2.0~km\,s$^{-1}$ for the 
narrow and broad component respectively.
These values are significantly narrower than the widths obtained
from the HIFI data, so our interpretation and analysis of these components is
different to that presented by \citet{Fuller&Ladd02}.
On the other hand, the bulk of C$^{18}$O line profiles (especially $J$=9--8 and $J$=10--9 transitions) 
are generally well 
fitted by a single Gaussian (for an example, see Fig.~\ref{fig:decomposition}b).
Therefore, in our analysis only the narrow C$^{18}$O component is 
considered for the three sources with two velocity component profile. 
Regarding the line intensity,  
the spectra of the observed high-mass YSOs have higher main beam temperatures  
than the spectra of the intermediate-mass objects, 
which in turn show stronger lines than the low-mass sources.  

Another result obtained when we extend this characterisation to the JCMT data is
the complexity of the $^{12}$CO~$J$=3--2 profiles compared to the $^{12}$CO~$J$=10--9 spectra  
(see Figs.~\ref{fig:Spec12CO10-9} and \ref{fig:Spec12CO3-2} for comparison across the studied sample).
The HIFI data probe warmer gas  
from inner regions of the molecular core and present 
simpler emission line profiles (with no deep absorptions and foreground emission features)
than the lower-$J$ spectra. However, similar to the $^{12}$CO~$J$=10--9 line profiles, 
the $^{12}$CO~$J$=3--2 spectra can be decomposed into different velocity components.
A broad velocity component is identified in 39 out of 47 sources, ranging
from $\sim$7.4 to 53.5~km\,s$^{-1}$ in width.
For C$^{18}$O, the shape of the $J$=3--2 lines  
are very similar to those of the $J$=9--8 lines 
(see Figs.~\ref{fig:HIFIvsJCMTLM} to \ref{fig:HIFIvsJCMT} for examples).

The FWHM of the $^{12}$CO~$J$=3--2 broad component for most of
the high-mass sources is approximately double the width obtained for the $^{13}$CO~$J$=10--9 broad component
(values in Tables~\ref{fig:Spec13CO10-9} and \ref{tbl:jcmt_data}).
In the case of the one source for which a $^{12}$CO~$J$=10--9 observation was performed as part of WISH (W3-IRS5), the
width of the broad component is similar to that calculated for the $J$=3--2 spectrum (factor of 1.2$\pm$0.1) and is
twice the width of the $^{13}$CO~$J$=10--9 emission line (see Fig.~\ref{fig:1213co}).
Similar ratios were found by \citet{vanderWiel13sub} for the high-mass source AFGL2591 
as part of the CHESS (``Chemical \textit{HErschel} Survey of Star-forming regions'') key program observations.
For the intermediate-mass object NGC\,2071, 
the $^{12}$CO~$J$=10--9 broad component is 1.7$\pm$0.1 larger than 
the width of the $^{13}$CO~$J$=10--9 broad component. 
This ratio is 1.5$\pm$0.4 for the one low-mass YSO for which 
a decomposition of the line profile can be performed in both transitions simultaneously (Ser\,SMM1). 
Thus, it appears that the $^{12}$CO~$J$=10--9 profile becomes increasingly broader compared to the 
$^{13}$CO~$J$=10--9 profile with increasing protostellar mass. 
The average ratio of the width of the
broad component of the $^{12}$CO~$J$=10--9 line divided by the width of the broad component of the 
$^{12}$CO~$J$=3--2 line is approximately 1.0$\pm$0.1 for the intermediate-mass sources and 1.3$\pm$0.2 for the 
low-mass protostars. 

In order to compare the broad velocity component of the $^{12}$CO data with the 
narrow C$^{18}$O line profiles across the
entire studied luminosity range, 
the FWHM of the $^{12}$CO~$J$=3--2 spectra is used as a proxy for the
FWHM of the $^{12}$CO~$J$=10--9 profiles for the high-mass sample.
The widths of the fits obtained for the $^{12}$CO broad velocity components and the narrow  
C$^{18}$O~$J$=9--8 and $J$=3--2 lines are plotted versus their bolometric luminosities 
(Figs.~\ref{fig:HistogrBroad} and \ref{fig:HistogrC18O-HIFI}).  
From the figure of the broad velocity component of the $^{12}$CO data we infer 
that the line-wings become broader from low- to high-mass. Low-mass Class~0 protostars 
characterised by powerful outflows, such as L1448, BHR\,71 and L\,1157, 
are the clear outstanding sources in the plot. The median FWHM of this 
component for  
each sub-group of protostars together with the calculated median of the FWHM values for the C$^{18}$O~$J$=3--2
and $J$=9--8 lines are summarised in Table~\ref{tbl:median_values}. 
Even though there are only six intermediate-mass sources and the results could be sample biased, 
the trend of increasing width from low- to 
high-mass is consistent with the result obtained for intermediate-mass objects.

Regarding the C$^{18}$O lines, Fig.~\ref{fig:HistogrC18O-HIFI} and Table~\ref{tbl:median_values} 
show a similar behaviour to that observed for the broad component of the $^{12}$CO but with less dispersion, i.e.,
the profiles become broader from low- to high-mass. This trend is statistically stronger for the $J$=3--2 transition 
(the Pearson correlation coefficient is higher than that calculated for the $J$=9--8 line widths) since 
the number of detections is higher for the low-mass sample. 
The C$^{18}$O~$J$=3--2
spectra show slightly narrower profiles than the $J$=9--8 line for the low- and intermediate-mass sources, 
with median values approximately half the values obtained for the $J$=9--8 line (see Table~\ref{tbl:median_values}).
On the other hand, for the high-mass sources the median values are practically the same, 
and similar widths are measured for the high-mass sub-sample in both transitions.
This result will be discussed further in Section~\ref{Discussion}.
 
\placeTableMedian
\placeFigHistogramaBroad
\placeFigHistogramaCdiecioHIFI

\subsection{Correlations with bolometric luminosity}\label{Correlations}

The analysis and characterisation of the line profiles continue with the calculation 
of the integrated intensity of the emission line, \textit{W}=${\int{T_{\rm{MB}}{\rm{d}}\varv}}$.
This parameter is obtained by integrating the intensity of each detected emission 
line over a velocity range which is defined using a $3\sigma$ rms cut. 

In order to obtain a more accurate value of \textit{W} 
for data with lower \textit{S/N}, 
such as for the high-$J$ C$^{18}$O lines 
from the low-mass sources, this parameter was approximated to the area
of the fitted single Gaussian profile. 
The calculated integrated intensities of some sources were compared with measurements from previous 
independent studies. In the case of NGC\,1333\,IRAS2A/4A/4B \citep{Yildiz10}, differences in $W$ 
are not larger than $10\%$. 
The obtained values from all lines are given in Tables~\ref{table:12CO10-9-W-Tpeak} 
to \ref{table:C18O10-9-W-Tpeak}.  

If the emission is optically thin, 
\textit{W} is proportional to the column density of the specific upper level.
In local thermal equilibrium (LTE), the variation of \textit{W} with $J_{\rm{u}}$ characterises 
the distribution of the observed species over the different 
rotational levels (see equation 15.28 from \citealt{Wilson&Rohlfs09}). 
In the case of the optically thin C$^{18}$O~$J$=9--8 line, the total C$^{18}$O column density, 
$N_{\rm{t}}$, is calculated for all
sources in order to obtain the H$_2$ column density, $N_{\rm{H_2}}$. The assumed excitation temperature, 
$T_{\rm{ex}}$, is 75~K based on the work of \citet{Yildiz10}, which shows that 90\% of the emission
in the $J$=9--8 transition originates at temperatures between 70 and 100~K. The column density $N_{\rm{H_2}}$ 
is then obtained by assuming an C$^{18}$O/H$_2$ abundance ratio of 5$\times$10$^{-7}$. This ratio is obtained by 
combining the $^{16}$O/$^{18}$O isotopologue abundance ratio equal to 540 (\citealt{Wilson&Rood94}), and
the $^{12}$CO/H$_2$ ratio as 2.7$\times$10$^{-4}$ (\citealt{Lacy94}). 
The calculated $N_{\rm{H_2}}$ values for C$^{18}$O~$J$=9--8
are presented in Table~\ref{table:C18O9-8-W-Tpeak}.

The integrated intensity is converted to line luminosity, $L_{\rm{CO}}$, in order to compare these results 
for sources over a wide range of distances. The CO and isotopologue line luminosities for each YSO is  calculated using 
equation 2 from \cite{Wu05} assuming a Gaussian beam and point
source objects. 
If the emission would cover the entire beam, the line luminosities would increase by a factor of 2.
The logarithm of this line luminosity,  $\log$\,($L_{\rm{CO}}$), is plotted versus the logarithm of the 
bolometric luminosity, $\log$\,($L_{\rm{bol}}$), 
for $^{12}$CO~$J$=10--9, $^{13}$CO~$J$=10--9 and C$^{18}$O~$J$=9--8 in Fig.~\ref{fig:Correlation12CO}. 
The errors are calculated 
from the rms of the spectrum and considering $\sim$20\% distance uncertainty.  
A strong correlation is measured (Pearson correlation coefficient r $>$0.92) between 
the logarithms of $L_{\rm{CO}}$ and $L_{\rm{bol}}$ for all observed CO lines.
The top plot of Fig.~\ref{fig:Correlation12CO} shows $L_{\rm{CO}}$ for $^{12}$CO~$J$=10--9 
emission for all the observed sources versus their $L_{\rm{bol}}$. Only one high-mass source 
was observed as part of WISH in this line with HIFI (W3-IRS5) with the value of the integrated intensity for AFGL2591 
obtained from \citet{vanderWiel13sub}. 
Even though this plot is mainly restricted to low- and intermediate-mass sources, a strong 
correlation is still detected over 5 orders of magnitude in both axes. 
Both low-mass Class~0 and Class~I YSOs follow the same correlation, though the uncertainties of the calculated 
$L_{\rm{CO}}$ for these sources are higher than for the other types of protostars because the $S/N$ is lower.
All high-mass objects were observed in $^{13}$CO~$J$=10--9, so the correlation between $\log$\,($L_{\rm{CO}}$) 
and $\log$\,($L_{\rm{bol}}$) (Fig.~\ref{fig:Correlation12CO}, middle) is 
confirmed and extends over almost 6 orders of magnitude in both axes.
This correlation is also seen for C$^{18}$O~$J$=9--8 but with higher dispersion
(Fig.~\ref{fig:Correlation12CO}, bottom), and in the other observed transitions of this isotopologue. 

The values of the correlation coefficient
and the fit parameters for all these molecular transitions are presented in Table~\ref{tab:linear_fit}.
The correlation prevails for all transitions even if the values of integrated intensity are not converted to line luminosity.
Therefore, $\log$\,($W$) still correlates with $\log$\,($L_{\rm{bol}}$) over at least 3 and 6 orders of magnitude on the 
y and x axis, respectively.
Similar correlations are obtained when plotting the logarithm of $L_{\rm{CO}}$ versus the logarithm of the 
source envelope mass, $M_{\rm{env}}$, for all the targeted lines  
(see Fig.~\ref{fig:CorrelationC18OMenv} for an example using the \mbox{C$^{18}$O~$J$=9--8 line}). 
In these representations, the modelled envelope mass of the source is directly compared to $L_{\rm{CO}}$, 
a tracer of the warm envelope mass. 
Therefore, the correlation is extended and probed by another proxy of the mass of the protostellar system.

Since the index of the fitted power-law exponents is $\sim$1 
within the uncertainty of the fits (see Table~\ref{tab:linear_fit}), 
these correlations show 
that $\log$\,($L_{\rm{CO}}$) is proportional to $\log$\,($L_{\rm{bol}}$). In the optically thin case,
this correlation implies that the column density of warm CO increases proportionally with the 
mass of the young stellar object.
This result can be applied to C$^{18}$O because the emission lines of this isotopologue are 
expected to be optically thin. Assuming that $^{13}$CO~$J$=10--9 is optically thin as well, 
the column density would increase 
proportionally with the luminosity of the source, and practically 
with the same factor as the studied C$^{18}$O transitions. 
Therefore, even though the conditions in low-, intermediate- and high-mass star-forming regions 
are different and distinct physical and chemical processes are expected to be more significant 
in each scenario (e.g., ionising radiation, clustering, etc.), the column density of CO seems to depend on 
the luminosity of the central protostar alone, showing a self-similar behaviour from low- to high-mass. 

To test the optically thin assumption for $^{12}$CO~$J$=10--9 and especially for $^{13}$CO~$J$=10--9, 
the line luminosities for the $^{13}$CO~$J$=10--9 and the C$^{18}$O~$J$=10--9 data were multiplied by 
a $^{12}$C/$^{13}$C ratio of 65 \citep{Vladilo93}\footnote{ The $^{12}$C/$^{13}$C ratio varies with galactocentric 
radius by up to a factor of 2 but this effect is minor and is ignored.},
and by a $^{16}$O/$^{18}$O ratio of 540 \citep{Wilson&Rood94}, respectively.
Therefore, the observed and predicted values of $L_{\rm{CO}}$ for $^{12}$CO~$J$=10--9 and $^{13}$CO~$J$=10--9,
together with those of C$^{18}$O~$J$=10--9 can be compared across the studied luminosity range
(see Fig.~\ref{fig:Correlation1213CO}).
In the case of the $^{13}$CO~$J$=10--9 line, the values of the observed and predicted line luminosity
are similar ($\lesssim$20\% in most of them), 
especially at lower luminosities. In addition, the slope of their fits are practically the same within the
uncertainty, so a similar behaviour is proved. From these results we can assume that in general $^{13}$CO~$J$=10--9 
is optically thin. 
For $^{12}$CO~$J$=10--9, the ratio of predicted to observed line luminosity  
{\mbox{($65\times L_{\rm{CO}}[^{13}$CO~$J$=10--9]/$L_{\rm{CO}}[^{12}$CO~$J$=10--9])}}
ranges from 0.8 to 12.5, for IRAS\,15398 and
W3-IRS5 respectively, and the average obtained is 3.3. 
Therefore, $^{12}$CO~$J$=10--9 is optically thick, at least at the line centre which dominates the intensity,
and the relative value of the optical depth, $\tau$, increases slightly with the mass of the protostar
($\tau$ $\sim$1.5 for the low-mass sources, 2.0 for the intermediate and $\sim$3.4 for the high-mass object). 
This is in keeping with the expectation that massive YSOs form in the densest parts of the giant molecular clouds, GMCs.

Correcting for optical depth, \mbox{$L_{\rm{CO}}[^{13}$CO~$J$=10--9]} can be used to derive 
\mbox{$L_{\rm{CO}}[^{12}$CO~$J$=10--9]} because both species present a similar behaviour across the
luminosity range (similar slopes in their fits). 
This relation can be used in the calculation of $L_{\rm{CO}}$ for those 
sources for which there are no $^{12}$CO~$J$=10--9 observations, that is, the high-mass sample.
As highlighted before, using $^{13}$CO as a proxy for $^{12}$CO is restricted to comparisons of integrated 
intensities of the emission lines across the studied 
mass spectrum, and cannot be extended to the analysis of the line profile of $^{12}$CO and 
{\mbox{$^{13}$CO~$J$=10--9}}.

\placeFigCorrelationCO
\placeFigCorrelationTreceCOMenv
\placeFigCorrelationdoceTreceCO
\placeTableCorrelationFits

\subsection{ Kinetic temperature}\label{Tkin}

The ratio of the $^{12}$CO~$J$=10--9 and $J$=3--2 line wings can be
used to constrain the kinetic temperature $T_{\rm{kin}}$ of the
entrained outflow gas if the two lines originate from the same gas.  
Y{\i}ld{\i}z et al.\ (2012) and (2013, submitted) have
determined this for the sample of low-mass YSOs. Here we consider two
sources to investigate whether the conditions in the outflowing gas
change with increasing YSO mass: the intermediate-mass YSO NGC\,2071 and
the high-mass object W3-IRS5. 
The critical densities, $n_{\rm{cr}}$, of the $^{12}$CO~$J$=3--2 and
$J$=10--9 transitions at 70~K are $\sim$2.0$\times$10$^4$ and 
4.2$\times$10$^5$~cm$^{-3}$, respectively. 
The values were calculated using equation 2 from \citet{Yang10}, the CO rate coefficients presented
in their paper and considering only para-H$_2$ collisions.
The densities inside the HIFI beam for $^{12}$CO~$J$=10--9
(20$\arcsec$) of both sources are higher than $n_{\rm{cr}}$.
Therefore, the emission is thermalised and $T_{\rm{kin}}$ can be directly constrained by the
$^{12}$CO~$J$=10--9/$J$=3--2 line wing ratios. 

The observed ratios of the red and blue wings for these two sources as a function of absolute 
offset from the source velocity are presented in Fig.~\ref{fig:line_ratio}. 
These ratios are compared with the values
calculated by \citet{Yildiz13sub} from the $^{12}$CO~$J$=10--9 and $J$=3--2 
averaged spectra for the low-mass Class~0 sample (shaded regions).
Since the emission is optically thin and we can assume LTE, the kinetic temperatures are calculated
from the equation that relates the column density and the integrated intensity \citep{Wilson&Rohlfs09}.
The obtained $T_{\rm{kin}}$ varies from 100 to 210~K.
Both the observed line ratios as well as the inferred kinetic 
temperatures are similar to those found for the low-mass YSOs, where
$T_{\rm{kin}}$ ranges from 70-200~K for Class~0. Although only a couple of higher-mass sources have been
investigated, the temperatures in the entrained outflow gas seem to be
similar across the mass range. Note that if part of the$ ^{12}$CO~$J$=10--9 emission originates from a 
separate warmer component, the above values should be regarded as upper limits. 

\placeFigTkin



\section{Discussion}\label{Discussion}   

The HIFI data show a variety of line profiles with spectra that can be decomposed into two different velocity 
components, such as the $^{12}$CO~$J$=10--9 lines, and spectra that show
narrow single Gaussian profiles (C$^{18}$O data). In addition, a strong correlation is found between 
the line and the bolometric luminosity for all lines. The next step is
to analyse these results to better understand which physical processes are taking place.

Section~\ref{BroadvsNarrow} compares the narrow C$^{18}$O lines and the 
$^{12}$CO broad velocity component in order to better understand the physics that
these components are tracing and the regions of the protostellar environment they are probing.
The dynamics of the inner envelope-outflow system is studied   
in Section~\ref{Origin}. Finally, the interpretation of CO as a dense gas tracer is discussed
in Section~\ref{SFT}.

\subsection{ Broad and narrow velocity components}\label{BroadvsNarrow}

The broad velocity component identified in most of the $^{12}$CO~$J$=3--2 and $J$=10--9 spectra 
is related to the velocity of the entrained outflowing material, so the wings of $^{12}$CO can be
used as tracers of the outflow properties (\citealt{Cabrit92}; \citealt{Bachiller99}). However, there are
different effects that should be taken into account when this profile component is analysed,
such as the viewing angle of the protostar and the \textit{S/N}. 
The former could alter the width of the broad component due to projection 
or even make it disappear if the outflow is located in
the plane of the sky. Low \textit{S/N} could also hide the broad component for sources with
weak emission. Moreover, the broad velocity component should be weaker if the emission lines
come from sources at later evolutionary stages since their outflows become weaker and less collimated 
(see reviews of \citealt{Bachiller99}; \citealt{Richer00}; \citealt{Arce07}). 

In order to compare the line profiles of all observed CO lines for each type of
YSO and avoid the effects of inclination and observational noise playing a 
role in the global analysis of the data,
an average spectrum of each line for each sub-type of protostar has been calculated and presented
in Figs.~\ref{fig:aver1213co} and \ref{fig:averc18o}. 
Regarding the low-mass sample, we observe a striking decrease 
in the width of the broad component from Class~0 to Class~I. 
This result shows the decrease of the outflow
force for more evolved sources in the low-mass sample 
is reflected in the average spectra \citep{Bontemps96}.  
\placeFigAveragedoceco
\placeFigAveragediecico

A narrow velocity component has been defined as a line profile
that can be fitted by a Gaussian function with a FWHM smaller than 7.5~km\,s$^{-1}$ 
(see Section~\ref{Decomposition} for more details).
Since C$^{18}$O lines are expected to trace dense quiescent envelope material, high-$J$ transitions 
are probing the warm gas in the inner envelope. 
The average C$^{18}$O spectra for each type of YSO are compared in Fig.~\ref{fig:averc18o}. 
The FWHM of the emission lines increases from low- to
high-mass protostars (see Fig.~\ref{fig:HistogrC18O-HIFI} and Table~\ref{tbl:median_values}). 
An explanation for this result could be that for massive regions,  
the UV radiation from the forming OB star ionises the gas, creating an
H{\sc{ii}} region inside the envelope which 
increases the pressure on its outer envelope.
This process may lead to an increase in the turbulent velocity
of the envelope material \citep{Matzner02}, thus broadening the narrow component. 
Therefore, our spectra are consistent with the idea that in general, turbulence in the protostellar envelopes 
of high-mass objects is expected to be stronger than for low-mass
YSOs (e.g. \citealt{Herpin12}).

Higher rotational transitions trace material at 
higher temperatures and probe deeper and denser parts of the inner envelope. 
For the low- and intermediate-mass sources, 
the FWHM of the C$^{18}$O~$J$=3--2 spectra are generally half that obtained for the C$^{18}$O~$J$=9--8 
and slightly smaller than those obtained for the $J$=5--4 transition. 
However, for the high-mass YSOs the values of the FWHM are similar for the $J$=3--2 and $J$=9--8 transitions.
In order to understand which kind of processes (thermal or non-thermal) dominate in the 
inner regions of the protostellar envelope traced by our observations, 
the contribution of these two processes to the line width is calculated. 
The aim is to explain whether the broadening of high-$J$ emission lines is caused by thermal
or non-thermal motions. 

In the case of the $J$=3--2 lines, the upper energy level is 31~K, so the thermal line width, 
$\Delta \varv_{\rm{th}}$, is 
0.12~km\,s$^{-1}$ for C$^{18}$O at this temperature. Comparing this value with the measured 
FWHM of the C$^{18}$O~$J$=3--2 spectra in Table \ref{tbl:jcmt_data},
we conclude that thermal motions contribute less than 5\% to the total observed line width, 
$\Delta \varv_{\rm{obs}}$. 
Therefore, the line width is dominated by non-thermal motions $\Delta \varv_{\rm{noth}}$.  
The C$^{18}$O~$J$=9--8 line profiles trace warmer gas (up to 300~K) with respect to $J$=3--2 
increasing the thermal contribution.
However, even at 300~K $\Delta \varv_{\rm{th}}$ is 0.68~km\,s$^{-1}$, which means that
$\Delta \varv_{\rm{noth}}/ \Delta \varv_{\rm{obs}}$ is larger than 0.93 even for the 
low-mass sources. Thus, non-thermal motions predominate over thermal ones in the studied regions of 
the protostellar envelopes. These motions are assumed to be
independent of scale and do not follow the traditional size-line width
relation (\citealt{Pineda10}). Therefore, these results are not biased by the distance of the source.

This analysis can be compared to pre-stellar cores, in which the line profiles are closer to 
being dominated by thermal motions.
For this purpose, the line width values calculated for our data 
are compared to those of \cite{Jijina99}.
In that work, a database of 264 dense cores mapped in the ammonia lines $(J,K)$=(1,1) and (2,2) is presented. 
Histograms in Fig.~\ref{fig:preLM} compare the values of the line widths observed for pre-stellar 
cores with the observed line width of the C$^{18}$O~$J$=3--2 and $J$=9--8 data for the WISH 
sample of protostars 
for low- (top) and for high-mass (bottom) objects. We observe
that also for pre-stellar cores, the line widths are larger for more massive objects.
From these histograms and following the previous discussion, we conclude that the broadening of 
the line profile from pre-stellar 
cores to protostars is due to non-thermal motions rather than thermal increase. Therefore, non-thermal 
processes (turbulence or infall motions) are crucial during the evolution of these 
objects and these motions increase with mass.
\placeFigPreLM


\subsection{ CO and dynamics: turbulence versus outflow}\label{Origin}

$^{12}$CO and C$^{18}$O spectra trace different physical structures 
originating close to the protostar (e.g. \citealt{Yildiz12}). 
The broad wings of the $^{12}$CO~$J$=10--9 and $J$=3--2 data are optically thin and 
trace fast-moving gas, that is, emission from entrained outflow material.
On the other hand, the narrow C$^{18}$O spectra probe the turbulent and infalling material 
in the protostellar envelope. The relationship between these two different components is still poorly understood. 

Following the discussion in $\S$~\ref{BroadvsNarrow}, we compare the
FWHM of the narrow component as traced by C$^{18}$O~$J$=9--8 with the FWHM of the broad velocity component
as traced by $^{12}$CO~$J$=10--9 or $J$=3--2 
for the sources detected in both transitions 
(see Fig.~\ref{fig:CorrelationDbDvDvC18O}).
The C$^{18}$O~$J$=9--8 data were chosen for this comparison because this transition 
has been observed for the entire sample of YSOs, in contrast to C$^{18}$O~$J$=5--4 and $J$=10--9 transitions.
A correlation is found (with a Pearson correlation coefficient $r >$0.6), 
indicating a relationship between the fast outflowing gas and 
the quiescent envelope material.

Considering the scenario in which the non-thermal component is dominated by turbulence,
the relation presented in Fig.~\ref{fig:CorrelationDbDvDvC18O} indicates that 
the increase of the velocity of the outflowing gas corresponds to an increase in the 
turbulence in the envelope material and this relation holds across the entire luminosity range. 
One option is that stronger outflows are injecting larger-scale movements into the envelope, 
which increases the turbulence. This effect is reflected as a broadening
of the C$^{18}$O line width. In addition, for the low-mass sources the width of the C$^{18}$O~$J$=9--8 lines is
larger than for the $J$=3--2 spectra, indicating that the hotter inner regions of the envelope are
more turbulent than its cooler outer parts.
In the case of the high-mass object, the FWHMs of the C$^{18}$O~$J$=9--8 and $J$=3--2 lines are comparable,
which could be partly caused because more luminous YSOs tend to be created in more massive and 
more turbulent molecular clouds \citep{Wang09}. 

Alternatively, infall processes could broaden the C$^{18}$O profiles. 
Indeed, an increase in FWHM by at least 50\% is 
found for C$^{18}$O~$J$=9--8 compared with $J$=3--2 in collapsing envelope 
simulations due to the higher infall velocities in the inner warm 
envelope \citep{Harsono13}. 
If the non-thermal component was dominated by these movements, 
the sources with larger infall rates should have broader C$^{18}$O line widths.
These objects generally have larger outflow rates, i.e., higher outflow activity, 
(see \citealt{Tomisaka98}; \citealt{Behrend&Maeder01}) which shows up as a broadening of the
$^{12}$CO wings since the amount of material injected into the outflow is larger. 
Therefore, we will observe the same result as in the previous scenario,
that is, a broadening of the $^{12}$CO line wings for sources with stronger infalling motions. This relation 
would hold across the studied luminosity range.

Theoretically, large differences in the dynamics of the outflow-envelope 
system between low- and high-mass are expected since the same physical 
processes are not necessarily playing the same relevant roles in these different types of YSOs. 
However, Fig.~\ref{fig:CorrelationDbDvDvC18O} shows
that the dynamics of the outflow and envelope are equally 
linked for the studied sample of protostars.
With the current analysis we cannot disentangle 
which is the dominant motion in the 
non-thermal component of the line profile and thus, we cannot conclude if the infalling 
process makes the outflow stronger 
or if the outflow drives turbulence back into the envelope or a combined effect is at play. 

\placeFigCorrelationDeltaVBandDeltaVCdieciocho

\subsection{ High-$J$ CO as a dense gas tracer}\label{SFT}

The strictly linear relationships between CO high-$J$ line luminosity and
bolometric luminosity presented in Figs.~\ref{fig:Correlation12CO} to 
\ref{fig:Correlation1213CO} require further
discussion. The bolometric luminosity of embedded protostars is
thought to be powered by accretion onto the growing star and is thus a
measure of the mass accretion rate. A well-known relation of
bolometric luminosity with the outflow momentum flux as measured from
$^{12}$CO low-$J$ maps has been found across the stellar mass range
\citep{Lada95,Bontemps96,Richer00}, so one natural explanation for our
observed $^{12}$CO~$J$=10--9 correlation is that it reflects this same
relation. However, the outflow wings contain only a fraction of the
$^{12}$CO~$J$=10--9 emission, with the broad/narrow intensity ratios varying
from source to source (Table~\ref{table:12CO10-9-W-Tpeak}, see also \citealt{Yildiz13sub}, for
the low-mass sources). Together with the fact that the relations hold
equally well for $^{13}$CO and C$^{18}$O high-$J$ lines, this suggests the
presence of another underlying relation. Given the strong correlation
with $M_{\rm{env}}$ (Fig.~\ref{fig:CorrelationC18OMenv}), the most likely 
explanation is that the high-$J$ 
CO lines of all isotopologues trace primarily the amount of dense gas 
associated with the YSOs.

Can this relation be extended to larger scales than those probed here?
Wu et al.\,(2005, 2010) found a linear relation between HCN
integrated intensity and far-infrared luminosity for a set of galactic 
high-mass star forming regions on similar spatial scales as probed by
our data. They have extended this relation to include extragalactic
sources to show that this linear regime extends to the scales of
entire galaxies, as first demonstrated by \cite{Gao04}. In
contrast, the CO~$J$=1--0 line shows a superlinear relation with
far-infrared luminosity (sometimes converted into star formation rate)
and with the total (H I + H$_2$) gas surface density with a power-law
index of 1.4 (see \citealt{Kennicutt12} for review). The data presented in this paper 
suggest that it is the mass of dense molecular gas as traced by HCN
that controls the relation on large scales rather than the mass traced
by CO~$J$=1--0 (see also \citealt{Lada12}).

With the increased number of detections of $^{12}$CO~$J$=10--9 in 
local and high redshift galaxies with {\it Herschel}
\citep[e.g.,][]{vanderWerf10,Spinoglio12, Kamenetzky12, Meijerink13} and millimeter
interferometers \citep[e.g.,][]{Weiss07,Scott11}, the question arises whether
this line can serve as an equally good tracer of dense gas.  As an
initial test, we present in Fig.~\ref{fig:Correlationextra} our  
\mbox{$^{12}$CO~$J$=10--9 - $L_{\rm{FIR}}$} (assuming
$L_{\rm{bol}}=L_{\rm{FIR}}$) relation with these recent extragalactic detections
included. As can be seen, the relation does indeed 
extend to larger scales. Given the small number statistics, 
the correlation has only limited meaning and this relation
needs to be confirmed by additional data.

An alternative view has been presented by \cite{Krumholz07}
and \cite{Narayanan08} who argue that the linearity for dense gas
tracers is a coincidence resulting from the fact that these higher
excitation lines are subthermally excited and probe only a small
fraction of the total gas. They suggest that the relation should
change with higher critical density tracers and even become sublinear
with a power-law index less than 0.5 for transitions higher than $J$=7--6
for CO. On the small scales probed by our data, we do not see this
effect and the linear relations clearly continue to hold up to the
$J$=10--9 transition.

\placeFigCorrelationCOextragalactic



\section{Conclusions}\label{Conclusions}

The analysis of the $^{12}$CO~$J$=3--2, $J$=10--9, $^{13}$CO~$J$=10--9, C$^{18}$O~$J$=3--2, $J$=5--4, 
$J$=9--8 and $J$=10--9 line profiles allows us to study several fundamental parameters of 
the emission line, such as the line width and the line luminosity, and to constrain the 
dynamics of different physical 
structures of WISH protostars across a wide range of luminosities. Complementing the HIFI data, 
lower-$J$ observations from the JCMT are included in order to achieve a uniform picture of the
interaction of YSOs with their immediate surroundings. 
Our results are summarised as follows:  

\begin{itemize}
  \renewcommand{\labelitemi}{$\bullet$}
  \item A gallery of line profiles identified in the HIFI CO spectra is presented.
    $^{12}$CO and $^{13}$CO~$J$=10--9 line profiles can be decomposed into two
    different velocity components, where the broader component is thought to trace the entrained outflowing
    gas. This broad component weakens from the low-mass Class~0 to Class~I stage. 
    Meanwhile, the narrow C$^{18}$O lines probe  
    the bulk of the quiescent envelope material. The widths are dominated by non-thermal motions including  
    turbulence and infall. The next step will be to constrain the contribution of 
    these non-thermal mechanisms on the C$^{18}$O line profiles by using radiative
    transfer codes such as RATRAN \citep{Hogerheijde00}.

  \item The narrow C$^{18}$O~$J$=9--8 line widths increase from low- to high-mass YSOs. Moreover, 
    for low- and intermediate-mass protostars, they are about twice the width of the C$^{18}$O~$J$=3--2 
    lines, suggesting increased turbulence/faster infall in the warmer inner envelope compared to the
    cooler outer envelope. 
    For high-mass objects the widths of the $J$=9--8 and $J$=3--2 lines are comparable, 
    suggesting that the molecular clouds in which these luminous YSOs form are more turbulent. 
    Extending the line width analysis to pre-stellar cores, a 
    broadening of the line profile is observed from these objects to protostars
    caused by non-thermal processes.
    
  \item A correlation is found between the width of the $^{12}$CO~$J$=3--2 and $J$=10--9 
    broad velocity component and  
    the width of the C$^{18}$O~$J$=9--8 profile. This suggests a link between entrained outflowing gas
    and envelope motions (turbulence and/or infalling) which holds from low- to high-mass 
    YSOs. This means that the
    interaction and effect of outflowing gas and  
    envelope material is the same across the studied luminosity range, indicative of the existence of an 
    underlying common physical mechanism which is independent of the source mass.
  
  \item A strong linear correlation is found between the logarithm of the line and 
    bolometric luminosities across six orders of magnitude on both axes, for all lines and isotopologues. 
    This correlation is also found between the logarithm of the line luminosity and envelope mass.
    This indicates that high-$J$ CO transitions (up to $J$=10--9) can be used as a dense gas tracer, 
    a relation that can be extended to larger scales (local and high redshift galaxies).

\end{itemize}


\begin{acknowledgements} 
The authors are grateful to the external referee Andr{\'e}s Guzm{\'a}n
for his careful and detailed report and to the editor Malcom Walmsley for his useful last comments.
These two reports helped to improve considerably this manuscript. 
We would also like to thank the WISH team for many inspiring
discussions, in particular the WISH internal referees Gina Santangelo
and Luis Chavarr{\'i}a. 
Astrochemistry in Leiden is supported by the Netherlands Research
School for Astronomy (NOVA), by a Spinoza grant and grant 614.001.008
from the Netherlands Organisation for Scientific Research (NWO), and
by the European Community's Seventh Framework Programme FP7/2007-2013
under grant agreement 238258 (LASSIE).
We would like to thanks the JCMT, which is operated by The 
Joint Astronomy Centre on Behalf of the Science and Technology Facilities 
Council of the United Kingdom, the Netherlands Organisation for Scientific 
Research, and the National Research Council of Canada.
This research used the facilities of the Canadian Astronomy 
Data Centre operated by the National Research Council of Canada with 
support of the Canadian Space agency.
HIFI has been designed and built by a consortium of 
institutes and university departments from across Europe, Canada and the 
United States under the leadership of SRON Netherlands Institute for Space
Research, Groningen, The Netherlands and with major contributions from 
Germany, France and the US. Consortium members are: Canada: CSA, 
U.Waterloo; France: CESR, LAB, LERMA, IRAM; Germany: KOSMA, 
MPIfR, MPS; Ireland, NUI Maynooth; Italy: ASI, IFSI-INAF, Osservatorio 
Astrofisico di Arcetri- INAF; Netherlands: SRON, TUD; Poland: CAMK, CBK; 
Spain: Observatorio Astron{\'o}mico Nacional (IGN), Centro de Astrobiolog{\'i}a 
(CSIC-INTA). Sweden: Chalmers University of Technology - MC2, RSS $\&$ 
GARD; Onsala Space Observatory; Swedish National Space Board, Stockholm 
University - Stockholm Observatory; Switzerland: ETH Zurich, FHNW; USA: 
Caltech, JPL, NHSC.
\end{acknowledgements}

\bibliographystyle{aa} 
\bibliography{bibdata}

\Online
\appendix

\section{ Characterisation of the HIFI data}\label{Data}

The main characteristics of the HIFI data, together with the spectra are presented in this appendix. 
The description of the observed lines focuses first on $^{12}$CO~$J$=10--9 
(Section~\ref{12CO (10--9)}), where the main characteristics 
for the low-, intermediate- and high-mass sources are listed in this order. 
Next we discuss the $^{13}$CO~$J$=10--9 spectra (Section~\ref{13CO (10--9)}), following the same structure,
and finally all observed C$^{18}$O lines (Section~\ref{C18O}) are presented.

\subsection{$^{12}$CO~$J$=10--9 line profiles}\label{12CO (10--9)}

The $^{12}$CO~$J$=10--9 line was observed for the entire sample of low- and intermediate-mass YSOs and
for one high-mass object (W3-IRS5). All the observed sources were detected and 
the emission profiles are the 
strongest and broadest among the targeted HIFI CO lines (see Fig.~\ref{fig:Spec12CO10-9} and 
Table~\ref{table:12CO10-9-W-Tpeak} for further information).
Within the low-mass Class~0 sample, the main beam peak temperature, 
$T_{\rm{MB}}^{\rm{peak}}$, ranges from 0.8 to 8.1~K. 
As indicated in Section~\ref{Decomposition}, the emission in some
sources (around 73~\% of the detected lines of this sub-group)
can be decomposed into two velocity components. The FWHM of the
narrower component varies from 2.3~km\,s$^{-1}$ to 9.3~km\,s$^{-1}$,
while the width of the broad component shows a larger variation, 
from 8.3~km\,s$^{-1}$ to 41.0~km\,s$^{-1}$.
For the low-mass Class~I YSOs, similar intensity ranges are found, 
with $T_{\rm{MB}}^{\rm{peak}}$ varying from $\sim$0.6 to 10.2~K.
The Class~I protostars present narrower emission lines than the Class~0 sources and
only the 27~\% of the emission line profiles can be decomposed
into two velocity components. The narrow component ranges from 
1.8 to 5.1~km\,s$^{-1}$ and the broad component varies from 10.0 to 16.1~km\,s$^{-1}$.

For the intermediate-mass protostars, the intensity increases and varies from $T_{\rm{MB}}^{\rm{peak}}$=2.4
to $\sim$28.0~K. The profiles are broader and the distinction 
between the two different components is clearer than for the low-mass sources, so all
profiles can be fitted with two Gaussian functions.
The FWHM of the two identified velocity components varies from 2.7 to 
7.6~km\,s$^{-1}$ for the narrow component and from 15.6 to 24.3~km\,s$^{-1}$
for the broad component.

Only W3-IRS5 was observed in $^{12}$CO~$J$=10--9 from the high-mass sample. 
The spectrum is presented in Appendix~\ref{JCMT_Data}, Fig.~\ref{fig:HIFIvsJCMT}, 
together with other lines of this source. 
The profile is more intense than any 
of the low- and intermediate-mass sources 
($T_{\rm{MB}}^{\rm{peak}}$=48.5~K) and has the largest FWHM: 8.4~km\,s$^{-1}$ for 
the narrower component and 28.6~km\,s$^{-1}$ for the broad component.

Self-absorption features have been detected in 5 out of 33 observed 
$^{12}$CO~$J$=10--9 emission lines 
(the sources are indicated in Table~\ref{table:12CO10-9-W-Tpeak}).
However, these features are weak and 
of the order of the rms of the spectrum, so no Gaussian profile has been fitted.
No specific symmetry can be determined, that is, there is no systematic shift in the emission of the broad
component relative to the source velocity (see Fig.~\ref{fig:Spec12CO10-9} for comparison). 
Overall, the data do not show any infall signature.
\placeSpectraCO
\placeTableCOWTpeak

\subsection{$^{13}$CO~$J$=10--9 line profiles}\label{13CO (10--9)}

The observed $^{13}$CO~$J$=10--9 emission lines for the low- and intermediate-mass sources are 
less intense, narrower and have a lower \textit{S/N} than the $^{12}$CO~$J$=10--9 spectra 
(Fig.~\ref{fig:Spec13CO10-9}). Table~\ref{table:13CO10-9-W-Tpeak} contains the 
parameters obtained from the one or two Gaussian fit to the detected line profiles.
In the case of the low-mass Class~0 spectra, three sources are not detected down to 17~mK rms
in 0.27~km\,s$^{-1}$ bins
and the profile of only two sources (Ser\,SMM1 and NGC1333\,IRAS4A) can be decomposed into two 
different velocity components. 
The $T_{\rm{MB}}^{\rm{peak}}$ ranges from 0.05 to 0.8~K and the FWHM of the narrow 
profiles varies from 0.7 to 6.8~km\,s$^{-1}$, with the largest values corresponding to the
broad velocity component being 13.2~km\,s$^{-1}$ for NGC1333\,IRAS4A.
In the case of the Class~I sample, four sources are not detected and
none of the emission line profiles can be decomposed into two velocity components.
The averaged intensity is lower than for the Class~0 objects, ranging from $T_{\rm{MB}}^{\rm{peak}}$=0.05
to 0.52~K. The value of the line width also drops and the interval varies from 
1.5 to 7.3~km\,s$^{-1}$.

For the intermediate-mass YSOs, a better characterisation of the line profile is possible since the lines 
are stronger and have higher \textit{S/N} than the low-mass objects with $T_{\rm{MB}}^{\rm{peak}}$
ranging from 0.1 to 2.7~K.
Compared to the $^{12}$CO~$J$=10--9 profiles, the $^{13}$CO~$J$=10--9 lines are more symmetric and 
only the emission profile of one source can be decomposed into two different 
Gaussian components. Regarding the FWHM of the lines fitted by the narrow Gaussian, 
the interval goes from 4.3 to 6.1~km\,s$^{-1}$.

Around 63$\%$ of the detected $^{13}$CO~$J$=10--9 emission lines (12 out of 19) for the high-mass YSOs can be 
decomposed into two distinct 
velocity components, whereas the decomposition of the profiles is only possible for 10\% of the 
detected low-mass objects (2 out of 20) and for $\sim$17\% of the detected intermediate-mass YSOs (1 out of 6). 
The reason for the lower percentage recorded for the low- and intermediate-mass sources could be 
caused by a lower \textit{S/N} compared with the bright 
high-mass sources. The weakest line from the high-mass sample has a $T_{\rm{MB}}^{\rm{peak}}$ of 0.7~K 
and the most intense a $T_{\rm{MB}}^{\rm{peak}}$ of 20.8~K. 
The FWHM of the narrower component varies from 3.3~km\,s$^{-1}$ to 7.2~km\,s$^{-1}$.
The width of the broad component presents a larger variation since the minimum value is 8.7~km\,s$^{-1}$ 
and the maximum 21.9~km\,s$^{-1}$. This component appears
either red- or blue-shifted.
There is no significant trend with
evolution stage as probed by the presence of IR-brightness or
ionising radiation (Fig.~\ref{fig:Spec13CO10-9}).
\placeSpectratreceCO
\placeTabletreceCOWTpeak

\subsection{C$^{18}$O line profiles}\label{C18O}

Three transitions of C$^{18}$O were obtained within  
WISH together with water observations: \mbox{C$^{18}$O~$J$=5--4}, \mbox{~$J$=9--8} and \mbox{~$J$=10--9}. 
Only Class~0 and intermediate-mass YSOs were observed in \mbox{C$^{18}$O~$J$=5--4},
tracing regions with an upper energy level of $\sim$79~K (see Fig.~\ref{fig:SpecC18O5-4}). 
This line is obtained in parallel with
a deep integration on the 548~GHz H$_{2}^{18}$O $1_{10}-1_{01}$ transition for 19 sources. Thus,
the spectra have very high \textit{S/N} with an rms of 9~mK for 
low-mass Class~0 sources and less than 20~mK for intermediate-mass YSOs in 0.27~km\,s$^{-1}$ bins.   
The main characteristic of this transition is the narrow profile seen in all 
the emission lines for the narrower component, with a FWHM of less than 2.0~km\,s$^{-1}$ for the 
low-mass sources, and 3.7~km\,s$^{-1}$ for the intermediate-mass objects. 
In addition, other features are detected thanks to the high \textit{S/N}, e.g. 
a weak broad velocity component for the low-mass objects NGC1333\,IRAS\,4A \citep{Yildiz10}, 
L483, Ser\,SMM1 and Ser\,SMM4. This component is also identified in the C$^{18}$O~$J$=5--4 line 
for the intermediate-mass sources NGC2071 (see Fig.~\ref{fig:c18o}) and Vela\,IRS19.
The values of the single or two Gaussian fit
of these lines are presented in Table \ref{table:C18O5-4-W-Tpeak}.
\placeSpectraCochoOcinco
\placeTableCochoOcincoWTpeak

For the $J$=9--8 transition, $\sim$55\% of the observed lines are detected, 
probably due to the lower $S/N$ caused by shorter exposure times than for the $J$=5--4 line.
The lines are detected in 5 out of 26 low-mass sources; 4 out of 6 intermediate-mass YSOs; 
and in all 19 high-mass protostars (see Fig.~\ref{fig:SpecC18O9-8}). 
The C$^{18}$O~$J$=9--8 emission lines appear weak with median 
$T_{\rm{MB}}^{\rm{peak}}$ values of 0.10, 0.14 and 0.83~K for the
low-, intermediate- and high-mass objects, respectively. 
Most of the emission line profiles 
of C$^{18}$O~$J$=9--8 can be fitted by a single Gaussian with 
a FWHM from 2.0~km\,s$^{-1}$ to 3.9~km\,s$^{-1}$ for the low-mass objects; from 2.8~km\,s$^{-1}$ to 
5.4~km\,s$^{-1}$ for the intermediate-mass sources; and from 3.1 to 6.4~km\,s$^{-1}$ 
for the high-mass YSOs (values summarised in Table~\ref{table:C18O9-8-W-Tpeak}). 
Only a two Gaussian decomposition has been performed for three ultra-compact H{\small II}
regions (G10.47+0.03, W51N-e1 and G5.89-0.39). 
For these objects, the broad velocity components are larger than 16~km\,s$^{-1}$, as 
is shown in Fig.~\ref{fig:SpecC18O9-8}. 
\placeSpectraCochoOnueve
\placeTableCochoOnueveWTpeak

Finally, the C$^{18}$O~$J$=10--9 transition was observed in 30 minute
exposures for all low-mass Class~0 protostars, one intermediate-mass
source and all high-mass YSOs.   
Additional deeper integrations of 300 minutes
were obtained for NGC1333\,IRAS\,2A (as part of WISH) and for
NGC1333\,IRAS\,4A, 
NGC1333\,IRAS\,4B, Elias\,29 and GSS\,30\,IRS1 (as part of open-time
programme OT2{\_}rvisser{\_}2) in parallel with deep H$_2^{18}$O searches.
The line was detected in five low-mass sources and in 
all 19 high-mass objects (Fig.~\ref{fig:SpecC18O10-9}). 
This line appears close to the 1097 GHz H$_2$O 3$_{12}-3_{03}$ transition.
For most of the high-mass objects, the line profile of this water transition shows broad wings which extend few km\,s$^{-1}$,
so the C$^{18}$O~$J$=10--9 emission line is found on top of the broad water red wing.
In order to properly analyse the emission of this CO isotopologue, the line wings of the 1097 GHz water
transition were fitted with a Gaussian profile, subtracted and the residuals plotted. With this method,
the C$^{18}$O~$J$=10--9 emission line for the high-mass sample has been isolated. 
The temperature of the gas that $J$=10--9 traces is likely similar to 
that traced by the $J$=9--8 transition, so the lines are also weak with median values of 
$T_{\rm{MB}}^{\rm{peak}}$ of 0.04~K for the low-mass protostars and
0.52~K for the high-mass objects.
The FWHM of the one single Gaussian profile which fits these lines are slightly larger, ranging 
from 3.4 to 7.5~km\,s$^{-1}$ for the high-mass sources (Table~\ref{table:C18O10-9-W-Tpeak} for more details). 
\placeSpectraCochoOdiez
\placeTableCochoOdiezWTpeak

\section{ JCMT data}\label{JCMT_Data}

The central spectrum of the $^{12}$CO and C$^{18}$O~$J$=3--2 spectral maps observed with 
the HARP instrument of the JCMT are presented in this section 
(see Figs.~\ref{fig:Spec12CO3-2} and \ref{fig:SpecC18O3-2}).
These central spectra were convolved to a 20$\arcsec$ beam
in order to compare them with the HIFI data.
The $^{12}$CO and C$^{18}$O~$J$=3--2 spectra for 
the low-mass sources BHR\,71, Ced110\,IRS4, IRAS\,12496 and HH\,46 
were observed with APEX because of their low declination. 
Similarly to the JCMT data, the central spectrum was convolved to a 20$\arcsec$ beam.
See \citet{Yildiz13sub} for more information about the low-mass protostar observations. 

The spectral maps are not presented in this paper because
only the values of FWHM from the broad component for the $^{12}$CO~$J$=3--2
central spectrum were used in the analysis and discussion, 
together with the width and integrated intensity of the C$^{18}$O~$J$=3--2 data. 
These values are presented in Table~\ref{tbl:jcmt_data}.
Some spectra from these species are also plotted together with the HIFI data in the figures shown in this 
appendix in order to make a direct comparison of the lines for different types of protostars.

The $^{12}$CO~$J$=3--2 data show more complex line profiles than the $J$=10--9, as indicated in 
Section~\ref{Line-profiles}, 
with intense self-absorption features and broad velocity components 
(see spectra from Fig.~\ref{fig:Spec12CO3-2}).
More than 82\% of the observed and detected lines (39 out of 47) present a broad velocity 
component, and the FWHM ranges from 7.4
to 53.3~km\,s$^{-1}$, values corresponding to a low- and high-mass YSOs respectively. 

On the other hand, the narrow C$^{18}$O~$J$=3--2 spectra show single Gaussian emission profiles, 
similar to those from higher-$J$ transitions observed with HIFI. The FWHM of these data varies from
0.6 to 7.3~km\,s$^{-1}$. The spectra is presented in Fig.~\ref{fig:SpecC18O3-2} and the constrained values
of FWHM and integrated intensity in Table~\ref{tbl:jcmt_data}.
\placeFigHIFIuvJCMTLM

\placeFigHIFIuvJCMTIM

\placeFigHIFIuvJCMT

\placeTableJCMT

\placeSpectraCOJCMT

\placeSpectraCochoOnueveJCMT

\end{document}